%% file: thesis.tex
\documentclass[12pt,a4wide,psfig]{report}

\usepackage{psfig} 
\begin{document}
\newcommand{\be}{\begin{equation}}
\newcommand{\ee}{\end{equation}}
\newcommand{\bea}{\begin{eqnarray}}
\newcommand{\eea}{\end{eqnarray}}
\def\atridot{\stackrel{...}{a}}
\input{cover}

\input{certificate}

\input{acknow}

\pagestyle{myheadings}
\pagenumbering{roman}
\tableofcontents
\newpage
%\listoffigures
\addcontentsline{toc}{part}{\normalsize{Preface}}
\include{preface}

\newpage
\pagenumbering{arabic}
\chapter{Introduction}
\markright{Introduction}
\newpage
The universe consists of everything that we can see through our 
best gadgets, have seen in the past and expect to see in any intelligible 
future. Cosmology deals with the physics of this most exhaustive collection 
of objects as a whole. Cosmology is concerned with the formation, the 
evolution, the future of the universe made of billions of galaxies spread 
over billions of light years. 
\par Cosmology really came of age as a science after the advent of general 
relativity in 1915 which made possible a systematic theoretical modelling of 
the universe and Hubble's observation in 1929 that the universe is in fact 
evolving and hence interesting as a physical science. Like all other branches 
of physics, cosmology is also an observational science, but until very 
recently, the data available had been limited and that too had been plagued by 
the lack of precision. The dramatic change in the scenario 
started with the COSMIC 
BACKGROUND EXPLORER (COBE) \cite{cobe1, cobe2, cobe3, cobe4} and 
over the past decade there had been an explosion 
of really high precision data, such as those from WILKINSON MICROWAVE 
ANISOTROPY PROBE (WMAP) \cite{wmap1, wmap2, wmap3, wmap4, dns}. 
One inherent problem in cosmology, that one cannot 
repeat experiments, will remain for ever, one cannot ask the universe to 
evolve afresh with various initial conditions. But the data available on the 
existing system holds the key to dictate the direction of research in this 
branch. 
\par Cosmology as a subject is as vast as the universe, 
and application of 
every branch of physics is indeed warranted. The recent advances in 
observational cosmology has led to many exciting discoveries and possibilities 
regarding the evolution of the universe, but arguably the most exciting 
and puzzling amongst them is that the universe now is expanding with an 
acceleration defying the properties of the known matter content  of the 
universe. For a systematic and lucid review of the observational results and 
their interpretations, we refer to the work by L. Perivolaropoulos \cite{peri}.
The present thesis endeavours to look at this problem from a 
very narrow angle.
\par As already mentioned, this acceleration is counter intuitive as gravity, which is 
the deciding interaction in the governance of the dynamics of the universe 
as a whole, is always attractive. Gravity is by far the weakest amongst 
the four basic interactions 
of nature, but the strong and weak interactions are short range and have hardly
 anything to do with the dynamics of the universe at a large scale, while 
electromagnetic interaction has almost no impact in view of the charge 
neutrality of the universe. So the galaxies should attract each other, and even
if the universe expands as suggested by the observations as well as the 
standard models of cosmology, the rate of expansion should be decreasing.
\vskip .2in
\section{Standard Cosmological Model :}
\markright{Standard Cosmological Model}
For an observationally realistic and logically viable model of 
the universe, one makes the following assumptions:
\\(1) General Relativity (GR) correctly describes gravity.
\\
(2) Cosmological principle is valid, i.e, universe on a large scale ($ > 10^6$ 
light years) is spatially homogeneous and isotropic. 
The typical size of a galaxy is 
$\sim ~10^5$ light years, an order of magnitude beyond which  
there is no preferred position or direction in the universe.\\
(3) Hydrodynamic approximation :  According to which the basic building 
blocks of the universe are galaxies and their distribution can be considered 
as a fluid distribution.

\par The most general metric satisfying the homogeneity and isotropy of the 
universe is given by 
\be 
ds^2 = dt^2 - a^2(t) [ \frac{dr^2}{1-kr^2} + r^2 d\theta^2 + r^2 \sin^2\theta 
                      d\phi^2]~,
\ee
where $t$ is the time and $r, \theta, \phi$ are space co-ordinates, $a(t)$ is 
the scale factor of the universe which gives the expansion history of 
the universe and $k$ is called the curvature index. This metric is called the 
Friedmann-Robertson-Walker (FRW) metric.

\par With the hydrodynamic approximation that the matter distribution of the 
universe can be approximated to be a perfect fluid, the energy 
momentum tensor for a perfect fluid distribution is taken as 
\be
T_{\mu\nu} = (\rho + p)v_{\mu}v_{\nu} - pg_{\mu\nu}~,
\ee
where $v_{\mu}$'s are the components of the fluid velocity vector, 
$\rho$ is the energy density 
and $p$ is the isotropic pressure of the perfect fluid.\\
With this input, the Einstein equations
\be 
G_{\mu\nu} = 8\pi G T_{\mu\nu}
\ee
lead to the differential field equations
\be\label{eqn:e1}
3\frac{ \dot{a}^2}{a^2} + 3\frac{k}{a^2} =  8 \pi G \rho,
\ee
\be\label{eqn:e2}
2\frac{\ddot{a}}{a} + \frac{ \dot{a}^2 + k}{a^2} = - 8 \pi G p,
\ee
where a dot denotes differentiation w.r.t. the cosmic time $t$. \\
Also we obtain a third equation, called the matter conservation equation, of 
the form
\be\label{eqn:consv}
\dot{\rho} + 3 \frac{\dot{a}}{a}(\rho + p) = 0~,
\ee
which is not an independent equation but can be derived from the two 
field equations or from the Bianchi identities. It deserves mention that the 
curvature index $k$ can not be determined from the field equations, 
and is rather put in by hand and can take values 0, +1 and -1 which 
correspond to flat, closed and open universes respectively.
\\ This system of equations can not be solved completely as there are 
three unknowns $a$, $\rho$ and $p$ and only two independent equations. So 
there is the need of a third equation for solving the system completely which 
is provided by the equation of state connecting the density and pressure 
of the cosmic fluid as 
\be
w = \frac{p}{\rho}~.
\ee

Normally this equation of state is taken as that of a barotropic fluid, i.e, 
$w$ is a constant. Thus, at present when the universe is expected to be 
matter dominated where 
there is no pressure ($p = 0$), $w = 0$. Similarly, at very early epoch, 
when the universe was very hot and was 
dominated by radiation, then $p = \frac{1}{3}\rho$, i.e, $w = \frac{1}{3}$. 
So, from these set of equations, $a(t)$ can be solved for and thus one 
can get an idea about the evolution of the universe.
\par To relate this FRW model with the observations, some useful parameters 
are defined.
\\
(i) Hubble constant $H$ is defined as
\be
H = \frac{\dot{a}}{a}
\ee
which is an observable parameter and gives the expansion rate of the universe. 
This parameter has the dimension of (time)$^{-1}$.\\

(ii) Although the universe is expanding, it is dominated by the gravitational 
interaction which gives rise to an attractive force. So this expansion 
is expected to be decelerated. To express this deceleration, a dimensionless 
deceleration parameter $q$ is defined as 
\be\label{eqn:qdef}
q = - \frac{\ddot{a}/a}{\dot{a}^{2}/a^2}~;
\ee

$q > 0$ indicates that $\frac{\ddot{a}}{a}$ is negative, i.e, universe 
is decelerating and $q < 0$ indicates that $\frac{\ddot{a}}{a}$ is positive, 
i.e, universe is accelerating.\\

(iii) A jerk parameter $r$ is defined as 
\be
r = \frac{{\stackrel{...}{a}}/ a}{{\dot{a}}^3 / a^3}~,
\ee
which is a measure of the rate of change of $q$. As present observations 
facilitate the study of the evolution of the deceleration parameter $q$, 
this jerk parameter has become useful. Sahni et al \cite{alam1} introduced 
 a pair of ``\emph{statefinder parameters}'' to characterize the quintessence 
models, particularly the interacting models. In addition to $r$, the other 
parameter $s$ is defined as 
\be
s = \frac{r - 1}{3(q - \frac{1}{2})}~.
\ee

(iv) The density of the universe is expressed in a dimensionless form by 
defining a density parameter given by
\be
\Omega = \frac{\rho}{\rho_{c}}~,
\ee
where $\rho_{c} = \frac{3 H^2}{8 \pi G}$ is called the critical density or 
closure density of the universe.
\par In terms of these parameters, equations (\ref{eqn:e1}) and (\ref{eqn:e2}) 
can be expressed as 
\be\label{eqn:eh1}
1 + \frac{k}{a^2 H^2} = \frac{\rho}{\rho_{c}} = \Omega~,
\ee
\be\label{eqn:eh2}
H^2 (1 - 2q) + \frac{k}{a^2} = -8 \pi G p~.
\ee
Now, if one considers a matter dominated universe where $p = 0$, equations
(\ref{eqn:eh1}) and (\ref{eqn:eh2}) leads to three possibilities :
\begin{figure}[!h]
\centerline{\psfig{figure=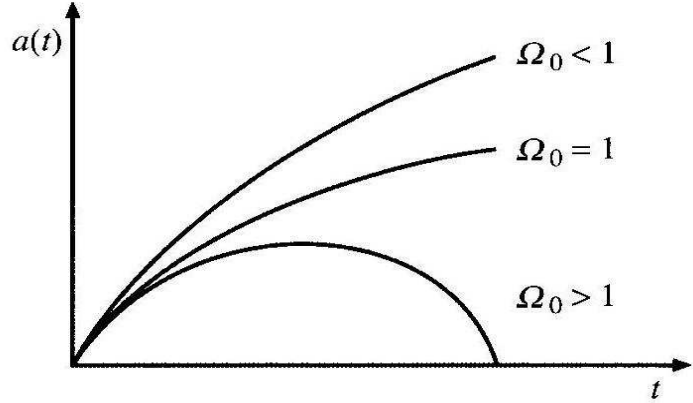,height=75mm,width=100mm}}
\caption{\normalsize{\em Evolution of the scale factor with time for open, flat and closed models}}
\label{fig1}
\end{figure}
\\
(a) When $k = -1$, $q < \frac{1}{2}$ and $\Omega < 1$, i.e, 
$\rho < \rho_{c}$. This means the matter density being less than the critical 
density, the universe will go on expanding for ever. Such models are called 
open universe models.
\\(b) When $k = +1$, $q > \frac{1}{2}$ and $\Omega > 1$, i.e,
$\rho > \rho_{c}$. So, the matter density being higher than the critical 
density, the universe will expand upto some maximum volume and then due to 
gravitational attraction it will re-collapse. Such models of the universe are 
called closed universe models.
\\(c) When $k = 0$, $q = \frac{1}{2}$ and $\Omega = 1$, i.e,
$\rho = \rho_{c}$, one has the limiting case between the above two. This is 
called a flat model as the space section has zero curvature.
\par In 1929, Hubble made a remarkable discovery regarding the motion of 
the galaxies. He observed that the galaxies are moving away from each other, 
and the velocity of separation $v$ between two galaxies is proportional to the 
distance $D$ between them, i.e, 
\begin{center}
$~~~~v ~\infty ~D~,$
\end{center}
\be
or,~ v = H D~,
\ee
where $H = \frac{\dot{a}}{a}$ is the Hubble constant defined earlier. This 
means the galaxies were closer together earlier. So, by tracking back one can 
arrive at a time when all the matter were concentrated at a single point. 
At that instant universe had a zero volume and an infinite density. That 
epoch, when $a(t) = 0$ and $H \rightarrow \infty$ corresponds to some 
violent activity and is given the name \emph{Big Bang Singularity}. The 
existence of singularity in a theory is an unwanted feature as laws of physics 
break down and naturally the features cannot be explained. But still the 
Big Bang model enjoys the status of a preferred theory as it has its own 
success stories :
\\(i) Can predict He abundance :~ One of the fundamental problems of cosmology 
     is to explain the primary creation of matter and to understand the 
observed abundances of different elements. At the beginning, when all 
the matter-energy 
of the universe was concentrated in a tiny volume, the spectrum of particles 
that we see today was surely absent. Although the singular stage is definitely 
out of our purview of explanation, the Big Bang theory can in fact trace the 
history of the universe when its size was around $10^{-33}$ cm. This is 
clearly much shorter than the de Broglie wavelength of most of the particles 
that we see today. Following Gamow's seminal work \cite{gamow}, one can 
explain the nucleosynthesis process that took place during the radiation 
dominated era. Particularly the abundance of lighter elements like Helium 
are well explained in Big Bang theory and the theoretical prediction has a 
close semblance with the observational results.\\
(ii) Could predict the relic thermal radiation at $\sim {2.7}~^0$K. 
This Cosmic Microwave Background Radiation (CMBR) was 
detected later in 1965 by Penzias and Wilson \cite{penzias}. This detection 
confirmed the 
assumption of isotropy of the universe and is considered to be the greatest 
triumph of the Big Bang theory and is arguably the strongest pillar of modern 
cosmology.\\
(iii) Age of the universe : The Friedmann models could provide a formula for 
the age of the universe which goes as 
\begin{center}
$ t \sim \frac{2}{3H}$ for a flat matter dominated universe.
\end{center}
Thus knowing the value of $H_{0}$ ( the subscript `0' indicates the present 
time ), $t_{0}$ can be easily calculated. Following this method, the standard 
Big Bang theory predicts the correct order of magnitude of the age of 
the universe $\sim 1.5 ~X~ 10^{10}$ 
years.\\
(iv) Formation of galaxies : Although the universe is homogeneous at a 
large scale, there are clumps ( galaxies and clusters of galaxies ) around us. 
So at a smaller scale, 
there should be inhomogeneity. The universe starts homogeneous, and still 
looks homogeneous at a scale more than $10^6$ light years, but at a smaller 
scale must have inhomogeneities, i.e, structures like galaxies. In the 
purview of standard Big Bang cosmology, this can also be explained as it 
has been shown that if there is some kind of perturbation, it can indeed 
give rise to some growing mode so that galaxies are formed.
\vskip .2in
\section{Problems of Standard Cosmological Model :}
\markright{Problems of Standard Cosmological Model}
Standard Big Bang cosmology has its share of problems too. 
A few of them are :\\
(1) Horizon problem : Given the present size of the universe ($\sim 10^{10}$ 
light years), it is quite possible to consider two far separated points  
such that there is no causal connection between these two points, 
i.e, their light cones never intersect even if one traces back to the last 
scattering surface ( LSS ) when matter and radiation decoupled and the 
universe became transparent. But even then the two points carry the same 
information at present as the universe is homogeneous and isotropic. This 
is known as the `\emph{horizon problem}'.
\\(2) Flatness problem : From equation (\ref{eqn:eh1}), one can arrive at a relation 
\begin{center}
$\Omega - 1 = \frac{k}{a^2 H^2}$~. 
\end{center}
If it is considered that the initial conditions including the density 
parameter $\Omega$ were set during the GUT epoch when temperature of the 
universe was $\sim ~10^{15}$ GeV, then 
\begin{center}
$\Omega = 1 \pm \delta$, where $\delta < 10^{-50}~.$
\end{center}

This means the departure from the $\Omega = 1$ value has to be very small. Any 
relaxation from this fine-tuning would have led to a much higher (or lower)
value of $\Omega$ at present which is not obtained observationally.
\par This fine-tuned value of $\Omega \approx 1$ leads to a $k = 0$ 
model, i.e, a spatially flat model of the universe. Without this extreme fine  
tuning, the universe would have either collapsed back within a time scale 
of $10^{-35}$ s (in a closed model) or would have expanded at a much higher 
rate (in an open model) than observed at present. Standard Big Bang model 
can not explain why $\Omega$ is so closely tuned to $1$ and the problem is 
termed the `\emph{flatness problem}' or equivalently the `\emph{fine tuning 
problem}'.
\\
(3) The monopole problem : Gauge field theories suggest that whenever there 
is any symmetry breaking, inevitably some particles are created which have 
the characteristics of magnetic monopoles. So, it is expected that during the 
phase transition of the universe, some monopoles must have been created which 
being highly stable particles, should have been  observed at present 
epoch also. But, 
in practice monopoles are not observed. This is known as the `\emph{monopole 
problem}'.
\vskip .2in
\section{The Inflationary Paradigm :}
\markright{Inflation}
The solution to this problem was suggested by Alan Guth in 1981 
\cite{guth} by introducing the so called \emph{inflationary model} of 
the universe. 
In this model, Guth suggested that during a very early epoch the universe 
had a very rapid phase of accelerated expansion having quite a 
number of e-foldings 
of the volume in a short span of time. At that 
time universe was dominated by vacuum energy and the equation of state was 
of the form 
\begin{center}
$\rho_{vac} + p_{vac} = 0~,$
\end{center}
where $\rho_{vac}$ is the vacuum energy density and $p_{vac}$ is the corresponding pressure. \\
Now, from the conservation equation (\ref{eqn:consv}), one obtains 
$\rho_{vac} = -p_{vac} =$constant. 
Thus~ $\Lambda = 8 \pi G \rho_{vac}$ serves as an 
effective cosmological constant. Then Einstein field equations (\ref{eqn:e1}) 
and (\ref{eqn:e2}) 
can be written as 
\be\label{eqn:lambda1}
3\frac{ \dot{a}^2}{a^2} + 3\frac{k}{a^2} =  \Lambda,
\ee
\be\label{eqn:lambda2}
2\frac{\ddot{a}}{a} + \frac{ \dot{a}^2 + k}{a^2} =\Lambda~.
\ee
\par Equations (\ref{eqn:lambda1}) and (\ref{eqn:lambda2}) can be combined to yield 
\begin{center}
$\ddot{a} = \frac{\Lambda}{3}a~.$
\end{center}
For a positive $\Lambda$, $\ddot{a}$ is thus positive and the universe has an accelerated expansion 
which is termed as the inflationary scenario. The deceleration parameter $q = -\frac{\ddot{a}a}{{\dot{a}}^2}$ 
is negative definite. From equation (\ref{eqn:eh1}) one can write
\be\label{eqn:flat}
\dot{\Omega} = 2 q H (\Omega - 1)~,
\ee
which clearly indicates that at $\Omega = 1$, one has $\dot{\Omega} = 0$, i.e, $\Omega = 1$ is the stable solution. 
For a negative value of $q$, $\dot{\Omega} < 0$ for $\Omega > 1$ and $\dot{\Omega} > 0$ for $\Omega < 1$ . So the 
density parameter decreases or increases to the stable value of $\Omega = 1$ when it is greater or less respectively 
than $\Omega = 1$. Hence the contribution from the spatial curvature, $\Omega_{k} = -\frac{k}{a^2 H^2}$, is washed out 
in view of equation (\ref{eqn:flat}). Hence, flatness problem was solved. 
\par The horizon problem was also solved in the following way. Two points, 
which were causally connected during a very early epoch, might have fallen so 
much apart during the inflationary expansion that at this epoch their past 
light cones do not have any intersection even if they are extended back to the 
last scattering surface.
\par Inflationary models could provide solution to the monopole problem also 
by considering that the monopoles, that were created during the symmetry 
breaking, were so diluted during this rapid expansion that the monopole 
density becomes hardly traceable at present.
\par So, it is seen that inflationary models could solve the problems of 
standard Big Bang cosmology more or less satisfactorily. But these models also 
suffer from some problems, the most famous one being the ``\emph{graceful exit 
problem}''. The problem originates from the fact that we see the galaxies and 
other structures around us. The  universe 
must have a decelerated expansion at some epoch of time in order to facilitate 
the galaxy formation. This demands that the universe has to come out of this 
inflationary phase. This problem that how the universe comes out of this 
rapid expansion phase and enters a decelerated phase of expansion is termed as 
the ``graceful exit problem''. A number of models have been suggested for the 
solution of this problem. All of them have their own merits and pitfalls 
the details of which are not discussed over here.
\par
So all the major problems in Standard Big Bang Cosmology were believed to be 
related to the early phase of the history of the universe. The present state of 
affairs in the universe were presumably competently taken care of, 
excepting some finer details like that regarding the structure formations. 
The recent observation that the present universe is accelerating came as jolt 
and thus the late time behaviour also 
warrants serious theoretical attention.    
\vskip .2in
\section{Observational Evidence of the Present \\Acceleration :}
\markright{Observational Evidences}
The evidence in support of an accelerating universe stems from 
the observations of the luminosity-redshift relation of type Ia supernovae. 
\begin{figure}[!h]
\centerline{\psfig{figure=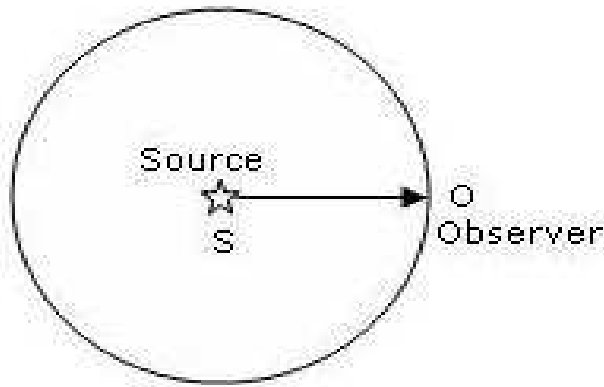,height=70mm,width=100mm}}
\caption{\normalsize{\em Measurement of luminosity distance $d_{L}$ from absolute and apparent luminosities}}
\label{fig2}
\end{figure}
\par In a static universe, if one considers a luminous object $S$ emitting 
a total power $L$ ( also called absolute luminosity ), then the intensity 
$l$ ( called apparent luminosity ) detected by an observer at $O$ at a radial 
distance $d_{l}$ from the luminous object (as shown in Figure \ref{fig2}) is given by 
\be 
l = \frac{L}{4 \pi {d_{l}}^2}~.
\ee

The quantity 
\be\label{eqn:dl}
d_{l} = \sqrt{\frac{L}{4 \pi l}}
\ee
is known as the luminosity distance. In a static universe, the 
luminosity distance is equal to the actual distance. In an expanding universe 
however, the intensity detected by the observer gets reduced because the 
energy of a photon emitted gets redshifted due to the cosmological 
expansion \cite{kolb}. Because of this expansion, the detected energy gets 
reduced by a factor of 
\be\label{eqn:redshift}
\frac{a(t_{0})}{a(t)} = 1 + z
\ee
where $a(t)$ is the scale factor of the universe at some cosmic time $t$, 
$t_{0}$ is the present time and $z$ is the redshift parameter , given by 
$z = \frac{\Delta\lambda}{\lambda}$, where $\lambda$ is the emitted wavelength and the wavelength 
received is $\lambda + \Delta\lambda$.
\par Thus in an expanding background, the observed apparent luminosity 
can be written as 
\be
l = \frac{L}{4 \pi {a(t_{0})}^2 {x(z)}^2 {(1 + z)}^2}
\ee
where $x(z)$ is the comoving distance of the luminous object, expressed as a function of the redshift 
$z$. This implies 
that in an expanding universe, the luminosity distance $d_{L}(z)$ is related 
to the comoving distance $x(z)$ as 
\be\label{eqn:dlz} 
d_{L}(z) = x(z)~ (1 + z)
\ee
where $a$ is normalized such that $a(t_{0}) = 1$.
\par As the light geodesics in a spatially flat expanding universe obey 
the relation
\be
c dt = a(z)~dx(z)~,
\ee
one can eliminate $x(z)$ using equation (\ref{eqn:dlz}) and express the expansion rate 
of the universe $H = \frac{\dot{a}}{a}$ in terms of $d_{L}(z)$ as
\be\label{eqn:hz}
H(z) = c \frac{1}{\frac{d}{dz}\left(\frac{d_{L}(z)}{(1 + z)}\right)}~.
\ee
Thus if the absolute luminosity of a distant object is known, its 
apparent luminosity can be measured as a function of $z$ and from equation 
(\ref{eqn:dl}), the luminosity distance $d_{L}$ can be calculated as a function 
of redshift $z$. The expansion history $H(z)$ can then be deduced by 
differentiating equation (\ref{eqn:hz}) with respect to $z$. 
On the other hand, if a theoretically predicted 
$H(z)$ is given, the corresponding $d_{L}(z)$ can be predicted 
by integrating equation 
(\ref{eqn:hz}) as 
\be
d_{L}(z) = c (1 + z)\int_{0}^{z}\frac{dz'}{H(z')}~.
\ee
Then this predicted $d_{L}(z)$ is compared with the measured $d_{L}(z)$ to 
test the consistency of a model. In practice, the astronomers do not use the 
ratio of absolute over apparent luminosity. Instead they use the difference 
between apparent magnitude $m$ and absolute magnitude $M$ given by the 
relation 
\be
m - M = 2.5 log_{10}~\left(\frac{L}{l}\right)~.
\ee
\par It is important to see how the deceleration parameter $q$, defined by equation (\ref{eqn:qdef}), 
can be estimated from this observed luminosity at a given redshift. For small values of the 
time scale and the distance compared to the respective Hubble scales, the scale factor $a(t)$ and 
hence $z$ can be written as power series, 
\be 
z = H_{0}(t_{0} - t_{1}) + \left(1 + \frac{q_{0}}{2}\right) H_{0}^{2} (t_{0} - t_{1})^2 + ~...........~, \ee
where the quantities with a suffix zero indicate their present values. 
Using this expression, 
the luminosity distance $d_{L}$ can be written as 
\be
d_{L} = H_{0}^{-1} \left[ z + \frac{1}{2}(1 - q_{0})z^2 + ~.............\right]~,
\ee
and hence the apparent luminosity $l$ becomes 
\be
l = \frac{L}{4 \pi d_{l}^2} = \frac{L H_{0}^2}{4 \pi z^2} \left[1 + (q_{0} - 1)z + ~.........\right]~. 
\ee
The apparent and absolute magnitudes will then be related as 
\be
m - M = 25 - 5 log_{10} H_{0}~(km/sec/Mpc) + 5log_{10} cz~(km/sec) + 1.086 (1 - q_{0})z + ...........
\ee
The apparent luminosity is smaller for a negative $q_{0}$ and thus 
the dimmer appearance of the supernovae calls for a negative $q_{0}$.
\begin{figure}[!h]
\centerline{\psfig{figure=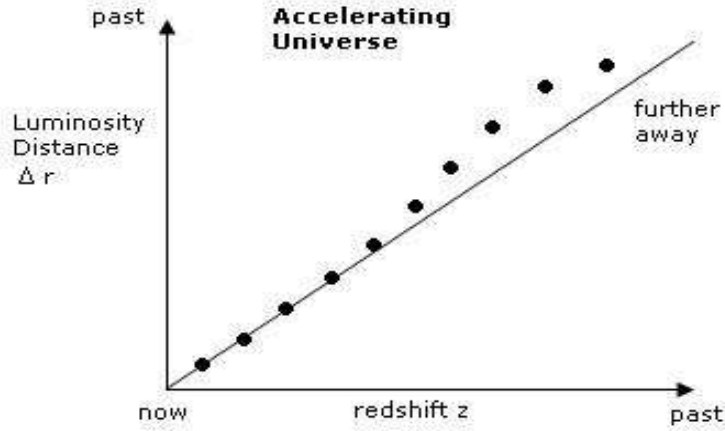,height=70mm,width=100mm}}
\caption{\normalsize{\em Hubble diagram for an accelerating universe}}
\label{fig3}
\end{figure}
The expansion history of the universe is very well depicted by the 
Hubble diagram where the x-axis shows the redshift $z$ of a luminous object 
and the y-axis shows the physical distance $\Delta r$ to those objects. 
This $\Delta r$ is in fact the luminosity distance. 
As the redshift $z$ is related to the scale factor $a(t)$ at the time of 
emission of radiation via equation (\ref{eqn:redshift}), whereas $\Delta r$ is 
related to the time in past when the emission was made,  
therefore the Hubble diagram provides information about the time dependence 
of the scale factor $a(t)$.
\par The slope of this diagram at a given redshift denotes the inverse of the 
rate of expansion $H(z)$, i.e, 
\be
\Delta r = \frac{1}{H(z)}c~z~.
\ee
In an accelerating universe, the  
slope $H^{-1}$ of the $\Delta{r}$ vs. $z$ curve is larger at high redshift.
Thus, at a given redshift, luminous objects appear to be at a greater 
distance, i.e, dimmer as compared to an empty universe expanding with a constant rate 
(see Figure \ref{fig3}).
\par
\begin{figure}[!h]
\centerline{\psfig{figure=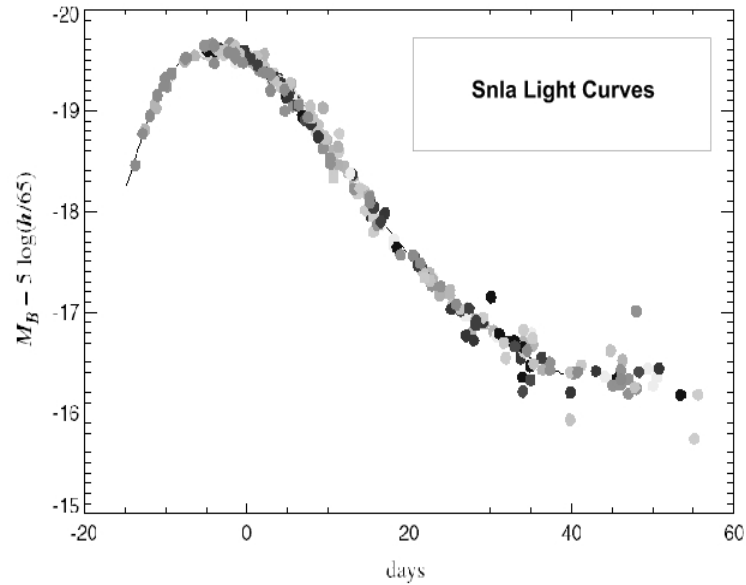,height=100mm,width=130mm}}
\caption{\normalsize{\em Light curve for a typical Supernova Ia}} 
\label{fig4} 
\end{figure}

 In the construction of Hubble diagram, those luminous objects are used 
whose absolute luminosity is known and therefore by measuring their apparent 
luminosity, their distances can be calculated. Those luminous objects are 
called standard candles or distance indicators. Type Ia supernovae serve as 
excellent standard candles for estimating the luminosity distance $d_{L}$. 
Type Ia supernovae are explosions believed to occur in binary star systems 
where one of the companions has a mass below the Chandrasekhar limit 
$1.4 M_{\bigodot}$ and thus become a white dwarf supported by 
degenerate electron   
pressure after hydrogen and helium contained in the star 
are burnt up. Once the other companion reaches the red giant phase, the white 
dwarf starts accreting matter from the companion star. Once the mass of the 
white dwarf star becomes equal to the Chandrasekhar limit, the gravitational 
pull overcomes the degeneracy pressure and the white dwarf starts to shrink. 
This increases the temperature and results in the carbon fusion. This leads 
to violent explosions, commonly called supernovae explosions. These explosions are 
detected by a light curve whose luminosity increases rapidly in a time scale of 
less than a month, reaches a maximum and disappears in a timescale of 1-2 months 
(see figure \ref{fig4}). Type Ia 
supernovae are characterised by the absence of hydrogen and abundance of 
silicon in the spectrum. 
\par Supernovae are preferred standard candles mainly for the following 
reasons : 
\\
i) These objects are highly luminous. This high absolute luminosity of 
supernovae Ia ($M = -19.5$) ensures that they can be seen from large 
distances ( $\sim 1000$ Mpc ) and thus are useful for measuring various 
cosmological parameters. 
\\
ii) The dispersion in supernovae luminosity at maximum light is extremely 
small and the corresponding change in intensity is 
$\sim 25 \%$.
\\
iii) Their explosion mechanism is fairly uniform and well understood. 
\par However, the major problem in using Sn Ia as standard candles is that 
they are rare events - for instance, in our galaxy they occur only 
a few times in a millennium. Also it 
is not easy to predict a supernovae explosion. However, the key feature of the 
high precision tools of observation is that one can now detect supernovae 
in other galaxies also. The basic strategy employed in observing supernovae 
Ia is as follows \cite{sperl, ariess, tonry, sp, barris, knop} :
\\
i) A number of wide fields of apparently empty sky are observed. With 
modern instruments on a 4 meter-class telescope, tens or thousands of 
galaxies are observed in a few patches of sky. \\
ii) three weeks later, the same galaxies are observed once again.\\
iii) The images are then subtracted to observe the supernovae explosions.\\
The result of this observation strategy is a set of Sn Ia light curves in 
various bands of spectrum. From the light curves, their peak apparent 
luminosity is used to construct the Hubble diagram.
\par The first project in which supernovae were used to determine the energy 
associated with the cosmological constant was carried out by Perlmutter 
et al. in 1997 
\cite{perl}. This project was named as Supernovae Cosmology Project (SCP). 
In one year they discovered seven supernovae at redshift $0.35 < z < 0.65$ 
and observed them with different telescopes from the earth. The Hubble 
diagrams they constructed were in good agreement with a standard decelerating 
Friedmann cosmology. However, one year later they updated their results 
by including the measurements of a very high redshift ($z \sim 0.83$) 
Supernovae Ia \cite{sperl}. This dramatically changed the scenario and a 
decelerating universe was ruled out at about 99\% confidence level. This 
result was confirmed independently by another pioneer group, viz, High-z 
Supernovae Search Team (HSST) \cite{ariess}. They had discovered 16 supernovae 
at a redshift $0.16 < z < 0.62$ and the results indicated an accelerating 
universe at a 99\% confidence level. 
\par In 2003, Tonry et al. \cite{tonry} reported the results of eight newly 
discovered supernovae in the range $0.3 < z < 1.2$. Their results reinforced 
the previous findings of accelerated expansion and also gave the confirmation 
of decelerated expansion at $z > 0.6$. So, obviously the universe must have 
had a transition from deceleration to acceleration in the past. This 
transition was confirmed and pinpointed by Riess et al. in 2004 \cite{riess}. 
They included 16 new high redshift supernovae and after analyzing all 
available data, constructed a reliable and robust data set consisting of 157 
points which is known as Gold data set. With this new data set, they could 
clearly identify the transition from decelerated to accelerated expansion 
at $z \approx 0.46 \pm 0.13$. However, it was not easy to conclude 
whether the data 
favour an accelerating or decelerating universe from only the Hubble diagram
corresponding to the Gold data set because of the error bars 
present in the data set. This would be easier if the Hubble diagram of 
figure \ref{fig3} where the distance is plotted vs redshift is superposed 
with the distance-redshift relation $d_{L}^{empty}(z)$ of an empty universe 
with $H(z)$ constant.  So, a more efficient plot was used for this purpose, 
viz, the logarithmic plot of $d_{L}(z)$ vs. $d_{L}^{empty}(z)$ which could 
easily distinguish between an accelerated and decelerated expansion. Such a 
plot for the Gold data set clearly indicated that the best fit is obtained by 
an expansion which was decelerated at the earlier times ($z > 0.5$) and 
accelerated at recent times ($z < 0.5$). Attempts have been made to explain 
the observed dimming of supernovae at high redshift by considering that this 
apparent dimming is due to the scattering of light by intergalactic dust or 
grey dust or even due to evolution of Sn Ia. However none of them has a firm 
footing \cite{riess, agu1, agu2} and thus strengthens the belief that 
the universe at 
present is undergoing an accelerated phase of expansion. This is also confirmed 
by the highly accurate Wilkinson Microwave Anisotropy Probe (WMAP) data 
\cite{wmap1, wmap2, wmap3, wmap4, dns}.
\vskip .2in
\section{Search for the Dark Energy :}
\markright{Dark Energy}
This observed acceleration of the universe brings in trouble as 
gravity is attractive and this acceleration can not be driven by the 
attractive gravitational properties of regular matter. So, obviously an 
additional component is required which can give rise to an effective 
repulsive gravity so that 
matter can move away from each other with an acceleration. The obvious 
question to address is therefore, ``\emph{What should be the properties of this 
additional component}?'' The answer can be obtained by comparing Newtonian 
gravity with Einstein's gravity.
\par In Newtonian gravity, the 
acceleration of a test particle having mass $m$ under the influence of 
gravity is given by 
\begin{center}
$m \ddot{a} = - \frac{G M m}{a^2}$
\end{center}
which gives
\be\label{eqn:newton1} 
\frac{\ddot{a}}{a} = -\frac{4 \pi G}{3}\rho
\ee
as $M = \frac{4}{3}\pi a^3 \rho$. \\
On the other hand, in Einstein's gravity, from equations (\ref{eqn:e1}) and 
(\ref{eqn:e2}) one can obtain
\be\label{eqn:newton2}
\frac{\ddot{a}}{a} = -\frac{4 \pi G}{3}(\rho + 3 p)~.
\ee
This can be written in terms of equation of state parameter $w$ as 
\be
\frac{\ddot{a}}{a} = -\frac{4 \pi G}{3}\rho(1 + 3w)
\ee
which gives a measure of the acceleration of the universe.\\
So, from equations (\ref{eqn:newton1}) and (\ref{eqn:newton2}), it is 
evident that unlike that in Newtonian gravity, 
in Einstein gravity, pressure also plays a major role along with the density 
in determining the space-time dynamics of the universe. In other words, 
\emph{``pressure carries weight in Einstein's gravity''}.\\
Now, as the universe at present is accelerating, $\ddot{a} > 0$. From equation 
(\ref{eqn:newton2}), a positive $\ddot{a}$ can be obtained only 
if $p$ is sufficiently 
negative such that $\rho + 3 p < 0$ or equivalently $w < -\frac{1}{3}$.  
So in order to explain the observed acceleration of the 
universe there is indeed the requirement of some form of matter which can 
generate sufficient negative pressure. This particular form of matter, now 
popularly referred to as ``\emph{dark energy}'', is believed to 
account for as much as 
70\% of the present energy of the universe.
\par A large number of possible candidates suitable as dark energy component 
have appeared in the literature. None of them has a clear advantage over 
all others. All have their merits, but none perhaps has a firm theoretical 
footing. Its effect has been, as discussed, is to provide a sufficient 
effective negative pressure. Nothing is known about its distribution vis-a-vis 
the dark matter, except that it does not cluster at any scale lower than the 
size of the universe. 
A few of these candidates are described below.
\vskip .2in
\subsection{Cosmological Constant Models :-}
The simplest dark energy candidate is the cosmological constant 
$\Lambda$ introduced by Einstein in 1917. With the introduction of the 
$\Lambda$-term, the Einstein's equation gets modified as 
\be 
G_{\mu\nu} = 8 \pi G T_{\mu\nu} + \Lambda g_{\mu\nu}~.
\ee
Einstein originally introduced the $\Lambda$-term on the left hand side of the 
field equation in order to obtain a static universe. But later on, when 
Hubble's observations suggested that the universe is expanding, he himself 
rejected the $\Lambda$-term. However, $\Lambda$ was again brought into being 
in early 1980's with the inflationary model of the universe \cite{guth}. 
During inflation, the 
universe was dominated by the vacuum energy $\rho_{vac}$ as discussed earlier. 
As the equation of state for such energy is 
\begin{center}
$\rho_{vac} + p_{vac} = 0$,
\end{center}
both $\rho_{vac}$ and $p_{vac}$ are constants, and the Einstein field 
equations will effectively look like 
\bea
3\frac{ \dot{a}^2}{a^2} + 3\frac{k}{a^2} =  \Lambda,\\
2\frac{\ddot{a}}{a} + \frac{ \dot{a}^2 + k}{a^2} =  \Lambda~,
\eea
where $\rho_{vac} = - p_{vac} = \frac{\Lambda}{8 \pi G}$.\\
Here $\Lambda$ is not introduced arbitrarily, but rather attains the 
significance of the vacuum energy density. By solving these equations one 
gets an exponentially expanding model of 
the universe at a very early epoch which provides solutions to the problems 
of Big Bang theory. As the simple inflationary model had its own problems, 
such as that of the `graceful exit', or the huge discrepancy in the 
theoretically predicted value of $\Lambda$ and the one suggested by 
observations, other more complicated models took over and $\Lambda$ had to 
hide itself into oblivion. In the late 90's, with the observation that 
universe is at present accelerating, 
$\Lambda$ again came back strongly after a brief period of hibernation.\\
For an FRW universe, the Einstein's equations with a cosmological constant
take the form
\be\label{eqn:l1}
3\frac{ \dot{a}^2}{a^2} + 3\frac{k}{a^2} =  8 \pi G \rho + \Lambda,
\ee
\be\label{eqn:l2}
2\frac{\ddot{a}}{a} + \frac{ \dot{a}^2 + k}{a^2} = - 8 \pi G p + \Lambda~.
\ee
From equations (\ref{eqn:l1}) and (\ref{eqn:l2}), one easily arrives 
at the relation
\be
\frac{\ddot{a}}{a} = -\frac{4 \pi G}{3}(\rho + 3 p) + \frac{\Lambda}{3}~,
\ee
from which it is clear that $\Lambda$ is capable of providing an 
acceleration ($\ddot{a} > 0$) if 
\begin{center}
$\Lambda > 4 \pi G (\rho + 3 p)$~.
\end{center}
In a very simple case, when the universe is spatially flat ($k = 0$) and 
is dust dominated ($p = 0$), equations (\ref{eqn:l1}) and (\ref{eqn:l2}) gives 
an exact analytic expression for the scale factor as 
\be
a(t) \propto {\left(\sinh \frac{3}{2} \sqrt{\frac{\Lambda}{3}}t \right)}^{2/3}~.
\ee
For a very small $t$ (early epoch), $a(t) \propto t^{2/3}$ and hence 
$q = \frac{1}{2}$, which gives a decelerated expansion in the early phase of 
dust dominated era as expected \cite{sahnirev}. On the other hand, for 
large $t$ (i.e, present epoch), 
$a \propto e^{\sqrt{\frac{\Lambda}{3}}t}$ which gives an accelerated expansion. 
This model is popularly known as $\Lambda$CDM model. 
However, there is a major problem related to the cosmological constant 
$\Lambda$. Field theory suggests that the lower limit of the value of 
vacuum energy density should be 
\begin{center}
$\rho_{vac} \approx 10^6~ {GeV}^4~,$
\end{center}
whereas the current observational upper limit on the cosmological constant is
\begin{center}
${\rho_{vac}|}_{0} \approx 10^{-29}~ g/{cm}^3 \approx 10^{-47}~{GeV}^4~,$
\end{center}
where the subscript `0' stands for present time. So, there is a discrepancy 
between the predicted value and the observed value by $10^{53}$ orders of 
magnitude.  
The cosmological constant $\Lambda$ is expected to have a large value during 
early epoch so as to resolve - via inflation - the horizon and flatness 
problems; at the same time it requires a low value at the present epoch 
to avoid conflict with the observations. This problem is referred to as the 
``\emph{cosmological constant problem}'' because of which $\Lambda$ has lost  
some of its ground. 
\par To avoid this problem, some dynamical models of dark energy have 
been proposed where $\Lambda$, instead of being a constant, is a slowly 
varying function of time.
\vskip .2in
\subsection{Varying $\Lambda$ Models :-}
The first candidate for a dynamical model of dark energy was the 
time varying cosmological parameter $\Lambda(t)$. A number of phenomenological 
models have been described to introduce this dynamical $\Lambda$-term as the 
new form of matter \cite{vishu1, vishu2, arbab, maia, sola1, sola2}    
and is taken as a  function of the scale factor $a(t)$ or the 
cosmic time $t$. However, there is no strong 
physical motivation behind this choice. In fact, the dynamical $\Lambda$-term 
leads to some problems. If $\Lambda$ is considered as a constant, then the 
introduction of the term $\Lambda g_{\mu\nu}$ in the Einstein field equations 
does not affect the conservation equation. It is so because the conservation 
equations come as a consequence of Bianchi identity
\begin{center}
$G^{\mu\nu}_{;\nu} = 0~,$
\end{center}
which remains the same as for a constant $\Lambda$, 
\begin{center}
$(\Lambda g^{\mu\nu})_{;\nu} = \Lambda_{,\nu}g^{\mu\nu} + 
 \Lambda g^{\mu\nu}_{;\nu} = 0~.$
\end{center}
However, in a varying $\Lambda$ model, the Bianchi identity leads to a 
modified conservation equation. If the normal matter is allowed to satisfy 
its own conservation equation of the form given in equation (\ref{eqn:consv}), 
$\Lambda$ automatically turns out to be a constant. 
The way out is to consider some interaction between 
$\Lambda(t)$ and matter such that one grows at the expense of the other. 
But in order to incorporate this interaction, one has to sacrifice the 
matter conservation equation. In most of the cases, the choice of 
the form of interaction does not have any physical background.  
So, the second natural 
choice for a dynamical $\Lambda$-term is a scalar field $\phi$ having some 
potential $V(\phi)$ and is popularly called the ``\emph{quintessence scalar 
field}''.
\vskip .2in
\subsection{Quintessence Models :-}
\par Inspired by the role of a scalar field with a suitable potential in 
providing an inflationary scenario in the early universe, attempts are made to 
formulate a ``dynamical dark energy" model based on scalar field cosmology. 
With the introduction of a scalar field minimally coupled to  
gravity, the action gets modified as 
\be
S = \int d^4x \sqrt{-g} \left[ \frac{R}{16 \pi G} + \frac{1}{2}\phi_
     {,\mu}\phi^{,\mu} - V(\phi) + {\mathit{L}}_{m} \right]~.
\ee
For a spatially flat FRW cosmology, Einstein's field equations take the form 
\be\label{eqn:quin1}
3\frac{ \dot{a}^2}{a^2}  =  \rho_{m} + \frac{1}{2}{\dot{\phi}}^2 + V(\phi)~,
\ee
\be\label{eqn:quin2}
2\frac{\ddot{a}}{a} + \frac{{\dot{a}}^2}{a^2} = - p_{m} -
     \frac{1}{2}{\dot{\phi}}^2 + V(\phi)~~.
\ee
Variation of the action with respect to the scalar field $\phi$ leads
to the wave equation for $\phi$ as
\be\label{eqn:wavephi}
\ddot{\phi} + 3\frac{\dot{a}}{a}\phi + V'(\phi) = 0
\ee
where a prime indicates differentiation with respect to $\phi$. 
\\In view of the 
equations (\ref{eqn:quin1}) - (\ref{eqn:wavephi}), the matter conservation 
equation 
\be\label{eqn:matter}
\dot{\rho_{m}} + 3 \frac{\dot{a}}{a}\left( \rho_{m} + p_{m} \right) = 0
\ee
is not an independent equation, and comes as a consequence of the Bianchi 
identities. Thus one is left with five unknowns, $a$, $\rho_{m}$, $p_{m}$, 
$\phi$ and $V(\phi)$ and only three independent equations to solve them. 
If the equation of state $p_{m} = p_{m}(\rho_{m})$ and the form of potential 
$V = V(\phi)$ are given, the system of equations is closed. Equations 
(\ref{eqn:quin1}) and (\ref{eqn:quin2}) reveal that the contribution to the 
effective energy density and pressure from the scalar field is 
\be\label{eqn:rhophi}
\rho_{\phi} = \frac{1}{2}{\dot{\phi}}^2 + V(\phi)~, 
\ee
\be\label{eqn:pphi}
p_{\phi} = \frac{1}{2}{\dot{\phi}}^2 - V(\phi)~,
\ee
where the first term on the right hand side represents the kinetic part 
whereas the second term represents the potential part.\\
In the matter dominated era the fluid pressure $p_{m} = 0$, and equation 
(\ref{eqn:newton2}) yields the result
\begin{center}
$$\frac{\ddot{a}}{a} = -\frac{4 \pi G}{3}\left[\rho_{m} + \rho_{\phi} + 
                      3 p_{\phi}
\right]$$
\end{center}
\be
~~~~~~~=- \frac{4 \pi G}{3}\left[\rho_{m} +2 {\dot{\phi}}^2 - 2 V(\phi)
\right]~.
\ee
If $V(\phi)$ grows to a sufficiently large value such that 
\be
2 V(\phi) > \rho_{m} +2 {\dot{\phi}}^2~,
\ee
$\ddot{a}/a$ has a positive value consistent with the present 
observation. These models have a further interesting possibility that 
if $V(\phi)$ has a smaller value compared to 
$\frac{\rho_{m} +2{\dot{\phi}}^2}{2}$ 
in the early matter dominated era and grows later to dominate the dynamics, 
the model would exhibit a signature flip in $q$ from a positive to a negative 
value at a recent past. A fine tuning of the parameters of the theory and the 
arbitrary constants of integration might lead to models consistent with the 
details of observational data. These features have made scalar field 
cosmologies a very active area of research, and now the scalar field with a 
potential giving rise to a negative pressure is specifically called a 
``\emph{quintessence field}" and the generic term for 
any field, that drives the late surge of accelerated expansion of 
the universe, has become `\emph{dark energy}'.
\par The system of equations suggest that if a particular dynamics of the 
universe, i.e, the temporal behaviour of the scale factor $a(t)$ is chosen, 
it is possible to find a potential consistent with that. Amongst the host 
of quintessence potentials available in the literature, some give an 
ever accelerating model, some give an acceleration in the large $t$ limit, and 
a few indeed allows for a smooth transition from a decelerated 
to an accelerated 
expansion in the matter dominated era itself consistent with the 
observation that the universe has entered the accelerated phase of expansion 
in the late matter dominated era. 
\par One typical example of the latter kind of a quintessence field is the 
one introduced by Sen and Sethi \cite{sethi}, where the potential is a double 
exponential, given by 
\be
V(\phi) = \alpha (e^{2\beta\phi} + e^{-2\beta\phi}) + V_{0}~.
\ee
where $\alpha$, $\beta$ and $V_{0}$ are constants. The solution of the scale 
factor comes as $\sinh ^{\beta} (t/t_{0})~$.
For a small $t$, the solution for the scale factor is like $\sim t^{\beta}$ 
whereas for a high value of $t$, $a \sim e^{\beta t/t_{0}}$. For $\beta < 1$, 
the model has a decelerated expansion in the early epoch, but indeed 
accelerates at a matured age. Similar kind of potentials have been used by 
Barreiro et al \cite{barreiro} and Rubano et al \cite{rubano} as well. 
\par
\begin{figure}[!h]
\centerline{\psfig{figure=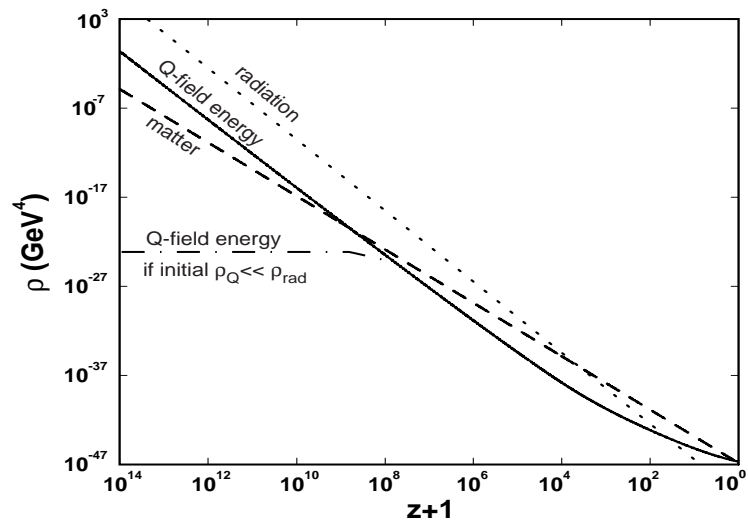,height=70mm,width=100mm}}
\caption{\normalsize{\em The quintessence field rolling down an inverse power law potential tracks first radiation, 
then matter and finally dominates the energy density of the universe at present. (From Zlatev, Wang 
and Steinhardt \cite{zlatev})}}
\label{tracker}
\end{figure}

 A particularly interesting class of quintessence fields are the so 
called ``\emph{tracker fields}" for which the potential is 
steep enough to satisfy the condition  
\be
\Gamma = \left(\frac{V''}{V}\right){\large{/}} \left(\frac{{V'}^2}{V^2} \right)\ge 1~.
\ee
All potentials satisfying this condition lead to a common evolutionary path 
for the fields starting from a wide range of initial conditions \cite{zlatev}. 
The energy density of the tracker fields has an evolution which mimics 
that of the background matter during most of its 
history (see Figure \ref{tracker}) and dominates over the matter density 
at later stages of the evolution to generate acceleration. Thus, these models 
can alleviate the coincidence problem which poses the question why the dark 
energy sector should dominate the dynamics of the universe at present.
\\
Some simple examples are the inverse power law potential \cite{ratra}
\be
V(\phi) = \frac{V_{0}}{\phi^{\alpha}}, 
\ee
with some restriction on $\alpha$, and the exponential potential \cite{ratra, 
wetterich, ferreira}
\be
V(\phi) = V_{0}~ e^{-\lambda\phi}~.
\ee
Within the tracker framework, a potential with a versatile ambition was 
given by Sahni and Wang \cite{wang}. This potential is given as  
\be
V(\phi) = V_{0} {[\cosh (\lambda\phi) - 1]}^p~.
\ee
For $\vline~\lambda\phi~\vline >> 1$ and $\phi < 0$, the 
potential is exponential,
$$V(\phi) \propto \exp(- p\lambda\phi)~,$$
and $w_{\phi} = \frac{p_{\phi}}{\rho_{\phi}}$ is very close to 
$w_{m} = \frac{p_{m}}{\rho_{m}}$.\\
On the other hand, when $\vline~\lambda\phi~\vline << 1, V(\phi) \propto 
{(\lambda\phi)}^{2p}$. Then, the average equation of state parameter is
given by 
\begin{center}
$<w_{\phi}> = \frac{p - 1}{p + 1}$. 
\end{center}
For $p \le \frac{1}{2}$, the model describes quintessence whereas for 
$p = 1$ it can play the role of a cold dark matter. So, this model is able to 
describe both dark matter and dark energy within a tracker framework (see 
\cite{arbey, urena} ).
\par Another useful potential with interesting features was proposed by 
Zlatev et al.\cite{zlatev} as
\be
V(\phi) = V_{0}\left[e^ {M_{p}/\phi} - 1\right]~.
\ee
The advantage of this potential is that it can significantly alleviate the 
fine tuning problem and $\rho_{\phi}$ can come to dominate over the present 
matter density from a large number of initial conditions. 
\par There are of course a lot of other examples. However, despite the 
many attractive features of these quintessence 
potentials, a degree of fine tuning does remain in fixing the parameters 
of the potentials. The generic problem of quintessence models is that 
the potentials are all taken arbitrarily and none of them has a sound 
physical basis. But indeed if we know the kind of acceleration required, 
more often than not, 
one can find out the form of potential which generates the desired 
acceleration. 
\vskip .2in
\subsection{Non-minimally Coupled Scalar Field Models :-}
In these models, the scalar field is non-minimally coupled to gravity 
such that the general action is of the form 
\be
S = \int d^4x \sqrt{-g}\left[\frac{f(\phi)R}{16 \pi G_{0}} - g(\phi)\phi^{,\mu}
          \phi_{,\mu} + \mathit{L}_{m}\right]~,
\ee
where $f(\phi)$ and $g(\phi)$ are some functions of the scalar field $\phi$, 
$G_{0}$ is the Newtonian constant of gravity. \\
Comparing this action with the Einstein's action, one gets 
\begin{center}
$G \sim \frac{1}{f(\phi)}~.$
\end{center}
So, in this type of scalar field models, $G$ is not a constant but is some 
function of the scalar field $\phi$. 
\par The simplest of the non-minimally coupled scalar field models is the 
Brans-Dicke theory. This theory was first introduced to incorporate Mach's 
principle in a relativistic theory of gravity. According to this principle, 
the inertia of a body is not an intrinsic property of its own, it rather  
depends on the mass distribution of the rest of the universe.\\
In Brans-Dicke theory
\begin{center}
$f(\phi) = \phi$~;  
\end{center}
\be
g(\phi) = \frac{\omega}{\phi}~,
\ee
$\omega$ being a dimensionless constant parameter. With these, the action 
for the Brans-Dicke theory takes 
the form \cite{brans}
\be\label{eqn:bdaction}
S = \int d^4x \sqrt{-g}\left[\frac{\phi R}{16 \pi G_{0}} - \frac{\omega}{\phi}
             \phi^{,\mu}\phi_{,\mu} + \mathit{L}_{m}\right]~.
\ee
If we consider the matter field to consist of a perfect fluid, then the 
field equations for a spatially flat Robertson - Walker spacetime become, 
\be\label{eqn:bd1}
3\frac{{\dot{a}}^2}{a^2} = 8 \pi G_{0}\frac{\rho_{m}}{\phi} + \frac{\omega}{2}
            \frac{{\dot{\phi}}^2}{{\phi}^2} - 3\frac{\dot{a}}{a}
             \frac{\dot{\phi}}{\phi}~,
\ee
\be\label{eqn:bd2}
2\frac{\ddot{a}}{a} + \frac{{\dot{a}}^2}{a^2} = -8 \pi G_{0}\frac{p_{m}}{\phi} 
        - \frac{\omega}{2}\frac{{\dot{\phi}}^2}{{\phi}^2} - 2 \frac{\dot{a}}{a}
             \frac{\dot{\phi}}{\phi} - \frac{\ddot{\phi}}{\phi}~.
\ee
Also, the wave equation for the Brans-Dicke field is 
\be\label{bdwave}
\ddot{\phi} + 3 \frac{\dot{a}}{a}\dot{\phi} = \frac{8 \pi G_{0}}{(2\omega + 3)}
              (\rho_{m} - 3p_{m})~.
\ee
Dicke \cite{dicke} in 1962 
framed an alternative version of their scalar tensor theory \cite{brans} by a 
simple redefinition of units. By effecting a conformal transformation
\be\label{eqn:confbd}
\bar{g}_{\mu\nu} = \phi g_{\mu\nu}~,
\ee
the action given by equation (\ref{eqn:bdaction}) becomes,
\be
S = \int\left[\frac{\bar{R}}{16 \pi G} + \frac{(2\omega + 3)}{2} 
       \Psi_{,\mu}\Psi^{,\mu} + {\bar{\mathit{L}}}_{m}\right]~,
\ee
where an overhead bar represents quantities in new frame 
and $\Psi$ = ln$\phi$.\\
Comparing this action with equation (\ref{eqn:bdaction}), it is seen that 
in this version $\phi$ is no longer coupled to $\bar{R}$. So, the effective 
constant of gravitation G, which is a function of the scalar field as 
\be
G = \frac{G_{0}}{\phi}~,
\ee
in the original version of the theory, now becomes a constant. 
\par The field equations (\ref{eqn:bd1}) and (\ref{eqn:bd2}) in the new frame 
look like 
\be\label{eqn:confbd1}
3\frac{{\dot{\bar{a}}}^2}{{\bar{a}}^2} = {\bar{\rho}}_{m} + \frac{(2\omega + 3)}
                      {4}{\dot{\Psi}}^2~,
\ee
\be
2\frac{\ddot{\bar{a}}}{\bar{a}} + \frac{{\dot{\bar{a}}}^2}{{\bar{a}}^2}
 = - {\bar{p}}_{m} - \frac{(2\omega + 3)}{4}{\dot{\Psi}}^2~,
\ee
and the wave equation becomes 
\be
\ddot{\Psi} + 3\frac{\dot{\bar{a}}}{\bar{a}}\dot{\Psi} = 
    \frac{{\bar{\rho}}_{m} - 3{\bar{p}}_{m}}{(2\omega + 3)}~,
\ee
where an overhead bar indicates quantities in new frame and 
${\bar{a}}^2 = \phi a^2$. The equations are written in units where 
$8 \pi G_{0} = 1$. 
The density and pressure of the normal matter in this version are related 
to those in the original version as 
\begin{center}
${\bar{\rho}}_{m} = {\phi}^{-2}\rho_{m}$ ~and~ ${\bar{p}}_{m} = {\phi}^{-2} p_{m}$.
\end{center}
The resulting field equations in the new frame look more tractable than the 
original version. Furthermore, the equations in this version give insight 
regarding the comparison of the energies of different components of matter. 
For example, it is clearly 
seen from equation (\ref{eqn:confbd1}) that the contribution to the energy 
density by the scalar field is given by 
\begin{center}
${\bar{\rho}}_{\phi} = \frac{(2\omega + 3)}{4}{\dot{\Psi}}^2.$
\end{center}
However, one has to pay some price for it. In the transformed version although 
$G$ becomes a constant, the rest mass of a test particle becomes a 
function of the scalar field \cite{dicke} and one has to sacrifice the equivalence 
principle. So, the geodesic equations are no longer valid and indeed the 
physical significance of different quantities in this version of the theory 
is somewhat obscure. However, because of its computational simplicity, it is 
easier to arrive at some solutions in the transformed version. And the 
problem can be resolved by transforming back to the original atomic units 
where one can talk about the various 
features more confidently. 
\par Although General Relativity (GR) is a better theory of gravity and it 
enjoys experimental evidences in its support, for various reasons 
Brans-Dicke (BD) theory continues to enjoy an alive interest. One 
important advantage of BD theory is that it becomes indistinguishable from 
GR in the limit $\omega \rightarrow \infty$. This limit has now shown to have 
only a restricted application \cite{nbss}, but the PPN parameters calculated in BD theory 
\cite{misner} clearly shows that at least in the weak field limit the 
predictions from local 
astronomical observations in this theory will be same as that in GR in 
the large $\omega$ limit. For these reasons, there is a popular notion that 
BD theory is perhaps the most natural generalization of GR. However, the local 
astronomical observations suggest that if BD theory has to be consistent, 
then $\omega \sim 10^3$ which renders the theory practically indistinguishable 
from GR \cite{omega}. 
\par BD theory proved to be useful in providing clues to the solutions for 
some of the  outstanding 
problems in cosmology. In 1981, Guth \cite{guth} proposed the inflationary 
model of cosmology. This model could provide solution to many of the 
cosmological problems as discussed earlier, but it suffered from the 
`\emph{graceful exit problem}'. A large number of models were 
introduced to solve 
this problem having their own merits and demerits.
\par In 1984, Mathiazhagan and Johri \cite{johri} addressed the problem in 
Brans-Dicke (BD) theory \cite{brans}. 
Under this framework, it was shown that along with a vacuum energy, the 
scale factor grows as a power function of time. Using a similar technique, 
La and Steinhardt \cite{la} presented the ``\emph{extended inflation model}'' 
in order to get a sufficient slow roll of the scalar field so that there is 
sufficient time for the completion of phase transition and thus the 
graceful exit problem could be resolved. However, this leads to unacceptable 
distortions of the microwave background \cite{weinberg}. To solve this problem, 
Steinhardt and Accetta \cite{accetta} developed the ``\emph{hyper-extended 
inflation model}''. 
Later Brans-Dicke theory was used for finding a solution to the graceful exit 
problem with a large number of potentials \cite{nb}, where the inflaton field 
oscillates during the later stages of evolution and the universe comes 
out of the inflationary phase. 
\par BD theory has also found applications in solving a few recent 
cosmological problems, such as, the quintessence problem. A number 
of models have been presented where the Brans-Dicke scalar tensor theory 
could potentially solve the problem of quintessence as it leads to 
non-decelerating solutions for the scale factor in the present matter 
dominated universe. In some of these models \cite{ss, berto}, the BD theory is 
modified by incorporating a potential $V(\phi)$ which is a function of the 
BD scalar field itself which could drive the acceleration. However, the 
problem with these models is that $V(\phi)$ is put in by hand and there is no 
physical motivation behind the choice of form of $V(\phi)$.
\par A few other models have also appeared in the literature where the 
cosmic acceleration is obtained with a quintessence field in BD theory 
\cite{nbdp}. However, this result hardly provides any improvement on the 
corresponding GR result. Recently Banerjee and Pavon \cite{pavon} have 
proposed a model where the 
BD theory could explain the present accelerated expansion of the universe 
without resorting to a cosmological constant or quintessence matter. This 
is better in the sense that one does not have to invoke any additional 
quintessence field to explain the acceleration. However, this model 
have problems in 
providing a decelerated expansion in the 
radiation-dominated epoch.
\par The general defect of all these models is that none of them could 
provide a smooth transition from the decelerated to accelerated phase 
of expansion and they rather provide an acceleration in some limit. 
Also in all of
these models, a consistent accelerated solution is 
obtained only for small negative values of the Brans-Dicke parameter 
$\omega$. This is in sharp contrast to the value obtained from 
local astronomical 
experiments which predict the value of $\omega$ to be of the order of 
a thousand \cite{omega}. Attempts have been made to overcome this problem by 
considering a modified version of Brans-Dicke theory, called Nordtvedt's 
theory, where the parameter $\omega$ is a function of the Brans-Dicke 
scalar field instead of being a constant \cite{nordtvedt}. In this case the 
field equations (\ref{eqn:bd1}) and (\ref{eqn:bd2}) remain intact, but the 
wave equation (\ref{bdwave}) gets modified as 
\be
\ddot{\phi} + 3 \frac{\dot{a}}{a}\dot{\phi} = \frac{\rho - 3 p}{(2\omega + 3)} 
           - \frac{\dot{\omega}{\dot{\phi}}}{(2\omega + 3)} ~.
\ee
Bartolo and Pietroni \cite{bartolo} have pointed out that a varying 
$\omega$ can 
indeed explain the late time behaviour of the universe. Also Banerjee and 
Pavon \cite{pavon} showed that a varying $\omega$ theory could give rise 
to a decelerating radiation model followed by an accelerating model in the 
matter dominated universe. However, none of these could provide smooth 
transition from deceleration to acceleration in the matter dominated era 
itself. So, a thorough survey of varying $\omega$ theory is indeed warranted 
to check if it gives rise  
to a model of the universe which can explain the transition from 
decelerated to accelerated phase of expansion in the matter dominated epoch 
itself with some high values of $\omega$ consistent with local astronomical 
experiments.
\vskip .2in
\subsection{Curvature Driven Accelerating Models :-}
The scalar field models or the cosmological constant models are amongst 
the most popular candidates of dark energy component. Recently, an attempt 
along a different direction is also gaining attention. This effort explores 
the possibility of whether geometry by itself can serve the purpose of 
providing late time acceleration of the universe.
\par The idea actually originates from the experience of inflationary models. 
It was shown by Starobinsky \cite{starobinsky} and Kerner et al \cite{kerner1, 
kerner2} that 
higher order modifications of the Ricci curvature $R$, in the form of $R^2$ or 
$R_{\mu\nu}R^{\mu\nu}$ in the Einstein - Hilbert action, could generate 
sufficient acceleration in the very early universe. However, with the 
evolution of the universe, $R$ is expected to fall off. This leads to the 
question whether the inverse powers of $R$, which becomes dominant during 
the late time, can help driving the recent acceleration.\\ 
The action gets modified as, 
\be
S = \int \left[\frac{1}{16 \pi G}f(R) + \mathit{L}_{m}\right]\sqrt{-g}d^4x
\ee
where the usual Einstein - Hilbert action is generalized by replacing $R$ with 
an arbitrary function $f(R)$. A variation of this action with respect to the 
metric yields the field equations as 
\be\label{eqn:curv}
G_{\mu\nu} = R_{\mu\nu} - \frac{1}{2}Rg_{\mu\nu} = T_{\mu\nu}^{c} 
               + T_{\mu\nu}^{M}~,
\ee
where $T_{\mu\nu}^{c}$ represents the contribution from the curvature and 
$T_{\mu\nu}^{M}$ denotes the energy momentum tensor components for the matter 
field scaled by a factor of 
$\frac{1}{f\prime(R)}$. Here the 
choice of units $8 \pi G = 1$ has been made. \\
$T_{\mu\nu}^{c}$ is explicitly given as, 
\be
T_{\mu\nu}^{c} = \frac{1}{f'(R)}\left[\frac{1}{2}g_{\mu\nu}(f(R) - Rf'(R))
 + {f'(R)}^{;\alpha\beta}(g_{\mu\alpha}g_{\nu\beta} - 
          g_{\mu\nu}g_{\alpha\beta})\right]~,
\ee
where a prime indicates differentiation with respect to the Ricci scalar $R$.\\
For a spatially flat Robertson-Walker spacetime, where 
\be
ds^2 = dt^2 - a^2(t) \left[dr^2 + r^2 {d\theta}^2 + r^2 {\sin^2 \theta} 
{d\phi}^2\right]~, 
\ee
the field equations (\ref{eqn:curv}) take the form
\be\label{eqn:curv1}
3\frac{{\dot{a}}^2}{a^2} = \frac{1}{f'(R)}\left\{\frac{1}{2}[f(R) - 
Rf'(R)] - 3\frac{\dot{a}}{a}\dot{R}f''(R)\right\} + \rho_{m}
\ee 
\be\label{eqn:curv2}
2\frac{\ddot{a}}{a} + \frac{\dot{a}^2}{a^2} = - \frac{1}{f'(R)}\left
\{2 \frac{\dot{a}}{a} \dot{R} f''(R) + \ddot{R} f''(R) + \dot{R}^2 f'''(R) 
- \frac{1}{2}\left(f(R)- R f'(R)\right) \right\}- p_{m}~.
\ee
It is evident that if $f(R) = R$, the field equations (\ref{eqn:curv1}) and 
(\ref{eqn:curv2}) take the form of usual Einstein field equations. \\
The Ricci scalar $R$ is given by 
\be
R = -6 \left[\frac{\ddot{a}}{a} + \frac{{\dot{a}}^2}{a^2} \right]~,
\ee
which involves a second order derivative of the scale factor $a$. As equation 
(\ref{eqn:curv2}) contains $\ddot{R}$, one actually has a system of fourth 
order differential equations. Depending on the functional form of $f(R)$, 
some of the terms on the right hand side of equation (\ref{eqn:curv2}) can 
provide an effective negative pressure and generate sufficient acceleration. 
\par A substantial amount of work has already been done along this line by 
choosing various functional forms of $f(R)$. Capozziello et. al \cite{capo1, capo2} 
considered $f(R) = R^n$ and showed that it leads to an accelerated expansion 
for $n = -1$ and $n = \frac{3}{2}$. The dynamical
behaviour of $R^n$ gravity has been studied in detail by Carloni et. al
\cite{carloni}. Carroll et. al \cite{carroll} used a combination 
of $R$ and $\frac{1}{R}$ in the action and a conformally transformed version 
of the theory where the effect of curvature is formally taken care of by a 
scalar field having some potential. They showed that it could generate a 
negative value for the deceleration parameter $q$. Vollick \cite{vollick} 
used $\frac{1}{R}$ term in the action and obtained an exponentially 
expanding and hence accelerating model for the universe. It deserves mention 
that Vollick actually employed a Palatini variation, so the field equations 
are different from equations (\ref{eqn:curv1}) and (\ref{eqn:curv2}). 
Nojiri and Odinstov \cite{nojiri} considered the Lagrangian of the form 
\begin{center}
$L = R + R^m + R^{-n}$ ~where $m, n$ are positive integers,
\end{center}
and showed that it is indeed possible to obtain an inflation at the early 
stage and a late time accelerated expansion from the same set of field 
equations. Other interesting investigations 
include the choice of $f(R)$ as $\sinh^{-1}(R)$ \cite{borowiec} or 
$\ln R$ \cite{odin}, 
which also could provide late time acceleration. However, all these models 
mentioned above have problems regarding the stability \cite{dolgov}. 
Furthermore, most of 
them either resort to a piecewise solution for large $R$ and small $R$, 
or provide acceleration in some limit or an eternally accelerating model. 
But none of them could show the transition from decelerated to 
accelerated phase of expansion in the same matter dominated regime. But still 
these investigations open up an interesting possibility for the search of 
dark energy in the non-linear contributions of the scalar curvature. 
\vskip .2in
\subsection{Chaplygin Gas Models :-}
A \emph{chaplygin gas model} is also one of the important candidates for 
solving the dark energy problem. The Born-Infeld lagrangian density 
\be
L = - V_{0}\sqrt{1 - \phi_{,\mu}\phi^{,\mu}}
\ee
leads to the chaplygin gas obeying the equation of state
\be\label{eqn:chap}
p = -\frac{A}{\rho}
\ee
where $A = {V_{0}}^2$. \\
The chaplygin gas can also be derived from a quintessence lagrangian
$$L = \frac{1}{2}{\dot{\phi}}^2 - V(\phi)$$
with the potential \cite{kamen}
\be
V(\phi) = \frac{\sqrt{A}}{2}\left(\cosh 3\phi + \frac{1}{\cosh 3\phi}\right)~.
\ee
The conservation equation
\be
d(\rho a^3) + p~d(a^3) = 0
\ee
immediately gives
\be\label{eqn:density1}
\rho = \sqrt{A + \frac{B}{a^6}}
\ee
where $B$ is a constant of integration if the equation of state is given by 
equation (\ref{eqn:chap}). \\
The more general form of equation (\ref{eqn:chap}) leads to the equation 
of state for the generalized chaplygin gas given 
by (see \cite{bento} and references therein ) 
\be
p = -\frac{A}{\rho^{\alpha}}
\ee
where $0 < \alpha < 1$ and $A$ is a positive constant. $\alpha = 1$ gives back 
the old chaplygin gas model. In the framework of FRW cosmology, this equation 
of state yields solution of the Einstein equations and leads to density evolving as 
\be\label{eqn:density2}
\rho ={\left( A + \frac{B}{a^{3(1 + \alpha)}}\right)}^{\frac{1}{1 + \alpha}}, 
\ee
where $a$ is the scale factor of the universe and $B$ is an integration 
constant. From equations (\ref{eqn:density1}) and (\ref{eqn:density2}), 
it is clear that for small $t$, 
i.e, for small $a$, one can obtain a dust dominated model 
$\left(\rho \sim \frac{1}{a^3}\right)$ whereas for large $t$, $\rho \sim 
\sqrt{A}$ or ($\rho \sim A^{\frac{1}{1 + \alpha}}$) and it behaves like a 
cosmological constant and provides acceleration. 
\vskip .2in
\subsection{Phantom Dark Energy Models :-} 
Caldwell \cite{caldwell}  pointed out that a very 
good fit to the luminosity-distance curve (Figure \ref{fig3}) can be 
provided by a dark energy component which violates the weak energy 
condition so that the equation of state $w < -1$. He dubbed  
this candidate as ``\emph{phantom dark energy}''. A study of high-z 
supernovae \cite{knop} also reveals that 
the dark energy equation of state has 99\% probability of having a value $<-1$ 
if no constraints are imposed on 
$\Omega_{m}$.  
\par A number of models appeared in the literature where the dynamical 
nature of phantom energy was constructed by taking a kinetic term with a 
`wrong' sign in equation (\ref{eqn:pphi}) so that it can give rise 
to the present 
acceleration of the universe \cite{mariusz, sun, singh, stefancic}. 
Phantom dark energy models have also been studied in Brans-Dicke 
theory \cite{smcarroll} or in an 
interacting scenario where the phantom field is coupled to some other 
field \cite{unified, majerotto}.
However, these models suffer from the 
problem of instability at the quantum level \cite{cline} as there is no proper 
ground state because of the negative kinetic energy. 
Also models with $w< -1$ suggest that the effective velocity of sound 
in the medium 
$v = \sqrt{~\vline~ dp / d\rho~\vline~}$ can become larger than the 
velocity of light. The phantom models imply a pathological behaviour for the 
cosmological model at a finite future.  
If $t_{eq}$ denotes the time when matter density and phantom energy 
density become equal, then 
the scale factor of the universe grows as 
\be
a(t) \approx a(t_{eq}){\left[(1 + w)\frac{t}{t_{eq}} - w\right]}^\frac{2}{3(1 + w)},~~w < -1~.
\ee
Therefore, when $t \rightarrow \left(\frac{w}{w + 1}\right)t_{eq}$, 
$a(t) \rightarrow \infty$~, i.e, 
the scale factor diverges in a finite time. At that epoch, Hubble parameter 
$H$ also diverges 
implying that the expansion rate of the universe reaches an infinite value 
in a finite time. This situation is termed as `\emph{Big Rip}'. 
Thus the universe 
dominated by phantom energy culminates to a 
future curvature singularity ( See 
\cite{staro, chiba, caldwell, caldwell2, mcinnes, hoffman, frampton1, 
frampton2, vbjohri, alcaniz, kaplinghat}).
However, some other models have also been investigated where $w < -1$ is 
attained without considering 
a negative sign for the kinetic term. These models are called 
`\emph{Braneworld models'} \cite{vsahni, alam} 
which has $w_{eff} < -1$ today, but does not run into a `Big Rip' 
in finite future. 
\vskip .2in
\subsection{Braneworld Models :-}
\begin{figure}[!h]
\centerline{\psfig{figure=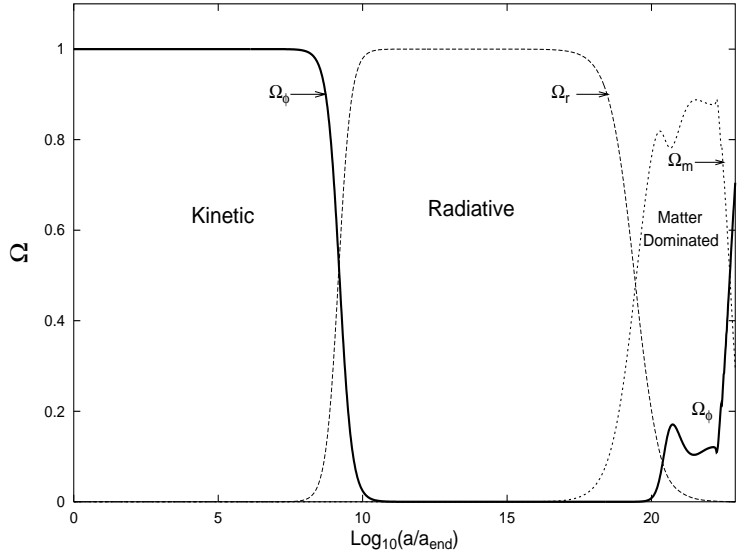,height=75mm,width=100mm}}
\caption{\normalsize{\em The post-inflationary density parameter $\Omega$ is plotted for the scalar field (solid line), 
radiation (dashed line), and cold dark matter (dotted line) in the quintessential inflationary model. 
(From Sahni, Sami and Souradeep \cite{sami})}}
\label{brane}
\end{figure}

Braneworld cosmology suggests that we could be living on a four  
dimensional `\emph{brane}' which is 
embedded in a five or higher dimensional `\emph{bulk}'. It is considered that 
matter fields are confined 
to the brane whereas gravity is free to propagate throughout the bulk 
(for a comprehensive 
discussion, we refer to the lectures by Roy Maartens \cite{roy}). In the 
Randall-Sundrum (RS) \cite{randall} 
scenario, the equation of motion of a scalar field propagating in the 
brane is given by
\be
\ddot{\phi} + 3 H \dot{\phi} + V'(\phi) = 0~,
\ee
where
\be\label{eqn:brane}
H^2 = \frac{8 \pi}{3 m^2}\rho\left( 1 + \frac{\rho}{2 \sigma} \right) + \frac{\Lambda_{4}}{3} + 
\frac{\varepsilon}{a^4}~,\\
\ee
\be
\rho = \frac{1}{2}{\dot{\phi}}^2 + V(\phi)~.
\ee
Here, $\varepsilon$ is an arbitrary constant and $\sigma$ is the brane tension which relates the 
four-dimensional Planck mass ($m$) and the five-dimensional Planck mass ($M$) as
\be
m = {\sqrt{\frac{3}{4 \pi}}}\left(\frac{M^3}{\sqrt{\sigma}}\right)~.
\ee
Also the four-dimensional cosmological constant $\Lambda_{4}$ on the brane and the five-dimensional 
cosmological constant $\Lambda_{b}$ on the bulk are related as
\be
\Lambda_{4} = \frac{4 \pi}{M^3}\left( \Lambda_{b} + \frac{4 \pi}{3 M^3}\sigma^2 \right)~.
\ee
Equation (\ref{eqn:brane}) contains an additional term $\frac{\rho}{2 \sigma}$ because of which the 
damping experienced by the scalar field as it rolls down the potential dramatically increases so that 
inflation can be sourced by potentials, such as $V \propto e^{-\lambda\phi}$, $V \propto \phi^{-\alpha}$ 
etc, which are normally too steep to produce slow-roll. This gives rise to the possibility that both 
inflation and quintessence may be obtained from the same scalar field. These models are called `\emph{quintessential inflationary models}' 
(see \cite{peebles, copeland, huey, sami, majumdar, liddle} and 
references therein). An example of quintessential inflation is shown in 
Figure (\ref{brane}).
\par A different way of obtaining an accelerating universe was suggested in 
the braneworld model developed by Deffayet, Dvali and Gabadadze (DDG) 
\cite{deffayet1, deffayet2}. Here both $\Lambda_{b}$ and $\sigma$ were 
set to zero, while 
a curvature term was introduced in the brane action so that it takes the form 
\be
S = M^3\int_{bulk}{\it{R}} +~ m^2\int_{brane} R + \int_{brane} L_{matter}~.
\ee
The resulting Hubble parameter in the DDG braneworld model is 
\be
H = \sqrt{\frac{8 \pi G \rho_{m}}{3} + \frac{1}{l_{c}^2}} + \frac{1}{l_{c}}~,
\ee
where $l_{c} = \frac{m^2}{M^3}$ is a new length scale. An important property 
of this model is that the acceleration of the universe is not obtained from 
any `\emph{dark energy}' component. Since gravity 
becomes five dimensional 
on length scales  $R > l_{c} = 2 H_{0}^{-1} ( 1 - \Omega_{m})^{-1}$, it is 
seen that the expansion of the universe is modified during late times instead 
of early times as in the RS model.
\par A more general class of braneworld models, which includes RS cosmology 
and DDG brane is described by the action \cite{collins, shtanov}
\be
S = M^3\int_{bulk}{\it{(R - 2\Lambda_{b})}} +~ \int_{brane} (m^2 R - 2\sigma) 
 + \int_{brane} L_{matter}~.
\ee
For $\sigma = \Lambda_{b} = 0$, it gives back the DDG model, whereas for 
$m = 0$ it reduces to RS model. 
Sahni and Shtanov \cite{vsahni} have shown that the Hubble parameter for 
this action comes out as 
\be\label{eqn:branegeneral}
\frac{H^2(z)}{H_{0}^2} = \Omega_{m}(1 + z)^3 + \Omega_{\sigma} + 2\Omega_{l} 
\mp 2 \sqrt{\Omega_{l}}\sqrt{\Omega_{m}(1 + z)^3 + \Omega_{\sigma} + 
\Omega_{l} + \Omega_{\Lambda_{b}}}~,
\ee
where $\Omega_{l} = \frac{1}{l_{c}^2 H_{0}^2}$, $\Omega_{m} = 
\frac{{\rho_{0}}_{m}}{3 m^2 H_{0}^2}$, $\Omega_{\sigma} = \frac{\sigma}
{3 m^2 H_{0}^2}$, $\Omega_{\Lambda_{b}} = -\frac{\Lambda_{b}}{6 H_{0}^2}$.
(The $\mp$ sign refers to two different ways in which the brane can be 
embedded in the bulk (for details see \cite{vsahni}). 
\par An important feature of the braneworld model given by equation 
(\ref{eqn:branegeneral}) is that it can lead to an effective equation of state of 
dark energy $w_{eff} \le -1$. Also it has been shown that in this model, the 
acceleration of the universe can be a transient phenomenon which ends once 
the universe returns to matter dominated expansion after the current 
accelerated phase of expansion and hence does not fall into the problem of 
`Big Rip'.
\vskip .2in
\section{Outline of the Present Thesis :}
\markright{Outline of the Thesis}
Whatever we have discussed so far indicates that our universe at present is
undergoing an accelerated expansion preceded by a decelerated one. This
means that the deceleration parameter $q$ must have a signature 
flip from a positive
value to a negative one in the recent past during the matter dominated era
itself. This indeed leads to the search for a candidate which can drive this
transition from deceleration to acceleration. This thesis is a collection of
papers which investigates some cosmological solutions where this transition is
obtained with or without the need of ``\emph{dark energy}''.\\
The investigations carried out will be presented in four chapters following 
the introduction. The first of them (chapter-2) consists of one paper entitled
``\emph{Acceleration of the universe with a simple trigonometric potential}''. 
In this paper we consider a minimally coupled scalar field $\phi$, 
with some potential $V(\phi)$, 
as the driver of the late time acceleration. The 
form of deceleration parameter $q$ as a function of scale factor $a$ has been chosen such that it has the desired property of signature flip and then from this the scalar field and required potential are found out.\\
The flat Robertson-Walker line element is written as  
\be
{ds}^2 = {dt}^2 - {a(t)}^2 [ {dr}^2 + r^2 {d\Omega}^2]~.
\ee 
The expression for the deceleration parameter $q$ is taken as,
\be\label{eqn:q}
q = -\frac{\ddot{a}/a}{{\dot{a}}^2/a^2} = -1 - \frac{p a^p}{(1 + a^p)}
\ee
where $p$ is a constant. It is found that for $-2 < p < -1$, the deceleration
parameter $q$ indeed has the property of a signature flip. 
Equation (\ref{eqn:q}) 
integrates to yield 
\be
H = \frac{\dot{a}}{a} = A(1 + a^p), ~~A > 0~,
\ee
where $A$ is an arbitrary constant of integration. 
\par In the present work, the problem is completely worked out 
for $p = -\frac{3}{2}$. The
reason for choosing this particular value of $p$ is that it yields $q = 0.5$ 
for a very low value of $a$. For a flat matter dominated FRW model without any 
quintessence matter indeed $q = 0.5$. So the matter dominated universe has a 
chance to evolve from the standard setting where the formation of structures 
is facilitated. Furthermore, 
the transition from the radiation dominated era to the matter dominated one is 
quite well understood for $q = 0.5$ on the matter-dominated side.
\par As we are interested in the matter dominated epoch, the fluid is taken in
the form of pressureless dust and the Einstein equations are written with a
minimally coupled scalar field $\phi$ having a potential $V(\phi)$. 
\par With the form of the deceleration parameter assumed in equation 
(\ref{eqn:q}), the
Einstein field equations are solved and it is found that the potential 
$V(\phi)$ is obtained as a trigonometric function of the scalar field $\phi$, 
viz, $ V(\phi) = \frac{9A^2}{2}\cot^2 \left(\frac{\sqrt{3}\phi}{4}\right) + 
3 A^2$.
Also the expressions for the dimensionless density parameter $\Omega_{m}$ 
for the normal matter and $\Omega_{\phi}$ for the quintessence matter 
are worked out for this model and it is found that the values obtained are 
well within the constraint range obtained from observations \cite{sahnirev, 
sahni, adam}.\\
The equation of state parameters $w$ for the total matter ($w_{t}$) 
and that for the quintessence matter ($w_{\phi}$) are calculated and are 
found to be consistent with observational requirements. 
\par This investigation shows that we can have a single analytical
expression for $q$ which gracefully transits from deceleration to
acceleration and thus has a better footing than the piecewise solutions. Also
the form of the potential obtained here is a simple trigonometric function
of $\phi$ and thus adds to the list of quintessence potentials that serve the
purpose of providing a presently accelerating universe \cite{varun}. \\
\par The chapter 3 consists of one paper entitled ``\emph{Spintessence : 
A possible candidate as a driver of the late time cosmic acceleration}''. In
this paper we work with a complex scalar field model, namely spintessence,
proposed by Boyle et al. \cite{boyle}. The scalar field is taken as,
\be
\Psi = \phi_{1} + i \phi_{2}~,
\ee
which can also be written as
\be
\Psi = \phi e^{i\omega t}
\ee
such that the complex part is taken care of by a phase term. 
This kind of complex scalar fields are already 
known in the
literature with reference to the `\emph{cosmic strings}' \cite{vilenkin}. 
However, cosmic
strings have a very specific form of potential whereas in quintessence models,
the form of potential has to be found out. 
\par In the present work it has been shown completely analytically that the
deceleration parameter $q$ indeed has a signature flip from positive to
negative values which indicates an early deceleration and a 
present acceleration.
 In the spintessence model proposed by Boyle et al. \cite{boyle}, it has been
considered that $\omega$ is a slowly varying function of time which indeed
can be approximated to be a constant over 
the entire period of dust dominated era.
In the present work, we also choose $\omega$ to be a constant. This choice
along with the field equations leads to an equation
\be
a^3 \frac{dH}{dt} = - l - mH^2~,
\ee
where $a$ is the scale factor of the universe, $H = \frac{\dot{a}}{a}$ is the
Hubble parameter and $l$, $m$ are positive constants, which involve the
parameters of the theory and constants of integration.
\par Now, we make a transformation of time co-ordinate as,
\be
\frac{dx}{dt} = \frac{1}{a^3}
\ee
which being positive definite indicates that $x$ is a monotonically increasing
function of $t$. We now express the various parameters of the model as
functions of new cosmic time $x$ without any loss of generality or distortion
of events.
\par With these choices, the Hubble parameter $H$ and the deceleration 
parameter $q$ comes out in terms of the new variable $x$ as, 
\be
H = n \tan (\beta - m n x)~
\ee
and
\be\label{spinq}
q = -1 + \frac{1}{a^3}\left(\frac{l}{H^2} + m \right)~.
\ee
It is evident from equation (\ref{spinq}) that $q$ has a zero when 
\be
{a_{1}}^3 = \frac{l}{{H_{1}^2}} + m~,
\ee
the suffix `1' indicates the values of the quantities for $q = 0$.\\
Also from equation(\ref{spinq}) it has been shown that 
${\frac{dq}{da}\vline}_{~1}$ is negative definite which indicates that $q$ is 
a decreasing function of $a$, at least when $q = 0$. Thus, $q$ definitely 
enters a negative value regime from a positive value at $a = a_{1}$ and 
$H = H_{1}$. So, it has been shown completely analytically that a
spintessence model can very well serve the purpose of providing the signature
flip in $q$.  The important result achieved here is that if $q$ changes its 
sign in course of evolution, this change will be in the right direction, i.e, 
from positive to negative. Also it has been shown that the flip occurs 
irrespective of the particular form of potential which is a bonus as
the form of the quintessence potential is yet to be specified.  
Another important feature of this work is 
that the value of the scale factor
$a$, where this flip in $q$ takes place, can be expressed in terms of
parameters of the theory and various constants of integration. The latter are
arbitrary constants and thus can be adjusted to fit into the observational
results.
\par The chapter 4 consists of two papers where attempts have been made to
explain the late time acceleration of the universe in the framework of scalar
tensor theories. In the first paper we work out the problem in Brans-Dicke
theory \cite{brans} whereas in the second paper we work in Nordtvedt's theory 
\cite{nordtvedt} which is a generalized version of Brans-Dicke theory.
\par The first paper in this chapter is entitled ``\emph{A late time 
acceleration of the universe with two scalar fields : many possibilities}'' and 
addresses the `\emph{graceful entry}' problem of how 
the universe transits from a
decelerating phase of expansion to an accelerated one. The basic motivation
for using two scalar fields stems from existing literature on inflation.
Mazenko, Wald and Unruh \cite{mazenko} showed that a classical slow roll is
invalid if the scalar field driving inflation is self interacting. Also it has 
been shown that a single scalar field with a slow roll puts generic restrictions
on the potentials driving inflation \cite{adams}. These problems led to the
belief that a successful inflationary model requires two scalar
fields \cite{linde}. The present work uses this idea of introducing two scalar
fields. One of these fields is responsible for the present acceleration of the
universe, called the quintessence field, which interacts with the other 
such that the quintessence field has an oscillatory behaviour
at the beginning of the matter dominated epoch and grows later so as to
dominate the dynamics of the universe. None of the 
quintessence fields already there in literature has a proper physical
motivation and it will be even more embarrassing to introduce a second field
without any underlying physics. So, the better arena is provided by 
a scalar tensor theory, such as Brans-Dicke theory, where one scalar 
field is already there
in the purview of the theory and is not put in by hand.
\par The relevant action in Brans-Dicke theory is given by
\be
S = \int \left[\frac{\psi R}{16 \pi G_{0}} - \omega \frac{\Psi_{,\mu}
        \Psi^{,\mu}}{\Psi} - \frac{1}{2}\phi_{,\mu}\phi^{,\mu} - U(\Psi, \phi)
         + L_{m}\right] \sqrt{-g} d^4x~,
\ee
where $\Psi$ is the Brans-Dicke scalar field, $R$ is the Ricci scalar, $G_{0}$
is the Newtonian constant of gravitation, $\omega$ is the dimensionless
Brans-Dicke parameter, $\phi$ is the quintessence scalar field and
$U(\Psi, \phi)$ is the potential via which the two fields interact amongst 
themselves. We choose the form of $U(\Psi, \phi)$ as $V(\phi){\Psi}^{-\beta}$
which was used by Banerjee and Ram \cite{nb} for finding a solution of
the graceful exit from inflationary paradigm in Brans-Dicke theory. 
With a slow roll
approximation $({\dot{\phi}}^2, \ddot{\phi} \approx 0)$, the conditions
for an initially oscillating $\phi$ which grows during later stages are
found out for two examples - a power law expansion and an exponential
expansion of the scale factor $a$.
\par For a power law expansion of the scale factor of the form
\be
a = a_{0} t^n, ~~n > 1,
\ee
the condition on the potential $V(\phi)$ comes out as
\be
{(ln V)''\vline}_{~i} \approx -\frac{1}{A} {t}^{3n - 2}_{i}
\ee
where $A$ is a constant involving the parameters of the model. The subscript
`$i$' stands for some initial time $(t = t_{i})$.\\
Similarly, for the exponential expansion of the form
\be
a = a_{0} e^{\alpha t},  ~~~~a_{0}, \alpha > 0~,
\ee
the condition on the potential comes out as
\be
{(\ln V)''\vline}_{~i} \approx 0~.
\ee
This indicates that for an exponentially expanding universe, the quintessence
potential $V(\phi)$ should be exponential at least in the beginning.
\par In this paper two simple examples have been considered. However, this
work can be extended for more complicated kind of accelerated expansion of
the universe. A key feature of this work is that here the numerical value
of the Brans-Dicke parameter $\omega$ is not severely constrained to very 
low values in order to drive the acceleration. For example,
in the exponential expansion case, $\omega$ is completely arbitrary whereas
in power law expansion case it has been found that value of $\omega$ can be
determined in terms of other free parameters of the model. So, in both the
cases $\omega$ can be adjusted to some high value which is compatible with the
limits imposed by solar system experiments \cite{will}. 
\par In the second paper in this chapter entitled ``\emph{An interacting scalar
field and the recent cosmic acceleration}'', the possibility of transition from
deceleration to acceleration has been studied in the framework of a generalised
scalar tensor theory.
\par The dark matter and dark energy components are usually 
considered to be non-interacting and their evolutions are considered to be
independent of each other. Recently Zimdahl and Pavon \cite{zimdahl1, 
zimdahl2} showed 
that the interaction between the dark matter and dark energy components 
could be useful in solving the coincidence problem and an interacting 
scenario may provide a more general and better framework for obtaining  
an accelerating universe (see also \cite{zimdahl3, Tsujikawa1, Tsujikawa2}) .
\par In this paper, we introduce an interaction between the dark matter and
the geometrical scalar field $\phi$. The Brans-Dicke field equations are
written in so called ``Einstein frame'' as discussed earlier. The field
equations in this version look tractable, although one has to sacrifice the
equivalence principle as the rest mass of a test particle becomes a function
of the scalar field $\phi$. However, in this work, the metric has been 
transformed back to the original atomic units and the conclusions are drawn 
only from this.\\
In place of the usual matter conservation equation
\be
\dot{{\bar{\rho}}_{m}} + 3\bar{H}{\bar{\rho}}_{m} = 0 ~,
\ee
we have chosen an interaction between the dark matter and the scalar field
$\phi$ of the form
\be{\label{eqn:int}}
\dot{{\bar{\rho}}_{m}} + 3\bar{H}{\bar{\rho}}_{m} = - \alpha \bar{H} 
{\bar{\rho}}_{m}
\ee
where ${\bar{\rho}}_{m}$ is the matter density of the universe, 
$\bar{H}$ is the
Hubble parameter, $\alpha$ is a positive constant. The overhead bars
indicate that the equations are written in the conformally transformed version
(Einstein's frame) where the transformed metric components are related to the
original ones as
\be
\bar{g}_{\mu\nu} = \phi g_{\mu\nu}~.
\ee
The negative sign in equation (\ref{eqn:int}) indicates that the energy is
transferred from the dark matter component to the scalar field $\phi$.
It is found that this particular choice of interaction gives rise
to a simple power law solution of the scale factor $a$ - which can be either
ever accelerating or ever decelerating depending upon the choice of
parameters. Obviously we are not interested in such a scenario which does
not provide the transition from deceleration to acceleration. One way out
of this problem is to consider a generalization of Brans-Dicke theory where
the parameter $\omega$ is a function of the scalar field $\phi$ rather than
being a constant \cite{nordtvedt}.
\\
We make a choice of $\omega$ as
\be{\label{eqn:w}}
\frac{2\omega + 3}{4} = \frac{\alpha}{{(3 + \alpha)}^2}
              \frac{\phi}{{(\sqrt{\phi} - 1)}^2}~,
\ee
$\alpha$ being a positive constant.
\\
With this choice we obtain the solution for the scale factor in original
units as
\be{\label{eqn:a}}
a = \frac{A t^{\frac{2}{(3 + \alpha)}}}{(1 - \phi_{0} t)}~,
\ee
$\phi_{0}$, $A$ are constants.
\par From equation (\ref{eqn:a}) one can easily find out the expression for
$q$ in the original version as  
\be
q = -1 + \frac{\frac{2}{3 + \alpha}(1 - \phi_{0} t)^2 - {\phi_{0}}^2 t^2}
    {[\frac{2}{3 + \alpha}(1 - \phi_{0} t) + \phi_{0} t]^2}~.
\ee
The $q$ vs. $t$ plot shows the transition
from positive value to a negative value for small values of $\alpha$
and also the nature of the curve is not critically sensitive to small
changes in the value of $\alpha$.
\par However, from equation (\ref{eqn:a}) it is evident that at
$t \rightarrow \frac{1}{\phi_{0}}$, $a$, $H$ and $q$ all blow up. So the model
is not valid upto infinite future and has a future singularity. 
So the model mimics a 
phantom model and has a big rip. 
\par Also, it is evident from equation (\ref{eqn:w}) that $(2\omega + 3)$
has to be positive definite to allow a consistent model. But $\omega$ does not
have any stringent limit and thus it is possible to adjust $\omega$ to
some high value, compatible with the limits imposed by the solar system
experiments \cite{will}. Also equations (\ref{eqn:w}) and (\ref{eqn:a}) reveal
that in the $\omega \rightarrow \infty$ limit, $a \rightarrow t^{2/3}$ for
small values of $\alpha$. This is consistent with the notion that
Brans-Dicke theory yields General Relativity in the infinite $\omega$ limit.
Also for this model we have calculated the statefinder parameters \{$r, s$\}
introduced by Sahni et. al \cite{alam1, alam2} and found that the nature of 
the $r$ vs. $s$ plot for small values of $\alpha$ is similar to the one 
expected for scalar field quintessence models.
\par The salient feature of this model is that no dark energy component
is required here. The interaction between the geometrical scalar field
and matter component can drive the present acceleration of the universe.
Although a particular form of interaction has been chosen, this is by 
no means unique and other forms of interaction can also be tried. 
\par The chapter 5 consists of one paper entitled ``\emph{Curvature driven
acceleration : a utopia or a reality}?''. This work explores the possibility
whether geometry itself can serve the purpose of explaining the present
accelerated expansion without having to resort to some exotic scalar
field models. A substantial amount of work has been done along this direction 
which can provide an accelerating model for the universe \cite{capo1, capo2, 
carloni, carroll, vollick, nojiri, borowiec, odin} as discussed earlier. 
However, none of them could provide a smooth transition 
from deceleration to acceleration demanded by both theory \cite{paddy1, 
paddy2}  and observations \cite{ag}.
\par In the present work, we have written the field equations with a general 
$f(R)$ and studied the model for two specific cases, namely, $f(R) = 
R - \frac{\mu^4}{R}$ and $f(R) = e^{-R/6}$. In both the cases it is shown that 
the required transition from deceleration to acceleration can be obtained.\\
The relevant action for a general $f(R)$ is 
\be
A = \int\left[\frac{1}{16 \pi G}f(R) + L_{m}\right] \sqrt{-g} d^4x~.
\ee
For a spatially flat Robertson-Walker spacetime, where
\be
ds^2 = dt^2 - a^2(t) [dr^2 + r^2d\theta^2 + r^2\sin^2 \theta d\phi^2]~,
\ee
we write down the field equations for a vacuum universe taking $L_{m} = 0$.\\
In the first example we make a choice of $f(R)$ as 
\be
f(R) = R - \frac{\mu^4}{R},
\ee
where $\mu$ is a constant. This is exactly the form used by Carroll 
\cite{carroll} and Vollick \cite{vollick}. \\
For this particular choice, from the field equations we arrive at a relation 
\be\label{curv1}
2\dot{H} = \frac{1}{(R^2 + \mu^4)}\left[2\mu^4\frac{\ddot{R}}{R} - 6\mu^4
\frac{\dot{R}^2}{R^2} - 2\mu^4 H\frac{\dot{R}}{R}\right]~,
\ee
where the Ricci scalar $R$ is given by 
\be
R = -6\left(\frac{\ddot{a}}{a} + \frac{\dot{a}^2}{a^2}\right)~.
\ee
Equation (\ref{curv1}) is a highly nonlinear equation involving 
fourth order derivative of the scale factor $a$ and it is difficult to obtain 
a complete analytic solution for $a$. But as we are interested in the 
evolution of the deceleration parameter $q$, we translate equation 
(\ref{curv1}) into the evolution equation for $q$ by using the relation 
\be
q = -\frac{\ddot{a}/a}{\dot{a}^2/a^2} = -\frac{\dot{H}}{H^2} - 1~.
\ee
Also, the time derivatives are replaced by the derivatives w.r.t. $H$. The 
resulting equation takes the form 
\be
\frac{1}{3}q^{\dagger\dagger}(q^2 - 1)H^2 - \frac{2}{3}H^2{q^{\dagger}}^2 
(q + 2) - \frac{1}{3}q^{\dagger}H(q - 1)(4q + 7) - (2q + 1)(q - 1)^2 + 
H^4(q - 1)^4 = 0~,
\ee
where a $\dagger$ sign indicates differentiation w.r.t. $H$.\\
This equation is also highly nonlinear and cannot be solved analytically. But 
if one can provide two initial conditions for $q$ and $q^{\dagger}$, a 
numerical solution is on cards. So, we pick up sets of values for $q$ and 
$q^{\dagger}$ for $H = 1$ (i.e, the present values) from observationally 
consistent regions \cite{alam1, alam2} and plot $q$ versus $H$ numerically. 
The plot definitely shows a signature flip in $q$ from positive towards 
negative. Also, the nature of the curve is not critically sensitive to the 
initial conditions chosen.\\
In the second example, the form of $f(R)$ is chosen as 
\be
f(R) = e^{-R/6}~.
\ee
Following the same method as before, the evolution equation for $q$ is written 
as a function of $H$. With similar initial conditions for $q$ and 
$q^{\dagger}$ at $H = 1$, $q$ is numerically plotted against $H$ and it is 
found that the curve has features similar to the previous example, i.e, $q$ 
has a signature change from positive to negative in the recent past. The 
added feature of this latter example is that $q$ has another 
signature flip from 
negative to positive direction indicating a decelerated expansion again in 
near future. 
\par So, in the present work both the examples indicate that gravity in its 
own right can lead to a late surge of accelerated expansion having a past 
deceleration which is essential for explaining nucleosynthesis and the 
structure formation of the universe. An added advantage of the second example 
is that in this case the universe re-enters a decelerated expansion phase 
in the near future and thus `\emph{phantom menace}' is avoided, 
i.e, the universe 
does not attain infinite rate of expansion in a finite future. 
\par There is already some criticism of $\frac{1}{R}$ gravity regarding its 
stability \cite{dolgov}, but still there are reasons to be optimistic about 
a curvature driven acceleration. It deserves mention that although the 
$\frac{1}{R}$ term has been quite widely used in the literature, it suffers 
from the drawback that at $R = 0$, i.e, in the late stage of the evolution, 
the model has a singularity. The other example, i.e, $e^{-R/6}$ is 
regular everywhere. The present work deals with a vacuum universe, but in a 
more general framework one has to either put in matter or should derive the 
relevant matter from curvature itself. Thus the modified theory of gravity 
provides a platform for more ambitious and detailed work in future.
\newpage
\addcontentsline{toc}{part}{\normalsize{Bibliography}}

\markright{}
\chapter{A Simple Quintessence Field}
\markright{A Simple Quintessence Field}
\newpage
\section{Acceleration of the Universe with a Simple Trigonometric Potential \\(Journal reference : N. Banerjee and S. Das, Gen. Rel. Grav., {\bf 37}, 1695 (2005); astro-ph/0505121.)}
\newpage
\subsection{Introduction}
Over the last few years, there are growing evidences in favour
of the scenario that the universe at present is expanding with an
acceleration. The supernovae project \cite{ch2perl} and also the
Maxima \cite{ch2netter} and 
Boomerang \cite{ch2balbi} data on cosmic microwave background (CMB) strongly
suggest this acceleration. The very recent WMAP data \cite{ch2dns} also seem
to confirm this. This result is indeed counter-intuitive, as gravity holds
matter together and it might be expected that in the absence of any exotic
field, the universe should be decelerating. As this acceleration could be
brought about by an effective negative pressure, the first choice of candidate
for this  `dark energy' had been the `cosmological constant' or a time
varying cosmological parameter $\Lambda(t)$. Due to well-known
reasons, $\Lambda$ has fallen from grace (for an excellent uptodate review,
see \cite{ch2vs},\cite{ch2peddy}). A scalar field with a positive definite 
potential can indeed give rise to an effective negative pressure if the
potential term dominates over the kinetic term. The pressure to density
ratio for the scalar field, as required by the supernovae observations,
is given as $w_{\phi} < -\frac{2}{3}$ (See ref \cite{ch2macorra} and references therein).
This source of energy is called
the quintessence matter (Q-matter).
 In this context, non-minimally coupled scalar fields had been
investigated thoroughly to check if they could drive an accelerated expansion
\cite{ch2berto}. Brans-Dicke's scalar field appears to generate sufficient acceleration
in the matter era, but it has its problems in the earlier evolution \cite{ch2nb}.
A viscous fluid along with a Q-matter could also be a useful candidate and
this appears to solve the coincidence problem also \cite{ch2chimento}. This
coincidence problem, i.e, why the Q-matter dominates only recently,
was tackled  in the so called tracker
solutions \cite{ch2caldwell} where the scalar field energy density runs parallel to the matter energy density from below through the evolution and gets to dominate only during later stages. Very recently Chaplygin gas, which has a nonlinear
contribution of the energy density to the dynamics of the model, has also been invoked \cite{ch2aas}.
Most of these models do exhibit an accelerated expansion in the
matter-dominated regime.\\
\par  It deserves mention that the same model should have a deceleration
in the early phase of matter era 
in order to provide a perfect ambience for structure formation.
Furthermore, the accelerated phase is perhaps only a very recent one. There
are observational evidences too that beyond a certain value of the
redshift ($z \sim 1.7$), 
our universe had been going through a decelerated expansion \cite{ch2reiss}. 
This indication is indeed reassuring, as the formation of structure in the
universe is better supported by a decelerating model. This is because
local inhomogeneities will grow and become stable from the seeds of density
fluctuation only if the force field is attractive.\\

\par Amendola \cite{ch2amen} has argued that all the required structure
formation and other relevant observations regarding the supernovae could
well be explained even if the alleged acceleration of universe started
quite a long time back, even beyond $z = 5$. However, this work also shows
that the model requires both an accelerated and a decelerated phase of
expansion. But a more recent work by Padmanabhan and Roychowdhury
\cite{ch2ptr} shows a striking result. It indicates that if we take the
complete data set, i.e, acceleration upto a certain $z$ and deceleration
beyond that (i.e, for higher $z$), then only this conclusion of the
change of signature of
the deceleration parameter holds. On the other hand, the individual
data sets of the high and low redshift supernovae may well be consistent
with a decelerating universe without any `dark energy'.\\

       So indeed we are in need of some form of fields which governs the
dynamics in such a way that the deceleration parameter becomes negative
well into the matter era. One such Q-matter had been given by Sen and
Sethi \cite{ch2ss} where they include a potential which is a `double exponential'
of the scalar field. They obtained a scale factor which is a sine hyperbolic
function of time in the matter dominated regime. The deceleration parameter
$(q)$ indeed has a sign flip and with a little fine-tuning, the scale factor 
can grow during the early stages as $ t^{2/3}$ which is indeed the usual
solution for the Einstein equations for a flat FRW spacetime for
pressureless dust.\\

             In the present work, we adopt the following strategy. We choose a
form of $q$ as a function of the scale factor  $a$ so that it has the desired
property of a signature flip. Then with this input, the scalar field and the
required potential are found out. It turns out that a fairly simple
trigonometric
potential does the needful. The origin of the scalar potential, however, cannot
be indicated.
Surely this is not the ideal way to find out the dynamics of the universe, as
here the dynamics is assumed and then the fields are found out without any
reference to the origin of the field. But  in the
absence of more rigorous ways, this kind of investigations collectively might
finally indicate towards the path where one really has to search. This
{\it`reverse'} way of investigations had earlier been used extensively by
Ellis and Madsen \cite{ch2em} for finding out the potential driving inflation,
i.e, an accelerated phase of the universe at a very early stage of its
evolution.
\noindent
\subsection{Results}
\par For a spatially flat Robertson-Walker spacetime
\be\label{chap2eq1}
ds^2  = dt^2 - a^{2}(t) [dr^2 + r^2 d\Omega^2],
\ee 
the deceleration parameter $q$ is given by
\be\label{chap2eq2}
q = - \frac{ \ddot{a} a}{ \dot{a}^{2}}
\ee
where $a$ is the scale factor of the universe and is a function of
the cosmic time `$t$' alone.\\ In order to get a model consistent with
observations, one needs an expanding universe, i.e, a positive Hubble
parameter $ H = \frac{\dot{a}}{a} $ throughout the evolution, but a
deceleration parameter $q$, unlike being a positive constant throughout
the matter era at $ q = 1/2 $ as believed until the recent observations,
should be a function of the scale factor ( or that of $t$ ). Furthermore, this
functional dependence should be such that $q$ undergoes a transition from its
positive phase to a negative one in the matter dominated period itself. It is
thus imperative that the scale factor cannot have a simple power-law behaviour. 
If $ a \sim t^{n}$, the universe will have an accelerated or a decelerated
expansion for $ n > 1 $ or $ n < 1 $ respectively throughout the period.\\

  In the quest for a varying $q$ consistent with observations, in the same line
  as that floated by Ellis and Madsen, we propose the relation
\be\label{chap2eq3}
q = - \frac{ \ddot{a}/{a}}{ \dot{a}^2/ a^2} = - 1 - \frac{ pa^p}{1+a^p},
\ee
where $p$ is a constant. It is found that for a certain range of negative
values of $p$, this works remarkably well. 

 The equation (\ref{chap2eq3}) integrates to yield
\be\label{chap2eq4}
H = \frac{ \dot{a}}{a} = A(1+a^p)
\ee
where $A$ is an arbitrary constant of integration. $A$ is taken to be
positive, which ensures the positivity of the Hubble parameter (the expansion
of the universe is never denied!) irrespective of the signature or value of the
constant $p$.

\begin{figure}[!h]
\centerline{\psfig{figure=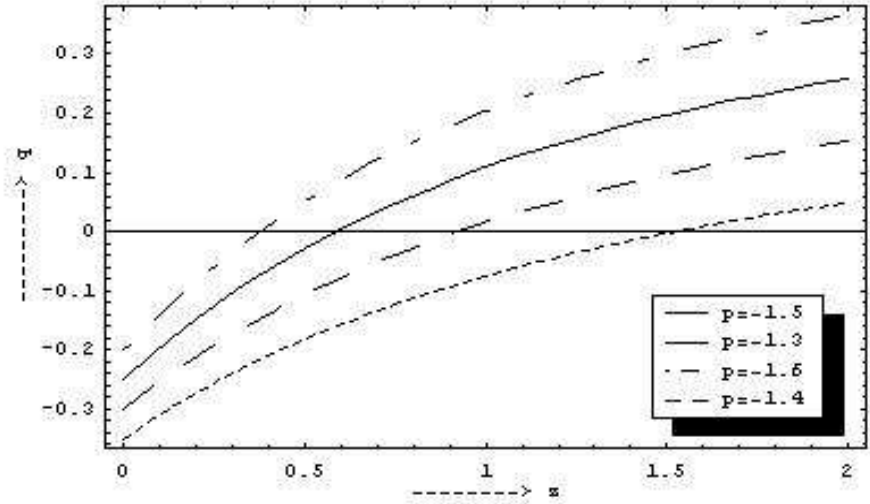,height=70mm,width=120mm}}
\caption{\em Plot of $q$ vs. $z$ for spatially flat dust dominated R-W model.}
\label{chap2first_fig}
\end{figure}

~It is found that for values of $p$ between -2 and -1, the model shows
exactly the behaviour which is desired (as shown in Figure \ref{chap2first_fig}). 

\par In what follows, we work out the problem completely
for $p=-3/2$, for which
the model works with a non minimally coupled scalar field with a potential
expressed as a simple trigonometric function of the scalar field.

 As the interest is in a matter dominated universe, the fluid is taken in
the form of a pressureless dust.
The Einstein equations for the space-time given by equation 
(\ref{chap2eq1}) are,
\be\label{chap2eq5}
3\frac{ \dot{a}^2}{a^2} = \rho + \frac{1}{2} \dot{\phi}^2 + V(\phi),
\ee
\be\label{chap2eq6}
2\frac{\ddot{a}}{a} + \frac{ \dot{a}^2}{a^2} = - \frac{1}{2} \dot{\phi}^2 + V(\phi)
\ee
where $\rho$ is the density of matter, $\phi$ is the scalar field and
$V(\phi)$ is a scalar potential.\\
The wave equation for the scalar field is
\be\label{chap2eq7}
\ddot{\phi} + 3\frac{\dot{a}}{a}\dot{\phi} + V'(\phi) = 0
\ee
In all equations an overhead dot implies differentiation w.r.t. time and 
a prime is that w.r.t. the scalar field $\phi$. The matter conservation
equation, which can in fact be obtained from these three equations in view
of the Bianchi identity, yields
\be\label{chap2eq8}
\rho = \frac{ \rho_{0}}{a^3},
\ee
$\rho_{0}$ being a constant. So we have three equations to solve for four unknowns.
\par We assume the deceleration parameter as given in equation (\ref{chap2eq3}), which can be
integrated twice to give  $H = \dot{a}/a$ as in equation (\ref{chap2eq4}) and the scale
factor as
\be\label{chap2eq9}
a = [e^{-Apt}-1]^{-\frac{1}{p}}
\ee
\\ Now, the system of equations (\ref{chap2eq5}), (\ref{chap2eq6}) and 
(\ref{chap2eq7}) is closed with the assumption of equation (\ref{chap2eq3}) 
or equivalently equation (\ref{chap2eq4}). So in order to solve the system 
completely, the parameter $`p$' should have a fixed value. We choose 
$p=-\frac{3}{2}$, as it yields $q = 0.5$ for a very low value of 
$\frac{a}{a_0}$, where $a_0$ is the present value of the scale factor. 
The physical motivation for choosing this value of $q$ for early matter 
dominated epoch is that for a spatially flat FRW model with $p = 0$ without 
any Q-matter indeed has $q = 0.5$ and that the transition from radiation to 
matter dominated epoch for this value of $q$ is well studied \cite{ch2coles}.

 ~ With $p=-\frac{3}{2}$, equations (\ref{chap2eq5}) and (\ref{chap2eq6}) are 
used to eliminate  $V(\phi)$, and
$\dot{\phi}$ can be calculated to be
\be\label{chap2eq10}
\dot{\phi} = \sqrt{3} A a^{\frac {-3}{4}} = \frac{\sqrt{3} A}{[e^{\frac{3At}{2}}-1]^{\frac 1 2}}
\ee
\\
 The scalar field is found out by integrating equation (\ref{chap2eq10}) as,
\be\label{chap2eq11}
\phi = \frac {4}{\sqrt{3}}  tan^{-1}(e^{+3At/2}-1)^{\frac 1 2},
\ee

The potential $V(\phi)$ can also be calculated from equations 
(\ref{chap2eq5}) and (\ref{chap2eq6})
first as a function of time and by the use of equation 
(\ref{chap2eq11}) as a function
of $\phi$ as,
\be\label{chap2eq12}
V(\phi)  = \frac{9A^2}{2} cot^2(\frac{\sqrt{3}\phi}{4}) + 3A^2
\ee

Now one has the complete set of the solutions, $a = a(t)$, $\phi = \phi(t)$,
$\rho = \rho(t)$ and $V = V(\phi)$  for $p = -3/2$.  The solutions, when
plugged in the field equations, namely (\ref{chap2eq5}), (\ref{chap2eq6}) 
and (\ref{chap2eq7}), satisfy all of
them provided
\be\label{chap2eq13}
\rho_{0} = 3A^2.
\ee
\par From equations (\ref{chap2eq5}) and (\ref{chap2eq6}), we note that 
the contribution from the quintessence field $\phi$ towards the density 
and effective pressure are given as
\be\label{chap2eq14}
\rho_{\phi} = {\frac 1 2}{\dot{\phi}^2} + V(\phi),
\ee
and
\be\label{chap2eq15} 
p_{\phi} = {\frac 1 2}{\dot{\phi}^2} - V(\phi)
\ee
respectively. From these, one can write down the expressions for the dimensionless density parameters $\Omega_{m} = \frac{\rho_{m}}{3H^2}$ and $\Omega_{\phi} =
\frac{\rho_{\phi}}{3H^2}$ respectively for the visible matter and the $Q$-matter.\\

 \par
\begin{figure}[!h]
\centerline{\psfig{figure=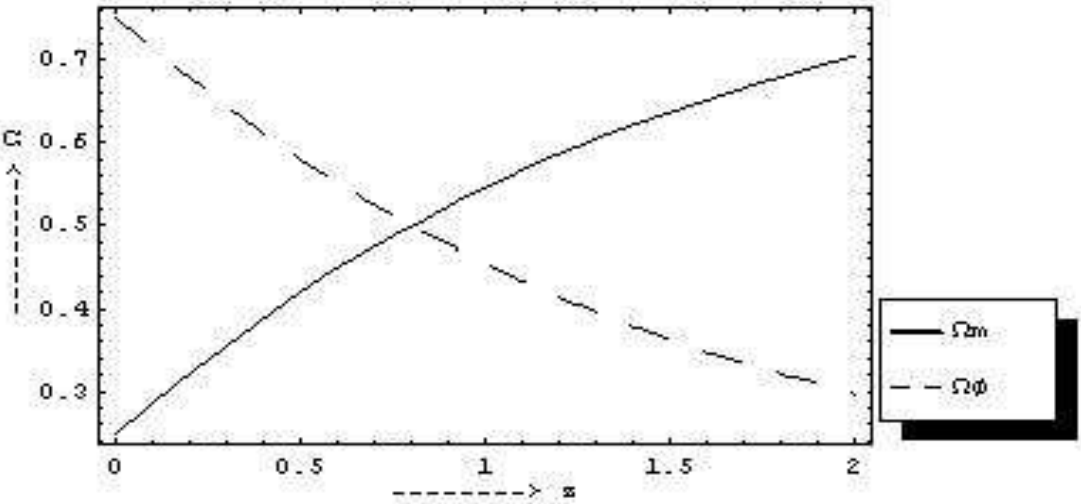,height=70mm,width=6.0in}}
\caption{\em Plot of $\Omega$ vs. $z$.}
\label{chap2second_fig}
\end{figure}

 With $ p = -3/2 $, the present model yields
\be\label{chap2eq16}
\Omega_{m} = \frac{(1+z){^3}}{[1+(1+z)^{-3/2}]^2} ~~,
\ee

and\\
\be\label{chap2eq17}
\Omega_{\phi} = 1 - \Omega_{m}~,
\ee
where $z$ is the redshift parameter given by
\be\label{chap2eq18}
1+z = \frac{a_{0}}{a}~,
\ee
 $a_{0}$ being the present value of the scale factor.  $\Omega_{m_{0}}$, the present
value of $\Omega_{m}$, comes out to be 0.25 and $\Omega_{\phi_{0}}$ = 0.75. 
These values are well within the constraints of 0.2 $\leq \Omega_{m_{0}} 
\leq$ 0.8  \cite{ch2ref},\cite{ch2vs}. Figure \ref{chap2second_fig} shows that 
$\Omega_{m}$ increases with $z$, i.e, decreases with the evolution of the 
universe. $\Omega_{\phi}$ starts dominating over $\Omega_{m}$ roughly at 
$z$ = 0.8. At the earlier epoch, i.e, at high $z$, $\Omega_{\phi}$ is very 
small, allowing a conducive matching onto the radiation era for the perfect 
ambience for nucleosynthesis. \\

\par Unlike Newtonian gravity, general relativity ensures that the 
pressure also contributes in driving the acceleration of the model. From 
equations (\ref{chap2eq5}) and (\ref{chap2eq6}), one has
\be\label{chap2eq19}
\frac{\ddot{a}}{a} = -\frac{1}{6} (\rho_{t} + 3 p_{t})
\ee
where $\rho_{t}$ and $p_{t}$ are the total effective density and pressure 
respectively. If the pressure is connected with the density by the relation
\be\label{chap2eq20}
p = w \rho,
\ee
the model will accelerate ( $\ddot{a} > 0$ ) only if
\be\label{chap2eq21}
w_{t} = \frac{p_{m} + p_{\phi}}{\rho_{m} + \rho_{\phi}} <  -\frac{1}{3}.
\ee

\begin{figure}[!h]
\mbox{\psfig{figure=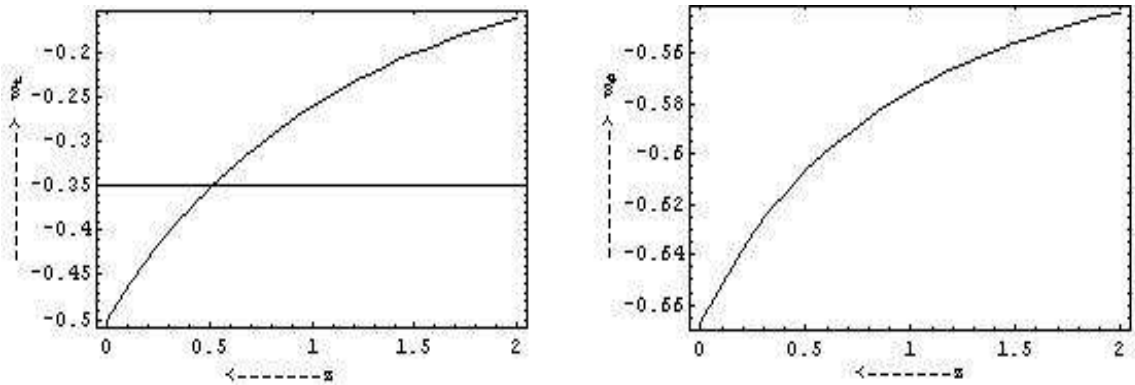,height=2.0in,width=6.0in}}
\caption{\em Plot of (a) $w_{t}$ vs. $z$  and (b) $w_{\phi}$ vs. $z$.}
\label{chap2third_fig}
\end{figure}

The subscripts `t', `m' and `$\phi$' stand for total, normal fluid distribution
and the Q-matter $\phi$ respectively. For a matter dominated universe, 
$p_{m} = 0$ and hence $w_{m} = 0$. This particular model gives
\be\label{chap2eq22}
w_{\phi} = \frac{p_{\phi}}{\rho_{\phi}} = \frac{1 + (1 + z)^{\frac {3} {2}}}{-1 - 2(1 + z)^{\frac{3}{2}}} 
\ee
\be\label{chap2eq23}
w_{t} = \frac{p_{\phi}}{\rho_{\phi} + \rho_{m}} = -[1 + {(1 + z)^{\frac{3}{2}}}]^{-1}
\ee
The evolution of $w_{t}$ versus $z$  and $w_{\phi}$ versus $z$ are shown in 
figure \ref{chap2third_fig}, which indicates that $w_{t}$ attains the 
required value of $-\frac{1}{3}$ or less only close to $z = 0.5$. Beyond 
that, $w_{t}$ is less negative, and the universe still decelerates.

\par The value of $w_{\phi0}$, i.e, the value of the equation of state 
parameter for the scalar field at $z = 0$ as given by the present model and 
as indicated by figure \ref{chap2third_fig}(b) is definitely within the 
constraint range \cite{ch2ref}.

\begin{figure}[!h]
\mbox{\psfig{figure=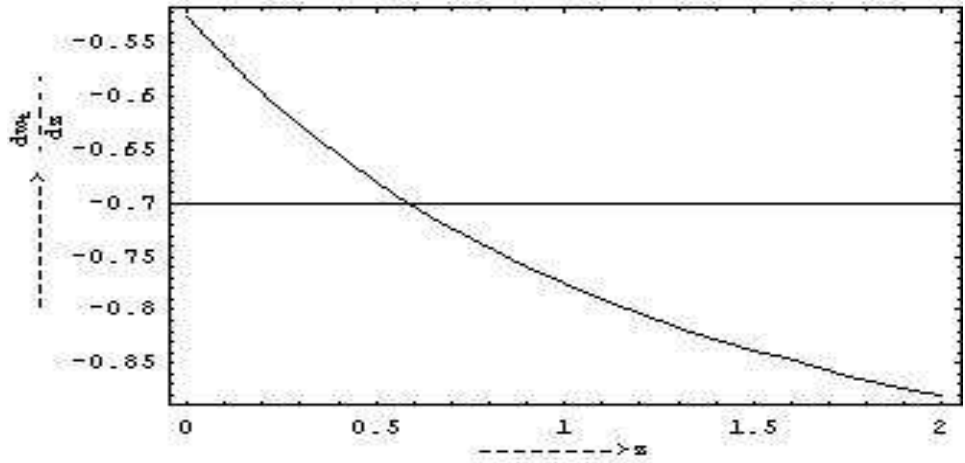,height=2.5in,width=5.0in}}
\caption{\em Plot of $\frac{dw_{t}}{dz}$ vs. $z$ for spatially flat dust 
dominated R-W model.}
\label{chap2fourth_fig}
\end{figure}
\par We also plot the rate of change of $w_{t}$ against $z$ ( as shown in 
figure \ref{chap2fourth_fig} ). It shows that $\frac{dw_{t}}{dz}$ is still 
negative at the present epoch, but the magnitude of $\frac{dw_{t}}{dz}$ is 
decreasing.
\par
\begin{figure}[!h]
\mbox{\psfig{figure=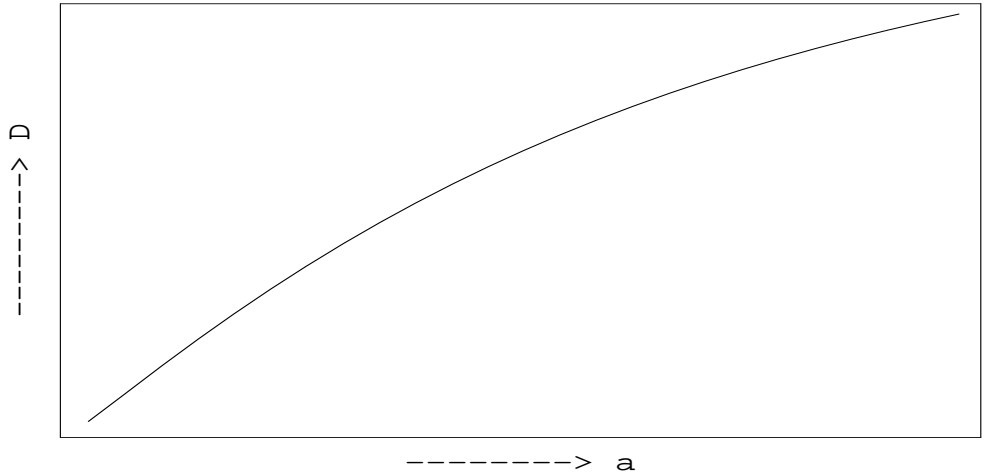,height=2.5in,width=5.3in}}
\caption{\em Plot of density contrast $D$ vs. $a$ where $ D = \frac{{\rho} -
     \bar{\rho}}{\bar{\rho}}$.}
\label{chap2fifth_fig}
\end{figure}

 The solution for the scale factor is good enough to allow the density 
contrast to grow favourably for the formation of large scale structure. 
Figure \ref{chap2fifth_fig} shows the growth of linearized density 
perturbation in this model and evidently indicates that it grows linearly 
with the scale factor during later stages as expected for the matter 
dominated epoch \cite{ch2coles}.

\par In the absence of the final form of the quintessence matter, search for 
the relevant form of potential will continue and the present investigation is
one of them. In view of the high degree of non-linearity of Einstein's 
equations, exact solutions always play a vital role as piecewise solutions 
have the problem of proper matching at different interfaces. The present 
model shows that inspite of the severe constraints imposed by observations,
one can still find an exact FRW model which gives values for the relevant
parameters like $q$, $w$, $\Omega_{\phi}$, $\Omega_{m}$ etc. safely within
the range given by observations. It also has the merit of having a single
analytical expression for $q = q(a)$, which gracefully transits from its 
positive phase to the negative one and adds to the list of quintessence
potentials that serve the purpose of modelling a presently accelerating
universe \cite{ch2v}. Definitely the model has problems, particularly that of
fine tuning, but infact, all quintessence models have some such problems.

\chapter{Complex Scalar Field as Quintessence}
\markright{Complex Scalar Field as Quintessence}
\newpage
\section{Spintessence : A Possible Candidate as a Driver of the Late Time Cosmic Acceleration \\ (Journal reference : N. Banerjee and S. Das, Astrophys. Space Sci., {\bf 305}, 25 (2006); gr-qc/0512036.)}
\newpage
\subsection{Introduction}
The stunning results of the observations on the luminosity - redshift 
relation of some distant supernovae \cite{ch3perl,ch3gar}, that the universe is 
currently undergoing an accelerated phase of expansion, poses a serious 
challenge for the standard big bang cosmology. As the standard gravitating 
matter gives rise to an attractive field only, this challenge is negotiated 
in the standard model by invoking some field which gives rise to an 
effective negative pressure. For many a reason, a dynamical `dark energy' 
is favoured against the apparently obvious choice of a cosmological 
constant $\Lambda$ for providing this negative pressure \cite{ch3varun}. 
This dynamical dark energy is called a quintessence matter. It cannot be 
overlooked that a successful explanation of the formation of structures in 
the early matter dominated era crucially requires an effectively attractive 
gravitational field and thus a decelerated phase of expansion must have 
been witnessed by an early epoch of matter dominated universe itself. Very 
recently Padmanabhan and Roy Chowdhury \cite{ch3padma} showed that the data 
set, only showing the accelerated phase of expansion, can well be 
interpreted in terms of a decelerated expansion in disguise. The 
acceleration only becomes meaningful if the full data set shows 
deceleration upto a certain age of the universe and an acceleration after 
that (see also ref \cite{ch3amen}). In keeping with such theoretical 
requirements, actual observations indeed indicate such a shift in the 
mode of expansion - deceleration upto a higher redshift regime 
(about $z \sim 1.5$) and acceleration in more recent era, i.e, for lesser 
values of $z$ \cite{ch3riess}.\\
\par This observation indeed came as a relief, as the formation of galaxies 
could proceed unhindered in the decelerated expansion phase. It also 
requires all quintessence models to pass through certain fitness tests, 
such as the model should exhibit a signature flip of the deceleration 
parameter $q$ from a positive to a negative value in the matter dominated 
era itself. Quite a few quintessence models do exhibit such a signature 
flip of the deceleration parameter. One very attractive model was that of 
~`spintessence' proposed by Boyle, Caldwell and Kamionkowski \cite{ch3boyle}. 
It essentially works with a complex scalar field and a potential, which is 
a function of the norm of the scalar field. The scalar field, 
\be\label{chap3eq1}
\psi = \phi_{1} + i \phi_{2},
\ee

can be written as

\be\label{chap3eq2}
\psi = \phi e^{i\omega t},
\ee

i.e, the complex part is taken care of by a phase term. This kind of a scalar 
field is already known in the literature in describing a `cosmic string' 
\cite{ch3vilen} although it has to be noted that a cosmic string has a very 
specific form of the potential $V(\phi)$, whereas, the relevant form has to 
be found out for a quintessence model. Boyle et al discussed the so called
spintessence model in two limits separately, namely for a high redshift region
and also for a very low redshift region. Apparently this model gives exactly
 what is required, an acceleration for the low $z$ limit whereas a deceleration
for a high redshift limit. In the present investigation, it is shown
completely analytically that the model indeed works. For high value of the
scale factor $a$, the deceleration parameter $q$ is negative whereas for a low
value of $a$, $q$ is positive. The most important feature is that the value
of $a$, where the sign flip of $q$ takes place, can be analytically expressed
in terms of the parameters of the theory and constants of integration.
\subsection{Field Equations and Results}
If we take the scalar field as given by equation (\ref{chap3eq2}), 
Einstein's field
equations become
\be\label{chap3eq3}
3\frac{ \dot{a}^2}{a^2} = \rho + \frac{1}{2} \dot{\phi}^2 + \frac{1}{2}
{\omega}^2 {\phi}^2 + V(\phi)~,
\ee
\be\label{chap3eq4}
2\frac{\ddot{a}}{a} + \frac{ \dot{a}^2}{a^2} = - \frac{1}{2} \dot{\phi}^2 
-\frac{1}{2} {\omega}^2 {\phi}^2 + V(\phi),
\ee

where $a$ is the scale factor, $\rho$ is the energy density of matter, $V$ is
the scalar potential which is a function of amplitude $\phi$ of the scalar field, and the phase $\omega$ is taken to be a constant. An overhead dot implies
differentiation w.r.t. the cosmic time $t$. In a more general case,
$\omega$ might have been a function of time.
\par The matter distribution is taken in the form of dust where the
thermodynamic pressure $p$ is equal to zero. This is consistent with the
`matter dominated' epoch. This leads to the first integral of the matter
conservation equation as
\be\label{chap3eq5}
\rho = \frac{ \rho_{0}}{a^3}.
\ee

Variation of the relevant action with respect to $\phi$ yields the 
wave equation
\be\label{chap3eq6}
\ddot{\phi} + 3\frac{\dot{a}}{a}\dot{\phi} + V'(\phi) = {\omega}^2 \phi ,
\ee
~~~where a prime is a differentiation w.r.t. $\phi$. \\
 As the scalar field actually has two components, a third conservation 
equation is found as 
\be\label{chap3eq7}
\omega = \frac{A_{0}}{{\phi}^2 a^3}~,
\ee
$A_{0}$ being a constant.\\
  In the spintessence model described by Boyle et al \cite{ch3boyle}, it has 
been considered that $\omega$ is a very slowly varying function of time 
which indeed can be considered as a constant
over the entire period of dust dominated era.
So for constant $\omega$, equation (\ref{chap3eq7}) yields
\be\label{chap3eq8}
{\phi}^2 a^3 = A~,
\ee
$A$ being a positive constant. \\

It deserves mention that only two equations amongst (\ref{chap3eq5}),
 (\ref{chap3eq6}) and (\ref{chap3eq8}) are independent as any one of 
them can be derived from the
Einstein field equations with the help of the other two in view of the
Bianchi identities. So, we have four unknowns, namely, $a$, $\rho$, $\phi$ and
$V(\phi)$ and four equations, e.g, (\ref{chap3eq3}), (\ref{chap3eq4}) and 
two from (\ref{chap3eq5})-(\ref{chap3eq8}). So this is 
a determined problem and an exact solution is on cards without any input. 
From equations (\ref{chap3eq3}), (\ref{chap3eq4}) and (\ref{chap3eq5}) 
one can write
\be\label{chap3eq9}
\frac{\ddot{a}}{a} - \frac{ \dot{a}^2}{a^2} = -\frac{\rho_{0}}{2 a^3} - 
\frac{1}{2} \dot{\phi}^2 -\frac{1}{2} {\omega}^2 {\phi}^2
\ee
which takes the form
\be\label{chap3eq10}
a^3 \frac{dH}{dt} = -l -m H^2 ,
\ee
where $H = \frac{\dot{a}}{a}$, the Hubble parameter, \\
      $l = \frac{1}{2}(\rho_{0} + {\omega}^2 A)$ and
      $m = \frac{9 A}{8}$ are positive constants.

~ In deriving this, the equation (\ref{chap3eq8}) has been used to 
eliminate $\phi$ and
$\dot{\phi}$ in terms of $a$ and $\dot{a}$. Now we make a transformation
of time coordinate by the equation
\be\label{chap3eq11}
\frac{dx}{dt} = \frac{1}{a^3}.
\ee

\par As $\frac{dx}{dt}$ is positive definite ( $a$ is the scale factor and
cannot take negative values ), we find that $x$ is a monotonically increasing
function of $t$. So one can use $x$ as the new cosmic time without any loss
of generality and deformation of the description of events. \\
In terms of $x$, equation (\ref{chap3eq10}) can be written as
\be\label{chap3eq12}
\frac{dH}{dx} = -l -m H^2~,
\ee
which can be readily integrated to yield
\be\label{chap3eq13}
H = n ~ tan(\beta - mnx)~,
\ee
where $n^2 = \frac{l}{m}$, is a positive constant, and $\beta$ is a constant
of integration.
\par Now we use this $H$ as a function of the new cosmic time variable $x$ 
and find out the behaviour of the deceleration parameter $q$, which is 
defined as
\be\label{chap3eq14}
q = -\frac{\dot{H}}{H^2} - 1 =  -\frac{H^\dagger}{a^3H^2} - 1~,
\ee 
where a dagger indicates differentiation w.r.t. $x$.\\
Equation (\ref{chap3eq12}) now yields
\be\label{chap3eq15}
q = -1 +\frac{1}{a^3} \left( \frac{l}{H^2} + m \right) ~.
\ee

\par Both $m$ and $l$ are positive constants. So $q$ has a `zero', when
\be\label{chap3eq16}
{a_{1}}^3 = \frac{l}{H_{1}^2} + m~.
\ee
The suffix 1 indicates the values of the quantities for $q = 0$. Furthermore, 
equation (\ref{chap3eq15}) can be differentiated to yield 
\be\label{chap3eq17}
{\frac{dq}{da}~\vline}_{~1} = -\frac{1}{a_{1}^4}\left
     [\frac{l}{H_{1}^2} + 3m\right]~,
\ee
at the point $q = 0$, $a = a_{1}$ and $H = H_{1}$. In deriving the
equation (\ref{chap3eq17}), equations (\ref{chap3eq10}) and (\ref{chap3eq16}) 
have been used. The last equation
clearly shows that $q$ is a decreasing function of $a$ atleast when $q = 0$.
So $q$ definitely 
enters a negative value regime from a positive value at $a = a_{1}$ and
$H = H_{1}$. 
\par For the sake of completeness, the solution for the scale factor $a$
can be found out by integrating equation (\ref{chap3eq13}) as 
\be\label{chap3eq18}
a =\left[{\frac{m}{ ln\vline\frac{1}
            {cos( \beta - mnx )}\vline^3}}\right]^{1/3}~.
\ee
\subsection{Discussion}
So evidently, the spintessence model proposed by Boyle et al \cite{ch3boyle}
passes the `fitness test', the deceleration parameter enters into a negative
value in a ``finite past". In view of the high degree of non linearity of
Einstein equations, exact analytic solutions are indeed more dependable, and
the present investigation provides that in support of a spintessence model. 
Also, the constants of the theory ( such as $\omega$ ) and the constants of
integration ( such as $l$ and $m$ ) are still free parameters and hence
provides the `comfort zone' for fitting into the observational results.
Another feature of this study is that the results obtained are completely
independent of the choice of potential $ V = V(\phi)$. This feature provides a 
bonus, as different potentials, used as the quintessence matter, are hardly
well-motivated and do not have any proper physical background. Recently quite
a few investigations show that a complex scalar field indeed serves
the purpose of driving a late time acceleration \cite{ch3je}. But most of these
investigations either invoke the solution in some limit ( such as for a
large $a$ ) or use some tuning of the form of the potential. The present
investigation provides a better footing for them as it
shows the transition of $q$ analytically. It also deserves mention that  Bento,
Bertolami and Sen \cite{ch3anjan} showed the effectiveness of a Chaplygin gas
as a quintessence matter. It is interesting to note that under some
assumptions this kind of a fluid formally resembles a complex scalar field
as discussed in this work.

\chapter{Acceleration of the Universe in Scalar - Tensor Theories}
\markright{Acceleration of the Universe in Scalar - Tensor Theories}
\newpage
\section{A Late Time Acceleration of the Universe with Two Scalar Fields : Many Possibilities \\ (Journal reference : N. Banerjee and S. Das, Mod. Phys. Lett. A, {\bf 21}, 
      2663 (2006); gr-qc/0605110.)}
\newpage
\subsection{Introduction}
The present cosmic acceleration is now generally believed to be a
certainty rather than a speculation. The recent data on supernovae of
type Ia suggested this possibility quite strongly \cite{ch4riess, ch4perl1,
ch4perl2, ch4tonry} and
the most trusted cosmological observations, namely that on Cosmic
Microwave Background Radiation \cite{ch4mel, ch4lange, ch4jaffe, ch4netter, 
ch4halver},
appear to be quite
compatible with an accelerated expansion of the present universe.
The natural outcome of these observations is indeed a vigorous search
for the form of matter which can give rise to such an expansion,
as a normal matter distribution gives rise to an attractive gravity
leading to a decelerated expansion. This particular form of matter,
now popularly referred to as ``dark energy", is shown to account for
as much as 70\%  of the present energy of the universe.
This is also confirmed by the highly accurate Wilkinson
Microwave Anisotropy Probe (WMAP) \cite{ch4bridle, ch4ben, ch4hin, ch4kog, 
ch4sper}.
A large number 
of possible candidates suitable as a dark energy component have
appeared in the literature. Excellent reviews on this topic are available
\cite{ch4sahni1, ch4sahni2, ch4peddy}. Most of the dark energy candidates
are constructed so as to generate an effective pressure which is sufficiently
negative driving an accelerated expansion. The alleged acceleration can
only be a very recent phenomenon and must have set in during the
late stages of the matter dominated expansion of the universe.
This requirement is crucial for the successful nucleosynthesis in
the radiation dominated era as well as for a perfect ambience for
the formation of structure during the matter dominated era. Fortunately, 
the observational evidences are also strongly in favour of a scenario
in which the expansion of the universe had been decelerated
(the deceleration parameter $q > 0$) for high redshifts and
becomes accelerated ($q < 0$) for low values of the redshift
$z$ \cite{ch4ag}. So the dark energy sector should have evolved in
such a way that the consequent negative pressure has begun dictating
terms only during a recent past.
\par This so easily reminds one about the inflationary universe models
 where an early accelerated expansion was invoked so as to
 wash away the horizon, fine tuning and some other associated
problems of the standard big bang cosmology. The legendary
problem, that the inflationary models themselves had, was that of a
``graceful exit"  -  how the accelerated expansion gives way to
the more sedate ($q < 0$) expansion so that the universe could
look like as we see it now. For a very comprehensive review, we
refer to Coles and Lucchin \cite{ch4coles} or Kolb and Turner \cite{ch4kolb}.
The ``graceful exit" problem actually stems from the fact that
for the potentials driving inflation, the phase transition to the
true vacuum  is never complete in a sizeable part of the actual
volume of the universe. The attempts to get out of this problem
involved the introduction of a scalar field which slowly
rolls down its potential so that there is sufficient time available
for the transition of phase throughout the actual volume of the universe.
 The current accelerated expansion thus poses the problem somewhat
complementary to the graceful exit - that of a ``graceful entry".
\par The present work addresses this problem, and perhaps provides some
clue regarding the solution of the problem. The basic motivation stems
from existing literature on inflation. Mazenko, Wald and Unruh \cite{ch4wald}
showed that a classical slow roll is in fact invalid when the
single scalar field driving inflation is self interacting. 
It was also shown that a slow roll with a single scalar field
puts generic restrictions on the potentials driving inflation \cite{ch4adams}.
 This kind of problems led to the belief that for a successful
inflationary model, one needs to have two scalar fields \cite{ch4linde}.
The present work uses this idea of utilising two scalar fields,
one of them being responsible for the present acceleration of the
universe and is called the quintessence field. The second one
interacts nonminimally with the former so that the quintessence
field has an oscillatory behaviour at the early matter dominated epoch
but indeed grows later to dominate the dynamics of the more recent 
stages of the evolution. If such a behaviour is achieved, some clue
towards the resolution of the graceful entry problem or the coincidence
problem may be obtained. There are quite a few quintessence potentials
already in the literature \cite{ch4sahni1} which drives a late time 
acceleration, but none of them really has an underlying physics
explaining their genesis. As one is already hard pressed to find
a proper physical background of the quintessence field, it will be even 
more embarassing to choose a second field without any physical motivation.
Naturally, the best arena is provided by a scalar tensor theory, such as
Brans-Dicke theory, where one scalar field is already there in the
purview of the theory and is not put in by hand. It deserves mention
that the Brans-Dicke scalar field was effectively used in
``extended inflation" in order to get a sufficient slow roll of
the scalar field \cite{ch4la, ch4johri}. Later Brans-Dicke theory was used
for finding a solution of the graceful exit problem with a large
number of potentials \cite{ch4nb}, where the inflaton field evolves
to an oscillatory phase during later stages.
\par In the next section we write down the field equations in 
Brans-Dicke (B-D) theory with a quintessence field $\phi$,
the potential $ V(\phi)$ driving acceleration being modulated
by the B-D scalar field $\psi$ as $ V(\phi)\psi^{-\beta}$.
With a slow roll approximation, the conditions for an initially
 oscillating $\phi$, which grows only during later stages, are
found out for two examples, a power law expansion and an 
exponential expansion of the scale factor. The particular form of
$V(\phi)$ is quite irrelevant in this context, the conditions
only put some restrictions on the constants of the theory and
the parameters of the model. So the form of the potential is
arbitrary to start with, only the conditions on the parameters
and the `value' of $V(\phi)$ has to be satisfied and hence many
possibilities are opened up to accommodate a physically viable
potential as the driver of the late time acceleration. However,
in some cases, this could restrict the form of $V(\phi)$ as well.
In the last section, we make some remarks on the results obtained.
\subsection{A Model with a Graceful Entry}
The relevant action in Brans-Dicke theory is given by
\be\label{chap41eq1}
S =\int [\frac{\psi R}{16\pi G_{0}} - 
\omega\frac{{\psi_{,\mu}}{\psi^{,\mu}}}{\psi} -
\frac{1}{2}{\phi_{,\mu}}{\phi^{,\mu}} - U(\psi, \phi) +
\it{L}_{m}]\sqrt{-g} d^4x
\ee
where $G_{0}$ is the Newtonian constant of gravitation, $\omega$ is the
dimensionless Brans-Dicke parameter, $R$ is the Ricci scalar, $\psi$ and
$\phi$ are the Brans-Dicke scalar field and quintessence scalar fields
respectively. If we now choose $U(\psi, \phi)$ as $ V(\phi)\psi^{-\beta}$ as
explained, the field equations, in units where  $8{\pi} G_{0} = 1$ , can 
be written as
\be\label{chap41eq2}
3 H^2 + 3 H \frac{\dot{\psi}}{\psi} - \frac{\omega}{2}\frac{\dot{\psi}^2}
    {\psi^2} = V(\phi)\psi^{-(\beta + 1)} + \frac{\rho}{\psi}~,
\ee
\be\label{chap41eq3}
 3H \dot{\phi} + V'(\phi)\psi^{-\beta} = 0~,
\ee
\be\label{chap41eq4}
(2\omega + 3) (\ddot{\psi} + 3H\dot{\psi}) = (\beta - 4) V(\phi) 
  \psi^{-\beta} + \rho~,
\ee
where a dot represents differentiation with respect to time $t$ and a
prime represents differentiation with respect to the scalar field $\phi$.
As we require the potential $U(\psi, \phi)$ to grow with time so that
the effective negative pressure dominates at a later stage, $\beta$ should
be negative for a $\psi$ growing with time or positive for a $\psi$ decaying
with time. The field equations are written in the slow roll approximation, i.e,
where $\dot{\phi}^2$ and $\ddot{\phi}$ are neglected in comparison to
others. $ H = \frac{\dot{a}}{a}$ is obviously the Hubble parameter.
As we dropped the field equation containing stresses, we can use the
matter conservation equation as the fourth independent equation which 
yields on integration
\be\label{chap41eq5}
\rho = \frac{\rho_{0}}{a^3}~,
\ee
$\rho_{0}$ being a constant. This is so as the fluid pressure is taken
to be zero as we are interested in the matter dominated era.
\par Using the expression for $V(\phi)\psi^{-\beta}$ from equation 
(\ref{chap41eq2}) 
in equation (\ref{chap41eq4}), we can write
\be\label{chap41eq6}
(2\omega + 3)[\frac{\ddot{\psi}}{\psi} + 3H\frac{\dot{\psi}}{\psi}]
 = (\beta - 4)[ 3H^2 + 3H\frac{\dot{\psi}}{\psi}-\frac{\omega}{2}
\frac{\dot{\psi}^2}{\psi^2}] - ( \beta - 5)\frac{\rho_{0}}{a^3\psi}~.
\ee
If the scale factor $a$ is known, this equation can be integrated
to yield the Brans-Dicke field $\psi$.\\
Equations (\ref{chap41eq3}) and (\ref{chap41eq4}) yield
\be\label{chap41eq7}
\frac{V(\phi)}{V'(\phi)} = -\frac{(2\omega + 3)(\ddot{\psi} + 3H\dot{\psi}) 
  - \rho}{3(\beta - 4) H\dot{\phi}}~.
\ee
Hence, if we define\\
\be\label{chap41eq8}
f(\phi) = \int \frac{V}{V'}d\phi~,
\ee                

then 
\be\label{chap41eq9}
f(\phi_{0}) - f(\phi_{i}) = \int_{t_{i}}^{t_{0}} F(t)dt~,
\ee
where
\be\label{chap41eq10}
F(t) = -\frac{(2\omega + 3) (\ddot{\psi} + 3H\dot{\psi}) - \rho}{3(\beta - 4)H}
\ee
and subscripts `o'  and `i' stands for the present value and some
initial value, such as the onset of the matter dominated phase of
evolution.
\par For a given $ a = a(t)$, therefore, equation (\ref{chap41eq6}) can be used
to find $\psi$, which in turn, with equations (\ref{chap41eq8}) and 
(\ref{chap41eq9}) determines
$f(\phi)$. Now, these equations can be used to put bounds on the
values of derivatives of the potential, which would ensure that
the dark energy has an oscillating phase in the early stages.
\par The complete wave equation for the quintessence field $\phi$ is
\begin{center}
$\ddot{\phi} + 3H\dot{\phi} + V'(\phi) \psi^{-\beta} = 0$. \\
\end{center}
The condition for a small oscillation of $\phi$ about a mean value is
$V'(\phi) = 0$. This provides a kind of plateau for the potential which 
hardly grows with evolution and hence the dynamics of the universe is
practically governed by the B-D field $\psi$ and the matter density $\rho$.
We find the condition for such an oscillation of $\phi$ at some initial
epoch by choosing $V'(\phi) \approx  0$ which yields
$$
{\frac{\ddot{\phi}}{3H\dot{\phi}}\vline~}_{i} \approx -1~.
$$
During later stages, the universe evolves according to the equations
(\ref{chap41eq2}) - (\ref{chap41eq4}) where $V'(\phi) \ne 0$, and the 
scalar field slowly rolls along
the potential so that the quintessence field takes an active role in the 
dynamics and gives an accelerated expansion of the universe.
\par Two examples, one for power law and the other exponential expansion,
will be discussed in the present work.\\
\\
I. Power law expansion :-
\\
If $a = a_{0}t^n$ ~where $n > 1$,
 the universe expands with a steady acceleration, i.e, with a constant
negative deceleration parameter ~$q = -\frac{(n -1)}{n}$.
\\With this,
\be\label{chap41eq11}
H = \frac{\dot{a}}{a} = \frac{n}{t}~,
\ee
and equation (\ref{chap41eq6}) has the form
\be\label{chap41eq12}
c_{1}\frac{\ddot{\psi}}{\psi} + c_{2}\frac{\dot{\psi}}{\psi}\frac{1}{t}
+ c_{3}\frac{\dot{\psi}^2}{\psi^2} + c_{4}\frac{1}{t^2} +
c_{5}\frac{1}{t^{3n}\psi} = 0 ~,
\ee
$c_{i}$'s being constants given by,
\be\label{chap41eq13}
c_{1} = (2\omega + 3),~~~
c_{2} = 3n(2\omega - \beta + 7 ),~~~
c_{3} = \frac{\omega}{2}(\beta - 4),~~
c_{4} = -3n^2(\beta - 4),~~
c_{5} = (\beta - 5) \frac{\rho_{0}}{a_{0}^3}~.
\ee
The simplest solution for $\psi$ in equation (\ref{chap41eq12}) is
\be\label{chap41eq14}
\psi = \psi_{0} t^{2 - 3n}~,
\ee
$\psi_{0}$ being a constant.\\
The consistency condition for this is,
$$
(2\omega + 3)(2 - 3n)(1 - 3n) + 3n(2\omega - \beta + 7)(2 - 3n) +
\frac{\omega}{2}(\beta - 4)(2 - 3n)^2 - 3n^2(\beta -4) +
(\beta - 5)\frac{\rho_{0}}{a_{0}^3\psi_{0}} = 0~.
$$
From equations (\ref{chap41eq9}) and (\ref{chap41eq10}),
\be\label{chap41eq15}
f(\phi_{0}) - f(\phi_{i}) = D~(t_{i}^{2 - 3n} - t_{0}^{2 - 3n})~,
\ee
where $D$ is a constant involving $c_{i}$'s, i.e, $n$, $\omega$,
$\psi_{0}$, $\rho_{0}$ etc. given by
$$
D = \frac{(2\omega + 3)\psi_{0}(2 - 3n) -
\frac{\rho_{0}}{a_{0}^3}}{3n(\beta - 4)(2 - 3n)} ~.
$$
From the form of potential $V = V(\phi)$, $f(\phi)$ can be found
from the relation (\ref{chap41eq8}). So, the different constants will 
be related by equation (\ref{chap41eq15}).
\par If the quintessence field oscillates with small amplitude
about the equilibrium at the beginning, i.e, at $t = t_{i}$ and
grows during the later stages of evolution, then 
${\frac{\ddot{\phi}}{3H\dot{\phi}}\vline~}_{i} \approx 1$, which
puts more constraints 
amongst different parameters of the theory.\\
Equations (\ref{chap41eq7}) and (\ref{chap41eq14}) yield
$$
\dot{\phi} = A~t^{(1 - 3n)}~(ln ~V)'~,
$$
where $A$ is a constant given by
$$
A = -\frac{(2\omega + 3)\psi_{0}(2 - 3n) -
\frac{\rho_{0}}{a_{0}^3}}{3n(\beta - 4)} = -D(2 - 3n)~.
$$
So,
$$
\frac{\ddot{\phi}}{3H\dot{\phi}} = -1 + \frac{1}{3n}
+\frac{A}{3n}~t^{2 - 3n} (ln ~V)'' ~.
$$
The condition for the small oscillation of $\phi$ close to 
$t = t_{i}$ is
$$
{\frac{\ddot{\phi}}{3H\dot{\phi}}\vline~}_{i} \approx -1~,
$$
which gives the condition on $(ln ~V)''$ as
\be\label{chap41eq16}
{(ln ~V)''\vline~}_{i} \approx -\frac{1}{A}~t_{i}^{3n - 2}.
\ee
\\
II. Exponential expansion :-
\\
\\Similarly for an exponential expansion at the present epoch,
the constraints can be derived. For such an expansion,
$$a = a_{0} ~e^{\alpha t}~,$$
where $a_{0}$, $\alpha$ are all positive constants.
\\Then,
$$
H = \frac{\dot{a}}{a} = \alpha ~~ and ~~
\rho = \frac{\rho_{0}}{a_{0}^3}e^{-3\alpha t}~.
$$
Equation (\ref{chap41eq6}) with this solution has the form
\be\label{chap41eq17}
b_{1}\frac{\ddot{\psi}}{\psi} + b_{2}\frac{\dot{\psi}}{\psi}
+ b_{3}\frac{\dot{\psi}^2}{\psi^2} + b_{4} +
b_{5}\frac{1}{e^{3\alpha t}\psi} = 0 ~,
\ee
where $b_{i}$'s are constants given by
\be\label{chap41eq18}
b_{1} = (2\omega + 3),~~~
b_{2} = 3\alpha(2\omega - \beta + 7 ),~~~
b_{3} = \frac{\omega}{2}(\beta - 4),~~~
b_{4} = -3\alpha^2(\beta - 4),~~
b_{5} = (\beta - 5) \frac{\rho_{0}}{a_{0}^3}~.
\ee
 A simple solution for $\psi$ is
\be\label{chap41eq19}
\psi = \psi_{0} e^{-3\alpha t}~,
\ee
$\psi_{0}$ being a constant.
The consistency condition for this is,
\be\label{chap41eq20}
9\alpha^2(\beta + \frac{\omega\beta}{2} - 2\omega - 4)
- 3\alpha^2(\beta - 4) + (\beta - 5)\frac{\rho_{0}}{a_{0}^3\psi_{0}} = 0~.
\ee
Equations (\ref{chap41eq9}) and (\ref{chap41eq10}) will put restrictions 
on the parameters of potential by
\be\label{chap41eq21}
f(\phi_{0}) - f(\phi_{i}) = \frac{\rho_{0}}{9\alpha^2a_{0}^3(\beta - 4)}
~[e^{-3\alpha t_{i}} - e^{-3\alpha t_{0}}]~.
\ee
The condition $\frac{\ddot{\phi}}{3H\dot{\phi}} \approx -1$ for the small
oscillation of the scalar field at $t = t_{i}$, with the help of equations
(\ref{chap41eq7}) and (\ref{chap41eq19}), 
leads to the interesting result
\be\label{chap41eq22}
{(ln ~V)''\vline~}_{i} \approx 0~,
\ee
which indicates that for an exponentially expanding present stage of evolution,
the quintessence potential behaves exponentially at least at the beginning.
\par Thus, if $V = V(\phi)$ is given, then equations (\ref{chap41eq15}) and 
(\ref{chap41eq21}) will 
help finding out the bounds on the values of the scalar field $\phi$ and
the constants appearing in $V(\phi)$ in terms of the initial and final epochs.
\par The conditions (\ref{chap41eq15}) or (\ref{chap41eq21}) put bounds on 
the potential so that the
quintessence field $\phi$ has an oscillatory behaviour in the beginning. As
the solutions in the examples considered has accelerated expansion,
the Q-field $\phi$ has a steady growth at later stages. For a power law
expansion, the growth has some arbitrariness  as $V(\phi)$ is not specified. 
For the exponential expansion, however, the growth of $\phi$ is governed
by an exponential potential.
\subsection{Conclusion}
It deserves mention that non-minimally coupled scalar fields were utilised by
Salopek et al\cite{ch4salopek} and Spokoiny \cite{ch4spokoiny1, ch4spokoiny2} 
in the
context of an early
inflation, where the field had an oscillation or a plateau at the end of
the inflation. In the present case, we need just the reverse for the Q-field,
and that is aided by the non-minimally coupled field $\psi$.
\par We see that for a wide range of choice of $V(\phi)$, a power
law acceleration is on cards, only the value of $V(\phi)$ at some
initial stage is restricted by equation (\ref{chap41eq16}). For an exponential
expansion of the scale factor, however, the potential $V(\phi)$
has to be an exponential function of $\phi$ ( in view of
equation (\ref{chap41eq22}) ). This investigation can be extended for more 
complicated kinds of accelerated expansion.
\par It is true that general
relativity is by far the best theory of gravity and the present
calculations are worked out in Brans-Dicke (B-D) theory, but this
should give some idea about how a second scalar field may be
conveniently used to get some desired results. Although B-D theory
lost a part of its appeal as the most natural generalization of
general relativity (GR) as the merger of B-D theory with GR for
large $\omega$ limit is shown to be somewhat restricted 
\cite{ch4ssen},
it still provides useful limits to the solution of the cosmological 
problems \cite{ch4la}. Another feature of the present work is that the
numerical value of $\omega$ required is not much restricted.
For power law expansion, $\omega$ is restricted by equation (\ref{chap41eq16}) 
which clearly shows that it can be adjusted by properly choosing
values of some other quantities, whereas for exponential expansion,
equation (\ref{chap41eq22}) shows that $\omega$ is arbitrary. 
This is encouraging
as it might be possible to get an acceleration even with a high value
of $\omega$, compatible with local astronomical observations \cite{ch4will}.
B-D theory had been shown to generate acceleration by itself \cite{ch4pavon},
although it had problems with early universe dynamics. B-D theory
with quintessence or some modifications of the theory 
\cite{ch4pav,ch4sesh, ch4aas, ch4sen, ch4ber}
were shown to explain the present cosmic acceleration, but all
these models, unlike the present work, required a very low value
of $\omega$, contrary to the local observations.

\vskip .2in

\newpage
\section{An Interacting Scalar Field and the Recent Cosmic Acceleration\\ (Journal reference : S. Das and N. Banerjee, Gen. Rel. Grav., {\bf 38}, 785 (2006); gr-qc/0507115.)}
\newpage 
\subsection{Introduction}
Over the last few years, the speculation that our universe is undergoing an
accelerated expansion has turned into a conviction. The recent observations
regarding the luminosity - redshift relation of type Ia supernovae 
\cite{ch4riess, ch4perl1, ch4perl2, ch4tonry}
and also the observations on Cosmic Microwave Background Radiation (CMBR)
\cite{ch4mel, ch4lange, ch4jaffe, ch4netter, ch4halver}  very 
strongly indicate this acceleration.
These observations naturally
lead to the search for some kind of matter field which would generate
sufficient negative pressure to drive the present acceleration. 
Furthermore, observations reveal that this unknown form of matter, popularly
referred to as the ``dark energy", accounts for almost 70\% of the present
energy of the universe. This is confirmed by the very recent Wilkinson
Microwave Anisotropy Probe (WMAP) data \cite{ch4bridle, ch4ben, ch4hin, 
ch4kog, ch4sper}. 
A large number of possible
candidates for this ``dark energy" component has already been proposed and
their behaviour have been studied extensively. There are excellent reviews 
on this topic \cite{ch4sahni2, ch42aas}.
\par It deserves mention that this alleged acceleration should only be a very
recent phenomenon and the universe must have undergone a deceleration
(deceleration parameter $q = -\frac{\ddot{a}/a}{\dot{a}^2/a^2} > 0$) in the
early phase of matter dominated era. This is crucial for the successful
nucleosynthesis as well as for the structure formation of the universe.
There are observational evidences too that beyond a certain value of the
redshift $z$ ( $z \sim 1.5$ ), the universe surely had a decelerated phase
of expansion \cite{ch4ag}. So, the dark energy component should have evolved
in such a way that its effect on the dynamics of the universe is dominant
only during later stages of the matter dominated epoch. A recent work by 
Padmanabhan and Roy Choudhury \cite{ch42paddy1, ch42paddy2} shows that 
in view of the error
bars in the observations, this signature flip in $q$ is essential for the 
conclusion that the present universe is accelerating.

\par So, we are very much in need of some form of a field as the candidate for
dark energy, which should govern the dynamics of the universe in such a
way that the deceleration parameter $q$
was positive in the early phases of the matter dominated era and becomes
negative during the later stages of evolution. One of the favoured
choices for the ``dark energy" component is a scalar field called
a quintessence field ( Q-field ) which slowly rolls 
down its potential such that
the potential term dominates over the kinetic term
and thus generates sufficient 
negative pressure for driving the acceleration. A large number of quintessence
potentials have appeared in the literature and their behaviour have been
studied extensively ( for a comprehensive review, see \cite{ch4sahni1} ). 
However,
most of the quintessence potentials do not have a proper physical background
explaining their genesis. In the absence of a proper theoretical plea for 
introducing a particular Q-field, non-minimally coupled
scalar field theories become attractive for carrying out
the possible role of the driver of the late time 
acceleration. The reason is simple; the required scalar field is already there
in the purview of the theory and does not need to be put in by hand.
Brans - Dicke theory is arguably the most natural choice as the
scalar - tensor generalization of general relativity (GR) 
because of its simplicity
and a possible reduction to GR in some limit. 
Obviously Brans - Dicke (BD) theory or its
modifications have already found some attention as a 
driver of the present cosmic acceleration \cite{ch4pav, ch4sesh, 
ch4aas, ch4sen, ch4ber, ch42eno} 
(see also \cite{ch42onemli1, ch42onemli2, ch42onemli3}).
It had
also been shown that BD theory can potentially 
generate sufficient acceleration in the matter dominated era even without any
help from an exotic Q - field \cite{ch4pavon}.
But this has problems with the required `transition' from 
a decelerated to an accelerated phase. Amongst other nonminimally
coupled theories, a dilatonic scalar field had also been considered as the
driver of the present acceleration \cite{ch42piazza}.
\par In most of the models the dark energy and dark matter components are
considered to be non-interacting and are allowed to evolve independently.
However, as the nature of these components are not completely known, the
interaction between them will indeed provide a more general
framework to work in. 
Recently, Zimdahl and Pavon \cite{ch42zimdahl1, ch42zimdahl2} have shown
that the interaction between 
dark energy and dark matter can be very useful in solving the coincidence
problem ( see also ref \cite{ch42soma1, ch42soma2, ch42soma3} ). 
Following this idea,
we consider an interaction 
or `transfer of energy' between the Brans - Dicke scalar field which is a
geometrical field and the dark matter. The idea of using a `transfer' of
energy between matter and the nonminimally coupled field had been used earlier
by Amendola \cite{ch42amen1, ch42amen2}. We do it specifically 
for a modified Brans - Dicke
theory.
The motivation for introducing this
modification of Brans - Dicke theory is the following. In the presence
of  matter and a quintessence field, with or without an interaction between
them, the evolution of net equation of state parameter $w$ plays a 
crucial role in driving a late surge of accelerated expansion. But WMAP
survey indicates that the time variation of $w$ may be very severely
restricted \cite{ch42jbp}. If the late acceleration is driven by an exchange of
energy between matter and a geometrical field $\phi$, the question of
the variation of $w$ would not arise.

\par We write down the Brans - Dicke field equations in the so called Einstein
frame. The field equations in this version look simpler and $G$ becomes a
constant. But one has to sacrifice the equivalence principle as the rest mass
of a test particle becomes a function of the scalar field \cite{ch42dicke}. 
So, the
geodesic equation is no longer valid and the different physical quantities
loose their significance. Nevertheless, the equations in this version of
the theory enables us to identify the energy contributions from different
components of matter. However, for final conclusions we go back to the
original atomic units where we can talk about the features with confidence.
We choose a particular form of the interaction and show that
a constant BD parameter
$\omega$ can not give us the required flip from a positive to a negative
signature of $q$ in the matter dominated era. We attempt 
to sort out this problem using a modified form of BD theory
where $\omega$ is a function of the scalar field $\phi$ \cite{ch42kn}. It has
been pointed out by  Bartolo and Pietroni \cite{ch42bartolo}
that a varying $\omega$ 
theory can indeed explain the late time behaviour of the universe.
 By choosing a particular functional form of $\omega$, we show that
in the interacting scenario, one can obtain a scale factor `$a$'
in the original version ( i.e, in atomic units ) of the theory so that the
deceleration parameter $q$ has the desired property of a signature flip
without having to invoke any quintessence field in the model. We also
calculate the statefinder pair  \{r,s\}, recently 
introduced by Sahni et al
\cite{ch42alam1, ch42alam2}, for this model. The statefinder 
probes the expansion dynamics of
the universe in terms of higher derivatives of the scale factor, i.e,
$\ddot{a}$ and $\atridot$. These statefinder parameters along with the SNAP
data can provide an excellent diagnostic for describing the properties of
dark energy component in future.

\subsection{Field Equations and Solutions}
 The field equations for a spatially flat Robertson - Walker spacetime in
Brans - Dicke theory are 
\be\label{chap42eq1}
3\frac{\dot{a}^2}{a^2} = \frac{\rho_{m}}{\phi} 
      + \frac{\omega}{2}\frac{\dot{\phi}^2}{\phi^2} - 3\frac{\dot{a}}{a}
                 \frac{\dot{\phi}}{\phi}~,
\ee
\be\label{chap42eq2}
2\frac{\ddot{a}}{a} + \frac{\dot{a}^2}{a^2} = -\frac{\omega}{2}
       \frac{\dot{\phi}^2}{\phi^2} - \frac{\ddot{\phi}}{\phi} -
              2\frac{\dot{a}}{a}\frac{\dot{\phi}}{\phi}~.
\ee

The field equations have been written with the assumption that at the present
epoch the universe is filled with pressureless dust, i.e, $p_{m} = 0$.
Here $\rho_{m}$ is the matter density of the universe, $\phi$ is the
Brans - Dicke scalar field, $a$ is the scale factor of the universe and
$\omega$ is the BD parameter. An overhead dot represents a 
differentiation with respect to time $t$.
\par The usual matter conservation equation has the form
\begin{center}
$\dot{\rho_{m}} + 3 H \rho_{m} = 0$ ~.
\end{center}
But here we consider an interaction between dark matter and the geometrical
scalar field and write down the matter conservation equation in the form
\be\label{chap42eq3} 
\dot{\rho_{m}} + 3 H \rho_{m} = Q~~,
\ee
such that the matter field grows or decays at the expense of the
BD field. The matter itself 
is not conserved here and the nature of interaction is determined by the
functional form of $Q$. We do not use the wave equation for the BD field
here because if we treat equations (\ref{chap42eq1}), (\ref{chap42eq2}) and 
(\ref{chap42eq3}) as independent equations,
then the wave equation comes out automatically as a consequence of the Bianchi
identity. It deserves mention that the wave equation will be modified to
contain $Q$ which will determine the rate of pumping energy from the BD field
to matter or vice-versa. This interaction term $Q$ is indeed a modification of
Brans - Dicke theory. But this interaction does not demand any nonminimal
coupling between matter and the scalar field $\phi$ and hence does not
infringe the geodesic equation in anyway. In this interaction, the rest 
mass of a test particle is not modified but rather a ``creation'' of matter
at the expense of the scalar field $\phi$ (or the reverse) takes place.
In a sense, it has some similarity with the ``C - field'' of the steady
state theory \cite{ch42hn}.  
\par In the Brans - Dicke theory, the effective gravitational constant is
given by $G = \frac{G_{0}}{\phi}$, which is indeed not a constant. Now, we
effect a conformal transformation 
\begin{center}
$\bar{g}_{\mu\nu} = \phi g_{\mu\nu}$~.
\end{center}
In the transformed version $G$ becomes a constant. However, this
transformation has some limitations which have been mentioned earlier. But
the resulting field equations look more tractable.
Equations (\ref{chap42eq1}) and (\ref{chap42eq2}) in the new frame look like
\be\label{chap42eq4}
3\frac{\dot{\bar{a}}^2}{\bar{a}^2} = \bar{\rho} + \frac{(2\omega + 3)}{4}
                        \dot{\psi}^2~,
\ee
\be\label{chap42eq5}
2\frac{\ddot{\bar{a}}}{\bar{a}} + \frac{\dot{\bar{a}}^2}{\bar{a}^2} = 
                   - \frac{(2\omega + 3)}{4} \dot{\psi}^2~,
\ee
and the matter conservation equation takes the form
\be\label{chap42eq6}
\dot{\bar{\rho}}_{m} + 3~\frac{\dot{\bar{a}}}{\bar{a}}~\bar{ \rho}_{m}
                                = \bar{Q}~~,
\ee
where an overbar represents quantities in new frame and $\psi = ln\phi$.
The scale factor and the matter density in the present version are related
to those in the original version as
\be\label{chap42eq7} 
\bar{a}^2 = \phi a^2~~~and ~~~\rho_{m} = \phi^2~\bar{\rho}_{m}~~.
\ee
Now, we choose the interaction $\bar{Q}$ of the form 
\be\label{chap42eq8}
\bar{Q} = -\alpha~\bar{H} \bar{\rho}_{m}~,
\ee
where $\alpha$ is a positive constant. This negative $\bar{Q}$ indicates a
transfer of energy from the dark-matter (DM) component to the geometrical
field $\phi$.\\
Equation (\ref{chap42eq6}) can be easily integrated with the help of 
equation (\ref{chap42eq8}) to yield
\be\label{chap42eq9}
\bar{\rho}_{m} = \rho_{0}~\bar{a}^{(-\alpha - 3)}~,
\ee
where $\rho_{0}$ is a constant of integration.\\
Then, equations (\ref{chap42eq4}) and (\ref{chap42eq5}) alongwith 
equation (\ref{chap42eq9}) has a solution
\be\label{chap42eq10}
\bar{a} = A \bar{t}^{2/(3 + \alpha)}
\ee
where $A$ is a constant given by
\begin{center}
$A = [\sqrt{\frac{\rho_{0}}{3 - \alpha}}(\frac{3 + \alpha}{2})]^
              {\frac{2}{3 + \alpha}}$~.
\end{center}
Some arbitrary constants of integration have been put equal to zero
while arriving at equation (\ref{chap42eq10}) for the sake of simplicity.\\
Using equations (\ref{chap42eq5}) and (\ref{chap42eq10}), one can easily 
arrive at the relation
\be\label{chap42eq11}
(\frac{2\omega + 3}{4})\dot{\psi}^2 = \frac{4\alpha}{(3 + \alpha)^2}
                        \frac{1}{\bar{t}^2}~.
\ee
\par If we consider a non-varying $\omega$, equation (\ref{chap42eq11}) 
will give rise to a
simple power law evolution of $\phi$. From equations (\ref{chap42eq7}) 
and (\ref{chap42eq10}), the scale
factor $a$ in atomic units will also have a power law evolution -
an ever accelerating 
or an ever decelerating model contrary to our requirement. This is consistent
with the exhaustive solutions in Brans - Dicke cosmology obtained by
Gurevich et al. \cite{ch42gurevich}, where the dust solutions are all power law.
This indicates that the chioce of constants of integration in
equation (\ref{chap42eq10}) does not generically change the model.
One way out of this
problem is to consider a generalization of Brans - Dicke theory where the
parameter $\omega$ is a function 
of the scalar field $\phi$ rather than a constant \cite{ch42kn}. An evolving
$\omega$ will be a contributory factor in determining the dynamics of the
universe.
\par We make a choice of $\omega$ as, 
\be\label{chap42eq12}
\frac{2\omega + 3}{4} = \frac{\alpha}{(3 + \alpha)^2}
                     \frac{\phi}{(\sqrt{\phi} - 1)^2}~.
\ee
Then, equation (\ref{chap42eq11}) can be integrated to yield 
\be\label{chap42eq13}
\phi = (1 - \phi_{0}\bar{t})^2~,
\ee
$\phi_{0}$ being a positive constant.
\par It deserves mention here that since $g_{00}$ and $\bar{g}_{00}$ are both
 equal to one, the time variable transforms as
\begin{center}
${d\bar{t}}^2 = \phi dt^2$~.
\end{center}
This along with equation (\ref{chap42eq13}) gives 
\be\label{chap42eq14}
\bar{t} = \frac{1}{\phi_{0}} \left[1 - \sqrt{2\phi_{0}(t_{0} - t)}~\right]~,
\ee
which is a monotonically increasing function of $t$ until $t = t_{0}$,
beyond which the model really does not work. So one can use $\bar{t}$
itself as the new cosmic time in the original version of the theory
without any loss of generality. So, for the sake of convenience, from
now onwards, we write $t$ in place of $\bar{t}$.
\par We transform the scale factor back to the original
units by equation (\ref{chap42eq7}), so that we 
are armed with the equivalence principle and can talk about the dynamics
quite confidently. \\
We have, 
\be\label{chap42eq15}
a = \frac{\bar{a}}{\sqrt{\phi}} = \frac{A~t^{2/(3 + \alpha)}}
  {(1 - \phi_{0}t)}~.
\ee

Also, the Hubble parameter and the deceleration parameter $q$ in the
original version comes out as,
\be\label{chap42eq16}
H = \frac{2}{3 + \alpha}\frac{1}{t} + \frac{\phi_{0}}{1 - \phi_{0}t}~,
\ee
\be\label{chap42eq17}
q = -1 + \frac{\frac{2}{3 + \alpha}(1 - \phi_{0}t)^2 - \phi_{0}^2t^2}
            {[\frac{2}{3 + \alpha}(1 - \phi_{0}t) + \phi_{0}t]^2}~.
\ee

From equations (\ref{chap42eq15}) and (\ref{chap42eq16}) it is 
evident that at $t \rightarrow
\frac{1}{\phi_{0}}$, both $a$ and $H$ blow up together giving a Big Rip. 
However, this rip has a different characteristic than that engineered by
a normal phantom field. In the latter, $\rho_{m}$ goes to zero but
$\rho_{DE}$ goes to infinity at the rip. In the present case, however, there
is no dark energy as such, and the scalar field is a part of geometry and
hence it is difficult to recognize its contribution to the energy density.
In the revised version, however, $\frac{2\omega + 3}{4}{\dot{\psi}}^2$ is
the contribution towards the stress tensor. It turns out that at
$t \rightarrow \frac{1}{\phi_{0}}$, this contribution remains quite finite.
So the big rip is brought into being by the interaction, and not by a
singularity in the stress tensor.
\begin{figure}[!h]
\label{first_fig}
\mbox{\psfig{figure=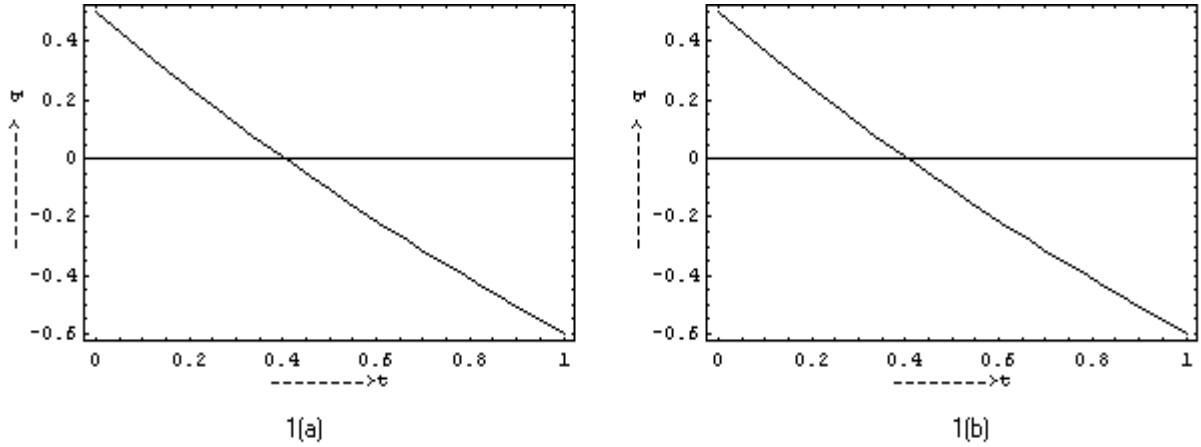,height=2.3in,width=6.333in}}
\caption{\em Figure 4.1(a) and 4.1(b) shows the 
plot of $q$ vs. $t$ for different values of $\alpha$. For figure 4.1(a)
we choose $\alpha = 0.0000003$  whereas for 
figure 4.2(b) we set the value as $\alpha = 0.0000001$.}
\end{figure}
\\
The plot of $q$ against $t$ ( figure 4.1 ) reveals that 
the deceleration
parameter indeed has a sign flip in the desired direction and indicates
an early deceleration ( $q > 0$ ) followed by a late time acceleration
( $q < 0$ ) of the universe. Also, the nature of the curve is not crucially
sensitive to the value of $\alpha$ chosen.
\par Equation (\ref{chap42eq5}) clearly indicates that 
$\ddot{\bar{a}}/\bar{a}$ is negative definite. So the signature 
flip in $q = -\frac{\ddot{a}/a}{\dot{a}^2/a^2}$ in the original Jordan 
frame must come from the time variation in $\phi$
via equation (\ref{chap42eq15}). Local astronomical experiments 
suggest that the
present variation of $G$ and hence that of $\phi$ has a very stringent
upper bound. It is
therefore imperative to check whether the present model is consistent with that
bound. For figure 4.1, the value of the constant of 
integration $\phi_{0}$ is
fixed at 0.3. With this value, and the age of the universe taken as $\sim$
15 Giga years, equation (\ref{chap42eq13}) yields
\begin{center}
${\vline~{\frac{\dot{\phi}}{\phi}~\vline}_{~0}}~\sim~10^{-10}$ per year, \\
\end{center}
which is consistent with the requirements of the local experiments 
\cite{ch4will}. 
The suffix $0$ indicates the present value. Also in this model, from equations
(\ref{chap42eq12}) and (\ref{chap42eq15}), we get as 
$\omega \rightarrow \infty$, $\phi \rightarrow 1$ and
$a \rightarrow t^{2/(3 + \alpha)}$. Therefore, for very small value of
$\alpha~ ( \sim 10^{-7})$, $a$ is indistinguishable from that in GR
$( a \sim t^{2/3})$. This is consistent with the notion that BD theory
yields GR in the infinite $\omega$ limit.
\subsection{Statefinder Parameters for the Model }
\par Recently Sahni et al. \cite{ch42alam1, ch42alam2} have 
introduced a pair of new
cosmological parameters \{r, s\}, termed as ``statefinder 
parameters". These parameters can effectively differentiate between
different forms of dark energy and provide a simple diagnostic
regarding whether a particular model fits into the basic observational
data. These parameters are 
\begin{center}
$r = \frac{\stackrel{...}{a}}{aH^3}$ and 
       $s = \frac{r - 1}{3(q - \frac{1}{2})}$~.
\end{center}
Accordingly, we find the statefinder parameters for the present model as
\be\label{chap42eq18}
r = 1 + \frac{3\phi_{0}^2 t^2 - 3\beta (1 - \phi_{0}t)^2}
         {[\beta + \phi_{0}t(1 - \beta)]^2} +
          \frac{2\beta (1 - \phi_{0}t)^3 + 2\phi_{0}^3 t^3}
              {[\beta + \phi_{0}t(1 - \beta)]^3}
\ee
and 
\be\label{chap42eq19}
s = \frac{3[\phi_{0}^2 t^2 - \beta(1 - \phi_{0}t)^2]
             [\beta + (1 - \beta)\phi_{0}t] + 2\beta(1 - \phi_{0}t)^3 +
                       2\phi_{0}^3t^3}
      {3[\beta + (1 - \beta)\phi_{0}t][-\frac{3}{2}{\lbrace \beta + (1 - \beta)
             \phi_{0}t \rbrace} + \beta(1 - \phi_{0}t)^2 - \phi_{0}^2t^2]^2}
\ee
where $\beta = \frac{2}{3 + \alpha}$.
\\
\begin{figure}[!h]
\begin{center}
\mbox{\psfig{figure=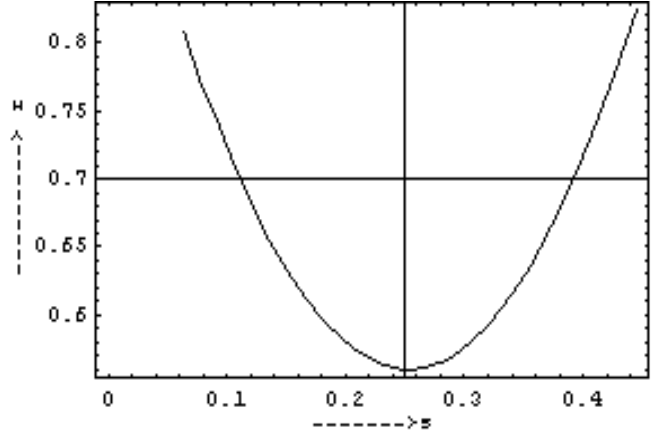,height=2.3in,width=3.5in}}
\caption{\em Plot of $r$ as a function of $s$ for $\alpha = .0000001$~.}
\label{second_fig}
\end{center}
\end{figure}
\\

\par If we now plot $r(s)$ for some small value of $\alpha$ ( $\alpha << 1$ ), 
we find that the nature of the curve is similar to the one expected for
scalar field quintessence models with equation of state parameter $w$ in
the range $-1 < w < 0$ \cite{ch42alam1, ch42alam2}.

\subsection{Discussion}
\par  Thus we see that for a spatially flat universe ($k = 0$), we can
construct a presently accelerating model in Brans - Dicke theory or more
precisely in a generalized version of it ( $\omega = \omega(\phi)$ )
if one considers an  interaction between 
dark-matter and the geometrical scalar field. It deserves mention that
for this interaction, the Lagrangian should also be modified by the
inclusion of an interference term between $\phi$ and $\it{L}_{m}$.
The salient feature of the
model is that no dark energy component is required. The nature
of the $q$ vs. $t$ curve 
is not crucially sensitive to small changes in the
 value of $\alpha$, the parameter which 
determines the strength of the interaction; only the time of `onset' of
acceleration would change by small amounts with $\alpha$. In revised unit,
this interaction can be switched off by putting $\alpha = 0$. However, in
Jordan frame it is not possible to switch off the interaction 
with this particular choice. If we put $\alpha = 0$, the scalar field itself
becomes trivial. In this frame, the conservation equation ( equation 
(\ref{chap42eq6}) along
with equation (\ref{chap42eq8}) )  transforms to
\be\label{chap42eq20}
\dot{\rho_{m}} + 3H\rho_{m} = - \left[\alpha H + 
               \frac{\alpha - 1}{2}\frac{\dot{\phi}}{\phi} \right]\rho_{m}~.
\ee
As one has both $H\rho_{m}$ and $\frac{\dot{\phi}}{\phi}\rho_{m}$ in the right 
hand side, the transfer of energy between matter and scalar field takes place
both due to the expansion of the universe and the evolution
of the scalar field. This equation shows that if
\begin{center}
$\phi = constant ~ a^{-2\alpha/(\alpha - 1)}$
\end{center}
the interaction vanishes. In this case, the transfer of energy due to
$H\rho_{m}$ and
$\frac{\dot{\phi}}{\phi}\rho_{m}$ cancel each other.
\par It is evident from equation (\ref{chap42eq11}) that if we 
consider the interaction
between dark matter and the geometrical field of the form considered in
equation (\ref{chap42eq8}), a constant $\omega$ will give rise to an 
ever accelerating 
or ever decelerating model and definitely we are not interested in that.
So, the idea of varying $\omega$ is crucial here
as it can very well serve the purpose of providing a signature flip in $q$
in this interacting scenario. It deserves mention that the specific
choice for the interaction in equation (\ref{chap42eq8}) and the choice 
of $\omega = \omega(\phi)$ in equation (\ref{chap42eq12}) are taken so as 
to yield the desired result.
This is indeed a toy model which simply shows that investigations regarding
an interaction amongst matter and the nonminimally coupled scalar field
is worthwhile.
\par From equation (\ref{chap42eq11}) it is also evident that 
$(2\omega + 3)$ has to be
positive definite, i.e, $\omega$ has to pick up some positive value or
at least $\omega$ should be greater than $-\frac{3}{2}$ in order to sustain a
consistent model. Also, the parameter $\omega$ does not have any stringent
limit and thus it may be possible to adjust the value of $\omega$ to some
higher value. Equation (\ref{chap42eq12}) indicates that if $\phi$ is 
very close to unity,
which is consistent with the present value of $G$, $\omega$ can attain a high
value at the present epoch. This is 
encouraging as it might be possible to obtain a model which exhibits early
deceleration and late time acceleration even with a high value of $\omega$,
compatible with the limit imposed on it by the 
solar system experiments \cite{ch4will}.
\par Also, from equation (\ref{chap42eq13}) it is evident that 
$\frac{\dot{\phi}}{\phi} < 0$
and $\frac{\ddot{\phi}}{\phi} > 0$. So, from equation (\ref{chap42eq2}), 
which is of
particular interest in studying the dynamics of the universe, we see that
the last term $2\frac{\dot{a}}{a}\frac{\dot{\phi}}{\phi}$ is negative and
is the key factor in driving the present acceleration of the universe.
This term basically provides the effective negative pressure and becomes
dominant during the later stages of evolution and drives an accelerated 
expansion. We have also calculated the statefinder parameters for the model
and show that the \{r, s\} pair mimics that of a quintessence model.
For this however, the constant $\alpha$ should be given a very small value
($\sim 10^{-7}$). In this
model we have considered a particular form of interaction which indeed is not
unique, and some complicated kind of interaction may lead to more viable
solutions of the various cosmological problems, particularly to a model
which is not restricted in future.
\par Although it is true that General Relativity (GR) is by far the best
theory of gravity and the natural generalisation of BD theory to GR for
large $\omega$ limit is shown to be restricted in some sense \cite{ch4ssen}, still
BD theory always seems to be ready to provide some useful
clues to the solution of various 
cosmological problems. The solution to the ``graceful exit" problem
of inflation
in terms of an `extended inflation' scenario \cite{ch4la, ch4johri} 
was first obtained in BD theory which provided hints towards the subsequent
resolutions of the problem in GR. Once again here, BD theory
in its own right could provide a model 
exhibiting the present cosmic acceleration without introducing any exotic
dark energy component and since nothing definite is known about the source
of this acceleration, this type of investigations may lead to some
track along which viable solutions may finally be arrived at. However, this is
a primitive model. This only shows that such investigations can be useful. It
remains to be seen if the solution is an attractor, and whether the model is
consistent with the structure formation.

\chapter{Curvature Driven Acceleration}
\markright{Curvature Driven Acceleration}
\newpage
\section{Curvature-driven Acceleration : a Utopia or a Reality?\\(Journal  reference : S. Das, N. Banerjee and N. Dadhich, Class. Quantum Grav.,{\bf23}, 4159 (2006).)}
\vspace{12cm}
[This work has been published in Classical and Quantum Gravity
[Class. Quantum Grav., {\bf 23}, 4159 (2006)]. A minor error in the published
version has been corrected in this version. The important conclusion that the
model shows a signature flip in $q$ at a finite past remains completely
unaltered.]
\newpage
\subsection{Introduction}
The search for a dark energy component, the driver of the present 
accelerated expansion of the universe, has gathered a huge momentum 
because the alleged acceleration is now believed to be a certainty, 
courtesy the WMAP data \cite{spergel}. As no single candidate enjoys 
a pronounced supremacy over the others as the dark energy component 
in terms of its being able to explain all the observational details 
as well as having a sound field theoretic support, any likely 
candidate deserves a careful scrutiny until a final unambiguous 
solution for the problem emerges. The cosmological constant $\Lambda$, 
a minimally coupled scalar field with a potential, Chaplygin gas or 
even a nonminimally coupled scalar field are amongst the most popular 
candidates ( see \cite{sahni} for a comprehensive review). Recently an 
attempt in a slightly different direction is gaining more and more 
importance. This effort explores the possibility of whether geometry in 
its own right could serve the purpose of explaining the present 
accelerated expansion. The idea actually stems from the fact that 
higher order modifications of the Ricci curvature $R$, in the form of 
$R^2$ or $R_{\mu\nu}R^{\mu\nu}$ etc. in the Einstein - Hilbert action
could generate an accelerated expansion in the very early 
universe \cite{kerner}.  As the curvature $R$ is expected to fall off 
with the evolution, it is an obvious question if inverse powers of 
$R$ in the action, which should become dominant during  the later stages, 
could drive a late time acceleration.
\par A substantial amount of work in this direction is already there 
in the literature. Capozziello et al. \cite{capoz} introduced an action 
where $R$ is replaced by $R^n$ and showed that it leads to an 
accelerated expansion, i.e, a negative value for the deceleration 
parameter $q$ for $n=-1$ and $n=\frac{3}{2}$. Carroll et al. \cite{carroll} 
used a combination of $R$ and $\frac{1}{R}$, and a 
conformally transformed version of theory, where the effect of the 
nonlinear contribution of the curvature is formally taken care of by 
a scalar field, could indeed generate a negative value for the 
deceleration parameter. Vollick also used this $1/R$ term in the 
action \cite{vollick} and the resulting field equations allowed an 
asymptotically exponential and hence accelerated expansion. The dynamical 
behaviour of $R^{n}$ gravity has been studied in detail by Carloni et.al 
\cite{carloni}. A remarkable result obtained by Nojiri and Odinstov 
\cite{nojiri} shows that it may indeed be possible to attain an inflation at 
an early stage and also a late surge of accelerated expansion from the same 
set of field equations if the modified Lagrangian has the form 
$\it{L} = R + R^m + R^{-n} $ where $m$ and $n$ are positive integers. 
However, the solutions obtained are piecewise, i.e, large and small values of 
the scalar curvature $R$, corresponding to early and late time behaviour of 
the model respectively, are treated separately. But this clearly hints 
towards a possibility that different modes of expansion at various stages of 
evolution could be accounted for by a curvature driven dynamics. Other 
interesting investigations such as that with an inverse $\sinh(R)$ 
\cite{borow} or with $\ln R$ terms \cite{odin} in the action are also 
there in the literature.
\par The question of stability \cite{dolgov} and other problems 
notwithstanding, these investigations surely open up an interesting 
possibility for the search of dark energy in the non-linear contributions 
of the scalar curvature in the field equations. However, in most of these 
investigations so far mentioned, the present acceleration comes either 
as an asymptotic solution of the field equations in the large cosmic 
time limit, or even as a permanent feature of the dynamics of the universe. 
But both the theoretical demand \cite{pt} as well as observations \cite{riess} 
( see also \cite{spergel} ) clearly indicate that the universe 
entered into its accelerated phase of expansion only very recently and 
had been decelerating for the major part of its evolution. So the deceleration parameter $q$ must have a signature flip from a positive to a negative value 
only in a recent past.
\par In the present work, we write down the field equations for a general 
Lagrangian $f(R)$ and investigate the behaviour of the model for two 
specific choices of $f(R)$, namely $f(R) = R - \frac{\mu^4}{R}$ and 
$f(R) = e^{-\frac{R}{6}}$. 
\par Although the field equations, a set of fourth order differential 
equations for the scale factor $a$, could not be completely solved 
analytically, the evolution of the `acceleration' of the universe could 
indeed be studied at one go, i.e, without having to resort to a 
piecewise solution. The results obtained are encouraging, both the 
examples show smooth transitions from the decelerated to the 
accelerated phase. In this work we virtually assume nothing regarding the 
relative strengths of different terms and let them compete in their own 
way, and still obtain the desired transition in the signature of the 
deceleration parameter $q$. This definitely provides a very strong support 
for the host of investigations on curvature driven acceleration, particularly 
those quoted in [5, 6, 7 and 8].
\par In the evolution equation, $q$ is expressed as a function of $H$, the 
Hubble parameter. This enables one to write an equation with only $q$ to 
solve for; as the only other variable remains is $H$ which becomes the 
argument. This method appears to be extremely useful, although it finds 
hardly any application in the literature. The only example noted by us 
is the one by Carroll et al. \cite{carr}, which, however, describes the 
nature only in an asymptotic limit.
\par In the next section the model with two examples are described and 
in the last section we include some discussion.
\noindent
\subsection{Curvature Driven Acceleration}
\par The relevant action is 
\be\label{chap5eq1}
\it A = \int \left[\frac{1}{16\pi G}f(R) + L_{m}\right]\sqrt{-g} d^4x ,
\ee
where the usual Einstein - Hilbert action is generalized by replacing $R$ 
with $f(R)$, which is an analytic function  of $R$, and $L_{m}$ is the 
Lagrangian for 
all the matter fields. A variation of this action with respect to the 
metric yields the field equations as 
\be\label{chap5eq2}
G_{\mu\nu} = R_{\mu\nu} - \frac{1}{2}Rg_{\mu\nu} = {T_{\mu\nu}}^{c}+ 
{T_{\mu\nu}}^{M} , 
\ee
where the choice of units $8\pi G = 1$ has been made. ${T_{\mu\nu}}^{M}$ 
represents the contribution from matter fields scaled by a factor of 
$\frac{1}{f'(R)}$ and ${T_{\mu\nu}}^{c}$ denotes that from the curvature 
to the effective stress energy tensor. ${T_{\mu\nu}}^{c}$ is actually given as
\be\label{chap5eq3}
{T_{\mu\nu}}^{c} = \frac{1}{f'(R)}\left[\frac{1}{2} g_{\mu\nu}
{(f(R) - R f'(R))} + 
      {f'(R)}^{;\alpha\beta} (g_{\mu\alpha} g_{\nu\beta} - 
        g_{\mu\nu} g_{\alpha\beta})\right]~.
\ee
The prime indicates differentiation with respect to Ricci scalar $R$.  
It deserves mention that we use a variation of (\ref{chap5eq1}) w.r.t. 
the metric tensor 
as in Einstein - Hilbert variational principle and not a Palatini variation 
where $A$ is varied w.r.t. both the metric and the affine connections. As 
the actual focus of the work is to scrutinize the role of geometry alone in 
driving an acceleration in the later stages, we shall work without any matter 
content, i.e, $L_{m} = 0$ leading to ${T_{\mu\nu}}^{M} = 0$. So for a 
spatially flat Robertson - Walker spacetime, where
\be\label{chap5eq4}
ds^2 = dt^2 - a^2(t) [ dr^2 + r^2 d\theta^2 + r^2 \sin^2\theta d\phi^2] ,
\ee
the field equations (\ref{chap5eq2}) take the form ( see \cite{capoz} ) 
\be\label{chap5eq5}
3\frac{\dot{a}^2}{a^2} = \frac{1}{f'}\left[{ \frac{1}{2}( f - Rf') - 
                       3\frac{\dot{a}}{a}\dot{R} f''}\right]~,~~~~~~~~~~~~~~~~
\ee
\be\label{chap5eq6}
2\frac{\ddot{a}}{a} + \frac{\dot{a}^2}{a^2} = -\frac{1}{f'}
     [{ 2\frac{\dot{a}}{a}
\dot{R} f'' + \ddot{R} f'' + \dot{R}^2 f'''  - \frac{1}{2}( f - Rf')}]~.
\ee
Here $a$ is the scale factor and an overhead dot indicates differentiation 
w.r.t. the cosmic time $t$. If $f(R) = R$, the equation (\ref{chap5eq2}) 
and hence (\ref{chap5eq5}) and (\ref{chap5eq6}) take the usual 
form of vacuum Einstein field equations. It should be 
noted that the Ricci scalar $R$ is given by 
\be\label{chap5eq7}
R = - 6\left[ \frac{\ddot{a}}{a} + \frac{\dot{a}^2}{a^2}\right]~,
\ee
and already involves a second order time derivative of $a$. As 
equation (\ref{chap5eq6}) 
contains $\ddot{R}$, one actually has a system of fourth order differential 
equations. 
\par It deserves mention at this stage that if $R$ is a constant, then 
whatever form of $f(R)$ is chosen except $f(R) = R$, equations 
(\ref{chap5eq5}) and (\ref{chap5eq6}) 
represent a vacuum universe with a cosmological constant and hence yield 
a deSitter solution, i.e, an ever accelerating universe. Evidently we are not 
interseted in that, we are rather in search of a model which clearly shows 
a transition from a decelerated to an accelerated phase of expansion of the 
universe. As we are looking for a curvature driven acceleration at late time, 
and the curvature is expected to fall off with the evolution, we shall take a 
form of $f(R)$ which has a sector growing with the fall of $R$. We work out 
two examples where indeed the primary purpose is served. 
\\
\\
(i) $f(R) = R - \frac{\mu^4}{R}.$
\\
\\
\par In the first example, we take 
\be\label{chap5eq8}
f(R) = R - \frac{\mu^4}{R}~,
\ee
where $\mu$ is a constant. Indeed $\mu$ has a dimension, that of 
$R^\frac{1}{2}$, i.e, that of $(time)^{-1}$.  This is exactly the form 
used by Carroll et al. \cite{carroll} and Vollick \cite{vollick}. 
Using the expression (\ref{chap5eq8}) in a 
combination of the field equations (\ref{chap5eq5}) and (\ref{chap5eq6}), 
one can easily arrive at the 
equation 
\be\label{chap5eq9}
2\dot{H} = \frac{1}{(R^2 + \mu^4)}\left[{ 2\mu^4 \frac{\ddot{R}}{R} - 
6\mu^4\frac{\dot{R}^2}{R^2} - 2\mu^4 H \frac{\dot{R}}{R}}\right]~, 
\ee
where $H = \frac{\dot{a}}{a}$, is the Hubble parameter. As both $R$ and $H$ 
are functions of $a$ and its derivatives, equation (\ref{chap5eq9}) looks 
set for 
yielding the solution for the scale factor. But it involves fourth order 
derivatives of $a$ ( $R$ already contains $\ddot{a}$ ) and is highly 
nonlinear. This makes it difficult to obtain a completely analytic solution 
for $a$. As opposed to the earlier investigations where either a piecewise 
or an asymptotic solution was studied, we adopt the following strategy. 
The point of interest is the evolution of the deceleration parameter 
\be\label{chap5eq10}
q = -\frac{a \ddot{a}}{\dot{a}^2} = -\frac{\dot{H}}{H^2} - 1 . 
\ee
So we translate equation (\ref{chap5eq9}) into the evolution equation 
for $q$ using 
equation (\ref{chap5eq10}) and obtain \\

$~~~~~~~~~~~~~~\mu^4\frac{\ddot{q}}{(q - 1)} - 3\mu^4 \frac{\dot{q}^2}{(q - 1)^2} 
- 6 \mu^4 H^2 {(q + 1)}^2 + 3\mu^4 H^2 (q + 1)$ 
\be\label{chap5eq11}
+~ 3 \mu^4 H \frac{(2q + 3)}{(q - 1)}\dot{q} + 36 H^6 (q - 1)^2 (q + 1) = 0 .
\ee
This equation, although still highly nonlinear, is a second order equation 
in $q$. But the problem is that both $q$ and $H$ are functions of time and 
cannot be solved for with the help of a single equation. However, they are not 
independent and are connected by equation (\ref{chap5eq10}). So we 
replace time derivatives 
by derivatives w.r.t. $H$ using equation (\ref{chap5eq10}) and write 
(\ref{chap5eq11}) as 
\be\label{chap5eq12} 
\frac{1}{3}(q^2 - 1)H^2 q^{\dagger\dagger} - \frac{2}{3} H^2 (q + 2) {q^
{\dagger}} ^2 - \frac{1}{3}H ( q - 1) (4q + 7) q^{\dagger} - (2q + 1)
{(q - 1)}^2 + H^4 ( q - 1)^4 = 0 . 
\ee
Here for the sake of simplicity $\mu^4$ is chosen to be 12 ( in proper units ), and a dagger represents a differentiation 
w.r.t. the Hubble parameter $H$. As $\frac{1}{H}$ 
is a measure of the age of the universe and $H$ is a monotonically decreasing 
function of the cosmic time, equation (\ref{chap5eq12}) can now be 
used as the evolution equation for $q$. The equation appears to be 
hopelessly nonlinear to give an 
analytic solution but if one provides two initial conditions, for $q$ and 
$q^{\dagger}$, for some value of $H$, a numerical solution is definitely 
on cards. We choose units so that $H_{0}$, the present value of $H$, is 
unity and pick up sets of values for $q$ and $q^{\dagger}$ for $H = 1$ 
( i.e, the present values ) from observationally consistent region \cite{alam} 
and plot $q$ versus $H$ numerically in figure 5.1. 
As the inverse of $H$ is the estimate 
for the cosmic age, `future' is given by $H < 1$ and past by $H > 1$. 
The plots speak for themselves. One has the desired feature of a 
negative $q$ at $H = 1$ and it comes to this negative phase only in the 
recent past. An important point to note here is that 
neither the nature 
of the plots, nor the values of $H$ at which the transition takes place, 
crucially depends on the choice of initial conditions, so the model is 
reasonably stable. 
\\
\begin{figure}[!h]
\centerline{\mbox{\psfig{figure=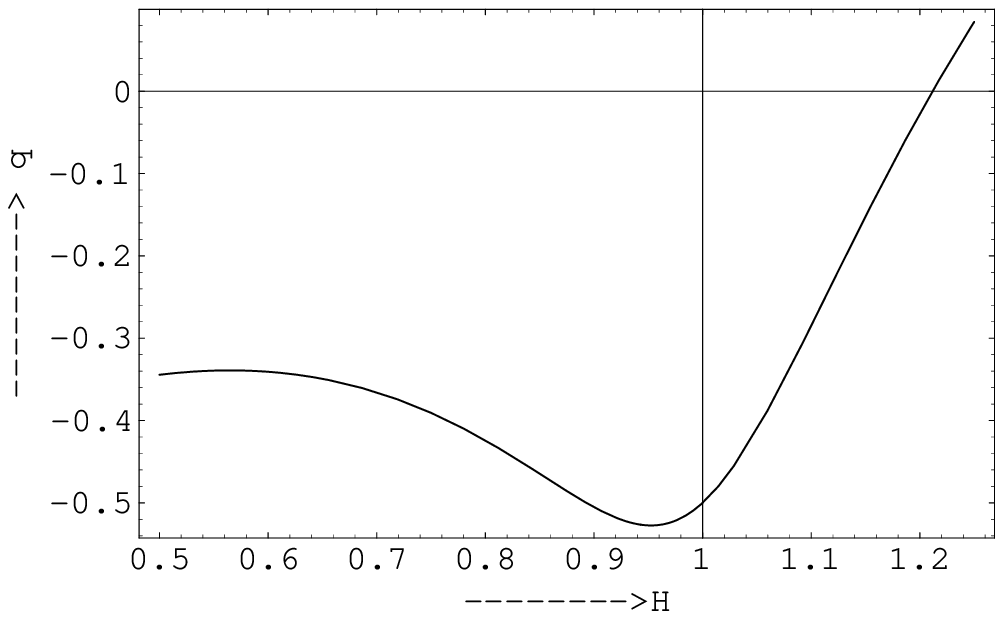,height=2.3in,width=3.0in}~~~ 
\psfig{figure=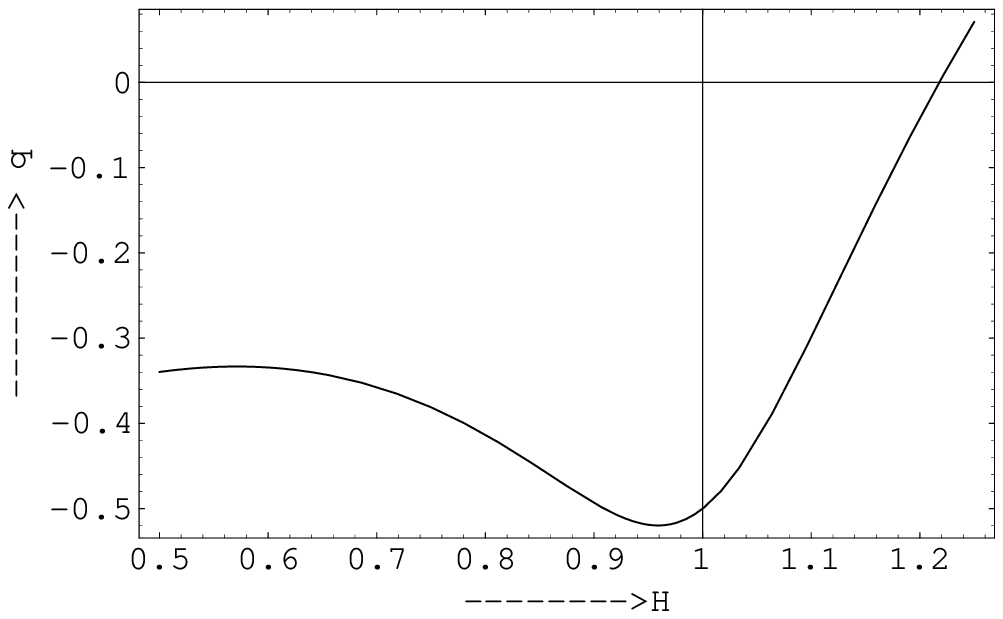,height=2.3in,width=3.0in}}}
\caption{\em Figure 5.1(a) and 5.1(b) shows the plot of $q$ vs. $H$ for
$f(R) = R - \frac{\mu^4}{R}$ for different initial conditions. For figure 
5.1(a) 
we choose the initial conditions as $q[1] = -0.5$, $q'[1] = 1.2$ whereas for 
figure 5.1(b) we set the initial conditions as $q[1] = -0.5$, $q'[1] = 1.0$.}
\label{first_fig}
\end{figure}
\\
(ii)  $f(R) = e^{-\frac{R}{6}}$ .
\\
\par In this choice, the function $f$ is monotonically increasing with $t$ 
as $R$ is decreasing with $t$. 
The field equations (\ref{chap5eq5}) and (\ref{chap5eq6}) have the form 
\be\label{chap5eq13}
3\frac{\dot{a}^2}{a^2} = -6\left[{ \frac{1}{2}(1 + \frac{R}{6}) - 
              \frac{1}{12}\frac{\dot{a}}{a}\dot{R} }\right]~,~~~~~~~~~~~~~~~~
\ee
\be\label{chap5eq14}
2\frac{\ddot{a}}{a} + \frac{\dot{a}^2}{a^2} = -6\left[{\frac{1}{2}(1 
                 + \frac{R}{6}) 
                        - \frac{1}{18}\frac{\dot{a}}{a}\dot{R} - 
                  \frac{1}{36}\ddot{R} + \frac{1}{216}\dot{R}^2 }\right] .
\ee
From these two equations it is easy to write 
\be\label{chap5eq15}
2\dot{H} = \frac{1}{6}\ddot{R} - \frac{1}{36}\dot{R}^2 - \frac{1}{6}H \dot{R} .
\ee
 Following the same method as before, the evolution of $q$ as a function of 
$H$ can be written as
\\

$H^4 (q + 1)q^{\dagger\dagger} + { [H^4 - (q + 1)H^6}] {q^{\dagger}}^2 
+ [ 2(q + 1)H^3 + 6qH^3 + 3H^3 - 4H^5( q^2 - 1)]q^{\dagger}$
\be\label{chap5eq16} 
+ 6 H^2 q^2 + 2 H^2 q - 4H^4 (q^2 - 1)(q - 1) - 8H^2 + 2 = 0~.
\ee

With similar initial conditions for $q$ and $q^{\dagger}$ at $H = 1$, 
the plot of $q$ versus $H$ ( figure 5.2 ) shows features similar to 
the previous example, the deceleration parameter $q$ has a signature 
change, from a positive to a negative phase in the recent past ( $H > 1$ ). 
This example has an additional feature that the universe re-enters a 
decelerated phase of expansion again in a near future ($H < 1$).
In this case also,  a small change in initial conditions hardly has any 
perceptible change in the graphs. 
\par In this case, a curvature singularity in a finite future is indicated. 
As $q = -\frac{\ddot{a}/a}{{\dot{a}}^2/a^2}$ and $H$ remains finite, 
$\frac{\ddot{a}}{a}$ and hence the curvature (via equation (\ref{chap5eq7})) 
has a 
singularity in a finite future. This is consistent with \cite{carroll} which 
indicates that a curvature quintessence may end up with three possibilities - 
an asymptotic de Sitter, a power law inflation or a curvature singularity in 
a finite future. The present case corresponds to the third possibility. The 
curves however show quite clearly that this singularity is not a `Big Rip' 
type, where due to continuous vigorous acceleration both $H$ and $a$ blow up 
in some finite future. Here the model clearly enters into a decelerating 
phase close to $H = 0.8$, as shown by the figure. 
\par From equation (\ref{chap5eq16}), one 
can also conclude that for very high value of $H$ (i.e, when the age of the 
universe was very small), $q \rightarrow -1$, which gives an early inflation. 
\begin{figure}[!h]
\centerline{\mbox{\psfig{figure=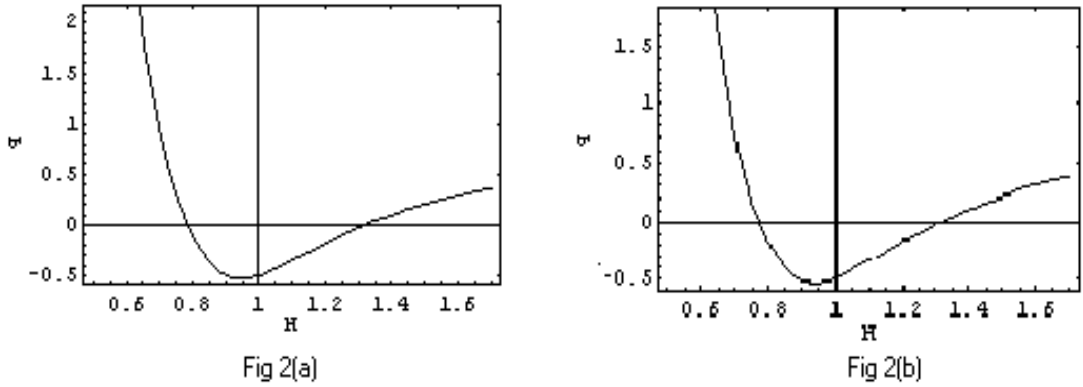,height=2.3in,width=6.0in}}}
\caption{\em Figure 5.2(a) and 5.2(b) shows the plot of $q$ vs. $H$ for
$f(R) = e^{-\frac{R}{6}}$ for different initial conditions. Here also for 
figure 5.2(a) and 5.2(b) we set the initial conditions as $q[1] = -0.5$, 
$q'[1] = 1.0$ and  $q[1] = -0.5$, $q'[1] = 1.2$ respectively.}
\label{second_fig}
\end{figure}

\par As the plots provide a sufficient data set, attempts could be made 
to find the closest analytical expression for $ q = q(H)$. These 
expressions are found to be polynomials. For example, a very close 
analytical expression for figure 5.2(b), within the accuracy of plots, is 
given as 
\be\label{chap5eq17}
  q = 47.95 H^6 - 335.73 H^5 + 991 H^4 - 1586.90 H^3 + 1459.20 H^2 
 - 729.73 H + 153.74  . 
\ee
This expression holds only when $H$ is reasonably close to one, and has 
nothing to do with other ranges of values of $H$.
\subsection{Discussion}
\par The present work indicates that by asking the question whether geometry 
in its own right can lead to the late surge of accelerated expansion, 
some feats can surely be achieved. Both the examples considered here 
indicate that one can build up models which start accelerating at the 
later stage of evolution and thus allow all the past glories of the 
decelerated model like nucleosynthesis or structure formation to remain 
intact. An added bonus of the second example is that the 
universe re-enters a decelerated phase in near future and the `phantom 
menace' is avoided - the universe does not have to have a singularity 
of infinite volume and infinite rate of expansion in a `finite' future.
\par It is of course true that a lot of other criteria have to be satisfied 
before one makes a final choice, and we are nowhere near that. Already 
there is a criticism of $\frac{1}{R}$ gravity that it is unsuitable 
for local astrophysics because of problems regarding stability \cite{dolgov}.
However, it was pointed out by Nojiri and Odinstov \cite{nojiri} that a 
polynomial may save the situation ( see also reference \cite{barrow} ). 
Our second example is exponential in $R$, 
i.e, a series of positive powers in $R$, and hence could well satisfy the 
criterion of stability. As already pointed out, although the choice of 
$f(R) = R - \frac{\mu^4}{R}$ is already there in the literature and 
served the purpose in a restricted sense than it does in the present work, 
the choice of $f(R) = e^{-\frac{R}{6}}$ has hardly any mention in the 
literature. The Lagrangian $f(R) = e^{-\frac{R}{6}}$ contains a cosmological 
constant as $f(R) \approx 1 - \frac{R}{6}$ for small $R$. So indeed one 
expects that it gives an accelerated expansion. But the interesting feature 
is that the same model gives an early inflation followed by a 
decelerated expansion, then an accelerated expansion around the present epoch 
and a decelerated phase once again in near future. 
\par It should also be noted that the present toy model deals with a
 vacuum universe and one has to either put in matter, or derive the relevant 
matter at the right epoch from the curvature itself. Some efforts 
towards this have already begun \cite{odinstov}.  
On the whole, there are reasons to be optimistic about a curvature-driven 
acceleration which might become more and more important in view 
of the fact that WMAP data could indicate a very strong constraint on the 
variation of the equation of state parameter $w$ \cite{paddy}.

\end{document}

%% file: cover.tex
\pagestyle{empty} 
	\begin{center} 
	\LARGE{\textbf{{ASPECTS OF \\QUINTESSENCE MATTER - \\THE DRIVER 
OF THE LATE TIME ACCELERATION OF THE UNIVERSE}}}\\
        \vspace{3.6cm} 
        \large{\textbf{THESIS SUBMITTED FOR THE DEGREE OF }}\\ 
	\large{\textbf{ DOCTOR OF PHILOSOPHY (SCIENCE) }}\\
        \large{\textbf{OF}}\\
        \large{\textbf{JADAVPUR UNIVERSITY}}\\
        \large{\textbf{2007}}\\
	\vspace{2.7cm}
	\Large{\textbf{SUDIPTA  DAS}}\\
	\normalsize{\textbf{DEPARTMENT OF PHYSICS}}\\
	\normalsize{\textbf{JADAVPUR UNIVERSITY}}\\
	\normalsize{\textbf{KOLKATA, INDIA}}
	\end{center} 
\newpage

%% file: certificate.tex
\pagestyle{empty} 
\begin{center}
\vspace*{3.8cm}
\Large{\underline{\textbf{CERTIFICATE FROM THE SUPERVISOR}}}\\ 
\end{center} 
\vspace{0.6in} 
\normalsize{This is to certify that the thesis entitled 
\emph{\bf {``Aspects of Quintessence Matter - The Driver of the Late Time 
Acceleration of the Universe"}}, submitted by {\textbf{Sudipta Das}} who 
got her name registered on {\textbf{23rd April, 2004}} for the award of 
{\textbf{Ph.D (Sciences)}} degree of {\textbf{Jadavpur University}}, is 
absolutely based upon her own work under the supervision of 
{\textbf{Dr. Narayan Banerjee}} and that neither this 
thesis nor any part of it has been submitted for any degree or any other 
academic award anywhere before. 
\vspace{.6in} 
\begin{flushright}
(Dr. Narayan Banerjee)\hspace*{2.0cm} \\ 
{Reader, Department of Physics}$~~~~~~~~$\hspace*{6.0cm}\\
Jadavpur University, Kolkata - 700 032~~~\\
India~~~~~~~~~~~~~~~~~~~~~~~~~~~~~\\
\end{flushright}  
\newpage

%% file: acknow.tex
{\huge\bf Acknowledgement}
\\
\\ 
\\
\\
At the very outset, I express my sincere gratitude to my supervisor Dr. 
Narayan Banerjee for his generous help, care, support and encouragement during 
my entire tenure as a research student. In the truest sense he is my 
friend, philosopher and guide. I owe to him whatever knowledge I have 
about General Relativity and Cosmology. 
I am also greatly indebted to him for teaching me the values of 
life and above all for guiding me to become a good human being. I shall always 
try to follow his advices. He has also coauthored in all the five papers. 
\par I am also grateful to Prof. A. Banerjee, Prof. S. B. Dutta Choudhury, 
Dr. S. Chatterjee and Dr. A. Sil of 
Relativity and Cosmology Research Centre, Jadavpur University for their help 
and valuable suggestions several times during my research period. I would 
also like to thank my co-researchers Sauravda, Mriganka, Mahuya, Koyel, Sumit 
and other members of Relativity and Cosmology Research Centre for their 
support and help. At this moment I should not forget my friends 
like Jyoti, Swapan, Vikram, 
Sujoy, Subhrojit, Apurba, Basab, Shibuda, Souvikda whose presence made my 
student life in Jadavpur University lively and cheerful. 
\par Special thanks are due to Prof. Naresh Dadhich of IUCAA, Pune and 
Dr. Anjan Ananda Sen of Jamia Milia Islamia, Delhi for useful discussions 
and suggestions. Prof. Dadhich has also co-authored in one of the papers. 
\par I would specially like to thank the authority and my colleagues of 
Ram Mohan Mission High School, Kolkata where I used to teach. They had been 
wonderful. I am greatly indebted to Sri Sujoy Biswas, Principal of the school, 
who has helped me a lot by granting me leave whenever I needed. Thanks are 
also due to my roommates for being so supportive.
\par I shall always cherish the wonderful moments I have spent in the 
University with my friends and colleagues. I shall remember their help and 
inspiration. 
\par I am also grateful to CSIR for providing financial support. 
\par I would like to express my deepest gratitude to my parents who are 
anxiously awaiting the completion of the thesis. It is their encouragement 
and blessings that enabled me to cross the hurdles of life. Thanks are also 
due to my sisters, my in-laws and all the members of my family for 
standing beside me.  
\par Last but not the least, thanks Pradipta for being so supportive and 
caring.
\\\\\\ 
Department of Physics, \hspace*{2.0in} \dotfill\\  
Jadavpur University, \hspace*{2.5in} {\bf Sudipta Das}\\
Kolkata 700032,\\
India.\\ 
%\hspace*{4.0in} \dotfill\\ 
%\hspace*{4.4in} {\bf Sudipta Das}  \\ 
\newpage

%% file: preface.tex
{\huge\bf Preface}
\\ 
\\
\par The last decade witnessed a radical change in cosmology - the science
of the universe. Cosmology now becomes an observation dependent science,
like other branches of physics. This is brought about by the high precision
observation techniques developed over the last few years. The great surprise 
that results from the high precision data is the inference 
that the universe is undergoing 
an accelerated expansion at present defying all intuitions. The search for 
the matter responsible for this unexpected behaviour of the universe 
provides one of the greatest excitements in contemporary theoretical physics. 
\par The present thesis is a collection of 
five papers based on my research on the 
problem of `dark energy', the driver of alleged present acceleration of the 
universe. All the five papers are published in international journals. 
\par The thesis is divided into five chapters. The first chapter is an 
introduction where the problem is defined and a survey of the work already 
in the literature has been made. A brief outline of the present work is 
also included in the introduction. 
\par The next four chapters include the actual work done. The reprints of 
the published papers are presented as the chapters or sections thereof. 
Only the last 
chapter contains a slightly improved version of the published paper. 
\\\\ 
\hspace*{4.0in} \dotfill\\ 
\hspace*{4.4in} {\bf Sudipta Das}  \\ 
\newpage